\documentclass[11pt,eqsecnum,preprint]{aastex}
\def\eq#1{\begin{equation} #1 \end{equation}}
\def\E#1{\hbox{$10^{#1}$}}
\def\dis      {\displaystyle}
\let \DS=\displaystyle
\def\x        {\hbox{$\times$}}
\def\about    {\hbox{$\sim$}}

\def\deg      {\hbox{$^\circ$}}
\newcommand{\call}	{{\cal L}}
\def\ltsimeq{\,\raise 0.3 ex\hbox{$ < $}\kern -0.75 em
 \lower 0.7 ex\hbox{$\sim$}\,}
\def\gtsimeq{\,\raise 0.3 ex\hbox{$ > $}\kern -0.75 em
 \lower 0.7 ex\hbox{$\sim$}\,}

\def\e{$^{-1}$}

\def\ee{$^{-2}$}

\def\eee{$^{-3}$}

\def\half {\hbox{$1 \over 2$}}
\let\ga=\gtsimeq
\let\la=\ltsimeq
\def\beq	{\begin{equation}}
\def\eeq	{\end{equation}}
\newcommand{\beqa}{\begin{eqnarray}}
\newcommand{\eeqa}{\end{eqnarray}}

\def\cc       {\hbox{cm$^{-3}$}}

\def\Lo       {\hbox{$L_\odot$}}
\def\kms      {\hbox{km~s$^{-1}$}}
\def\rate{{\rm cm^3~s^{-1}}}
\def\ac       {\hbox{$a_c$}}
\newcommand\aobs{A_{\rm obs}}
\def\Aobs     {\hbox{$A_{\rm obs}$}}

\def\asat     {\hbox{$a_{\rm sat}$}}
\def\dl       {\hbox{$d_\ell$}}
\def\dpar     {\hbox{$d_\|$}}
\def\dperp    {\hbox{$d_\perp$}}

\def\dthin	{\hbox{$d_{\rm thin}$}}

\def\Dv       {\hbox{$\Delta v_D$}}
\def\dvf      {\hbox{$\Delta v_{D5}$}}
\newcommand\fobs{F_{\rm obs}}

\def\hto      {\hbox{${\rm H_2O}$}}
\def\htwo      {\hbox{${\rm H_2}$}}
\newcommand{\liso}	{L_{\rm iso}}
\newcommand{\lisog}	{L_{{\rm iso},G}}

\def\Js       {\hbox{$J_s$}}

\def\ko       {\hbox{$\kappa_0$}}
\def\noo      {{n_{1,0}}}
\def\nto      {{n_{2,0}}}

\def\non{n_{1\nu}}
\def\ntn{n_{2\nu}}
\newcommand\oem{\Omega_{\rm em}}
\newcommand\oobs{\Omega_{\rm obs}}
\newcommand\osh{\Omega_{\rm shell}}
\newcommand\them	{\theta_{\rm em}}
\newcommand\thsd	{\theta_{\rm sd}}
\newcommand\thsh	{\theta_{\rm shell}}

\begin{document}
\title{ INTERSTELLAR H$_2$O MASERS FROM J SHOCKS}

\author{David Hollenbach}
\affil{NASA Ames Research Center, Moffett Field, CA 94035}
\author{Moshe Elitzur}
\affil{University of Kentucky, Lexington, KY 40506} \and
\author{Christopher F. McKee}
\affil{University of California, Berkeley, CA 94720}

\begin{abstract}
We present a model in which the 22 GHz H$_2$O masers observed in star-forming
regions occur behind shocks propagating in dense regions (preshock density
$n_0$ \about\ \E6--\E8 cm$^{-3}$).  We focus on high-velocity ($v_s
\ga 30$ \kms) dissociative J shocks in which the heat of H$_2$ re-formation
maintains a large column of \about\ 300--400 K gas; at these temperatures the
chemistry drives a considerable fraction of the oxygen not in CO to form
H$_2$O. The H$_2$O column densities, the hydrogen densities, and the warm
temperatures produced by these shocks are sufficiently high to enable
powerful maser action.
The observed brightness temperatures (generally
\about\ \E{11}--\E{14} K) are the result of coherent velocity regions that
have dimensions in the shock plane that are 10 to 100 times the shock
thickness of $\sim 10^{13}$ cm. The masers are therefore beamed towards the
observer, who typically views the shock ``edge-on", or perpendicular to the
shock velocity;  the brightest 
masers are then observed with the lowest line of sight velocities with respect to
the ambient gas.  We present numerical and analytic studies of the dependence
of the maser inversion, the resultant brightness temperature, the maser spot
size and shape, the isotropic luminosity, and the maser region magnetic field
on the shock parameters and the coherence path length; the overall result is
that in galactic H$_2$O 22 GHz masers these observed parameters can be
produced in J shocks with $n_0\sim $ \E6--\E8
cm$^{-3}$ and $v_s \sim 30$ --$200$ \kms. A number of key observables
such as maser shape, brightness temperature, and global isotropic luminosity depend
only on the particle flux into the shock, $j=n_0v_s$, rather than on $n_0$ and $v_s$ separately.
\end{abstract}


\section{Introduction}

Interstellar H$_2$O 22 GHz masers are associated with the earliest, most
embedded, phases of both low-mass and high-mass star formation, once the strong
protostellar outflows have commenced. In the low-mass case, Furuya et al (2001,
2003) find that while there are no masers in pre-protostellar cores, all class
0 protostars likely have water masers, with a lower fraction in class I and
none in class II. These masers often appear to be individual clumps, streaming
away from some center of activity  at velocities up to 200 km s$^{-1}$.
Individual features have apparent sizes of $\sim 10^{13}-10^{14}$ cm (Genzel
1986; Gwinn 1994a; Torrelles et al 2001a,b; Lekht et al 2007; Marvel et al 2008) and brightness
temperatures usually in the range $T_b\sim10^{11}-10^{14}$ K (Genzel 1986,
Gwinn 1994b). The brightness of the masers suggests they are saturated and
their observed linewidths ($\la 1$ km s$^{-1}$) suggest thermal temperatures
generally $\la 1000$ K (Liljestrom \& Gwinn 2000). The isotropic luminosity of
individual maser spots ranges from $\la 10^{-6}$ to 0.08 $L_{\odot}$ in the
Galaxy (Walker, Matsakis \& Garcia-Barreto 1982, Gwinn 1994a). The individual
maser spots are highly beamed toward the observer (Gwinn 1994c), so that the
observed flux of an individual spot measures an (assumed) isotropic luminosity
that is much higher than its actual luminosity.  Pumping by an external source
of radiation is ruled out by observations (e.g., Genzel 1986), and an internal
source of pump energy, such as the thermal energy produced in a shock, seems
required.

The development of powerful shocks in maser regions is inevitable in light of
the high velocities observed in the sources; Gwinn (1994a), Claussen et al
(1998), and Liljestrom \& Gwinn (2000) show, for example, that the vast
majority of maser spots in W49 and IRAS 05413-0104 have space velocities in
excess of 25 km s$^{-1}$.  The H$_2$O maser luminosity correlates with the
mechanical luminosity in the observed outflows or in the protostellar jets
(Felli, Palagi \& Tofani 1992, Claussen et al 1996, Furuya et al 2001), as
would be expected in a shock model. The source of excitation, then, appears to
be the interaction of the powerful outflows or jets from protostars in their
earliest, most embedded, phases of evolution with the dense gas that surrounds
them in this early stage --- either gas in disks or gas in the dense envelopes
that surround protostar/disk systems. Recent high angular resolution and proper
motion studies indicate groups of maser spots expanding away from the exciting
source, a geometry and dynamics highly suggestive of shocks (Gwinn 1994a;
Torrelles et al 2001a,b; Lekht et al 2007;  Marvel et al 2008; Goddi et al 2011; Moscadelli et al
2013).  In particular, the velocity vectors that Marvel et al find in several
sources are in close agreement with propagation in the plane of the sky, as
expected in shock-excited masers. The observed fluxes from the maser spots in
W49N that are within 10 km s\e\ of the systemic radial velocity can be up to an
order of magnitude greater than those from spots outside this velocity range
(Liljestrom \& Gwinn 2000), which again is consistent with excitation by shocks
propagating perpendicular to the line of sight.   Further evidence comes from
a recent survey by Walsh et al (2011), which finds that although maser emission
is spread over 350 km s$^{-1}$, 90\% of maser sites have a velocity spread of
less than 50 km s$^{-1}$.
The high-resolution observations of W49 and W3OH show that the maser features
outline the surfaces of elongated cocoons whose expansion is driven by twin
high-velocity (\about\ 1000 \kms), very young (a few hundred years) jets
(Elitzur 1995). In low-mass star-forming regions the ambient density can be
expected to be lower and the jets penetrate without creating complete shells,
leading to H-H objects and maser action only on working surfaces where they
generate local shocks. H$_2$O maser emission is also seen associated with jets
and their interaction with slower moving material around AGB stars (Imai et al
2002) and planetary nebulae (Miranda et al 2001, Uscanga et al 2008).  Recently,
a new class of six ``water fountain" pre-planetary nebulae have been found which
display bipolar structure with maser arcs aligned with high velocity outflows (Claussen et al
2009, Day et al 2010).   Finally, very strong H$_2$O
maser emission is observed in extragalactic sources where shocks may be
implicated such as the disks orbiting AGNs (Maoz \& McKee 1998 and references
therein), the jets from AGNs (Peck et al 2003, Tarchi et al 2011), and in
nearby star-forming galaxies (Darling et al 2008, Brogan et al 2010, Imai et al 2013).

Several authors besides ourselves have theoretically treated the possibility of
a shock origin for interstellar H$_2$O masers (Strelnitski 1973, 1980, 1984;
Schmeld, Strelnitski \& Muzylev 1976;  Kylafis \& Norman 1986; Tarter \&  Welch
1986). However, these models either lacked detail, required huge preshock
hydrogen densities ($> 10^9$ cm$^{-3}$) and therefore severe energy
requirements, or posited physically implausible electron and neutral
temperatures. We (Hollenbach, McKee \& Chernoff 1987, Elitzur, Hollenbach \&
McKee 1989, hereafter EHM, and Hollenbach, Elitzur \& McKee 1993) have proposed
a detailed shock model with calculated temperature and chemical structures and
with much less severe energy requirements. In our model the preshock gas has
hydrogen densities  $n_0 \ga 10^6$ cm$^{-3}$ and the masing occurs in {\it
postshock} gas with  densities $n_p\sim 10^8-10^9$ cm$^{-3}$.  One of our key
realizations was that the re-formation of molecular hydrogen in the postshock
gas provided a heating source that was essential in dissociative J shocks in
order to produce the columns of warm H$_2$O needed for observable interstellar
masers.  Our work has focussed on fast ($v_s\ga30-50$ km s$^{-1}$) dissociative
J shocks [see Draine (1980) for original definitions of J and C shocks], but we
have noted (EHM; Elitzur, Hollenbach, \& McKee 1992, hereafter EHM92) that
non-dissociative C shocks may also produce masers. Such shocks are limited to
velocities of $\la 40$ km s\e\ set by runaway ionization when relative velocity
between neutrals and ions in the shock becomes too large (Draine \& McKee 1993;
Flower \& Pineau des For{\^e}ts 2010). Just as in our J-shock models, in these
C shocks the H$_2$O is collisionally excited by warm neutral particles (H and
H$_2$) and the escaping H$_2$O infrared line radiation creates the inversions.
Kaufman \& Neufeld (1996) have modeled H$_2$O maser emission from such C
shocks, and applied their results to observations of 22 GHz and of other H$_2$O
maser transitions, especially in the submillimeter wavelength region (Menten,
Melnick, \& Phillips 1990; Cernicharo et al 1990; Menten et al 1990; Melnick et
al 1993). Detailed pumping calculations by Yates et al (1997) show that the 22
GHz masers have a broader range of physical conditions than the submillimeter
masers. According to these models there can be 22 GHz emission and no
accompanying 321 GHz masers in the same region, but not vice versa. Significant
submillimeter maser radiation requires emission regions at high temperatures
($\ga 1000$ K) that are more readily produced behind C shocks.

Because of observational challenges, the submillimeter masers have remained
relatively unstudied. Only recently, Patel et al (2007) imaged for the first
time the 321 GHz maser with the Submillimeter Array (SMA) in the Cepheus A
high-mass star-forming region, where they also mapped the 22 GHZ maser with the
Very Large Array (VLA). Nine submillimeter maser spots were detected and three
of them are associated with the centimeter masers spatially as well as
kinematically. In addition, there are 36 22 GHz maser spots without
corresponding submillimeter masers. These observations indicate that the
submillimeter masers are tracing significantly hotter regions (600--2000 K)
than the centimeter masers; 22 GHz masers that are not associated with the 321
GHz masers are likely to be arising in relatively cooler regions.

Overwhelmingly, the 22 GHz maser remains the most widely studied transition of
water, and in this paper we treat its production by J shocks. EHM demonstrated
numerically that J shocks with $n_0=10^7$ cm$^{-3}$, $v_s=100$ km s$^{-1}$
produce bright H$_2$O masers and gave analytic formulae for the dependence of
maser parameters on the shock parameters.  This paper extends that work by
providing a numerical study of the range of J shocks that produce H$_2$O
masers. We show that shocks in the range 10$^6$ cm$^{-3} \la n_0 \la 10^8$
cm$^{-3}$ and 30 km s$^{-1} \la v_s\la 200$ km s$^{-1}$  are likely to produce
the observed fast interstellar 22 GHz H$_2$O masers. EHM noted that this model
could apply to powerful extragalactic masers as well, and Maoz \& McKee (1998)
developed a detailed model for circumnuclear masers. In \S2 we model the H$_2$O
level populations  and the radiative transfer of the H$_2$O transitions in an
isothermal and isochoric slab such as is produced in dense, fast J shocks, and
we present analytic scalings and numerical results. In \S3 we describe the
physics incorporated into the shock code and present analytic and numerical
results for the shock structure.  In \S4 we apply the H$_2$O maser slab results
to the shock models in order to predict the 22 GHz brightness temperature $T_b$
of a single maser spot, its isotropic luminosity $L_{\rm iso}$, and the shape
and size of the maser spot as functions of the physical parameters $n_0$,
$v_s$, $v_A$ (the preshock Alfven speed -- a measure of the preshock magnetic
field), and the coherence pathlength $2\ell$ in the shock plane. We also
discuss the overall luminosity from a shocked masing region composed of many
maser spots (\S 4.3).  We compare the J-shock and C-shock masers in \S 4.4. In
\S5, we summarize our conclusions.

\section{H$_2$O MASER SLAB MODELS}

Fast, dense, dissociative J shocks produce a planar H$_2$ re-formation plateau
region that is nearly isothermal and isochoric. Here we model the H$_2$O maser
emission from a planar homogenous slab, deriving results applicable to {\it
all} collisionally pumped slabs, including C shocks, that may produce H$_2$O
maser emission (EHM; EHM92; Kaufman \& Neufeld 1996). These results are
combined with J-shock models in \S4.

A reasonably accurate description of maser emission under all circumstances is
provided by a standard formalism (see, e.g., Elitzur 1992; E92 hereafter) whose
essentials we reproduce in Appendix \ref{App:Basics} for completeness. In this
formalism the maser system is characterized by spatially constant effective
pump and loss rates that describe the interactions with all other levels, which
constitute the maser reservoir. The pump and loss rates are obtained from a
solution of the level populations for the full system (maser and reservoir) in
the absence of maser radiation (the unsaturated limit). We start with this
calculation, proceed to derive the parameters of the \hto\ maser pumping scheme
and apply these results to the planar geometry of shock-generated masers.

\subsection{Level Populations}

The populations per sub-level $n_i$ are conveniently expressed in terms of $y_i
= n_i/n(\hto)$, so that the overall population of level $i$ is $g_i y_i
n(\hto)$, where $g_i$ is the statistical weight. For each level $i$ included in
the pumping scheme, the steady state rate equation is
\eq{\label{eq:steady-state}
    g_iy_i\sum_j R_{ij} = \sum_j g_jy_j R_{ji},
}
where $R_{ij}$ is the overall rate coefficient ($\rm cm^3\,s^{-1}$) for
transitions from $i$ to $j$. Introduce the hydrogen nuclei density, $n$ =
$n({\rm H}) + 2n({\rm H_2})$. (We ignore the small contribution of collisions
with He.) Then the overall collision rate for the $i \to j$ transition is
$nq_{ij} = n({\rm H})q_{ij}^{{\rm H}} + n({\rm H_2})q_{ij}^{{\rm H_2}}$,
determined from the rate coefficients for collisions with H and H$_2$. Equation
(\ref{eq:steady-state}) is then converted into a relation between rates per
unit volume by multiplying both sides by $nn(\hto)$. While ignoring maser
radiation, we include the trapping of all other lines via a slab escape
probability formalism. External radiation, such as infrared continuum emission
from dust, is assumed negligible (recall that the dust is relatively cold in
shocks; see also \S3) and in this case
\eq{\label{eq:rates}
       R_{ij} = {\beta_{ij}A_{ij} \over n} + q_{ij}.
}
Here  $\beta_{ij}$ and $A_{ij}$ are, respectively, the escape probability and
$A$-coefficient for the $i \to j$ transition; the notation is such that
$A_{ij}=0$ if $i$ is a level below $j$ ($i < j$). The escape probability
$\beta_{ij}$ is a function of the transition optical depth at line center
\eq{\label{eq:tau}
  \tau_{ij} = {g_i A_{ij}\lambda^3 \over 8\pi^{3/2}}\,
              {n(\hto)d \over \Delta v_D}\,(y_j - y_i),
}
where $\lambda$ is the transition wavelength, $d$ is the source dimension that
controls photon escape (thickness in the slab geometry) and \Dv\ is the width
of the local (thermal and microturbulent) velocity field, which is assumed to
be of the form $\exp[-(v - v_0)^2/\Delta v_D^2]$; when the velocity field is
dominated by ordered motions across the slab, $\Dv/d$ is replaced by the
velocity gradient $dv_z/dz$. Therefore, for a given mix of H and H$_2$ the
distribution of populations $y_i$ depends only on the following three
parameters: $n$, $n(\hto)d/\Dv$ (which determines $\tau_{ij}$), and the
temperature $T$ (which determines $q_{ij}$). When a transition becomes
optically thick, $\beta_{ij} \propto 1/\tau_{ij}$ and the corresponding
radiative term in Equation \ref{eq:rates} does not depend separately on $n$ and
$\tau_{ij}$ but only on their product $n\tau_{ij} \propto n\,n(\hto)d/\Dv$
(EHM); that is, $R_{ij}$ depends only on temperature $T$ and the maser emission
measure $\xi$, defined as
\eq{\label{eq:xi}
    \xi \equiv {x_{-4}(\hto)n_9^2d_{13}\over\Delta v_{D5}}.
}
Here $n_9 = n/(\E9$ \cc), $x_{-4}(\hto)=x(\hto)/10^{-4}$ where $x(\hto)$ is the
abundance of H$_2$O molecules relative to hydrogen nuclei, $d_{13} = d/(\E{13}$
cm), and $\Delta v_{D5} = \Dv/(1\,\kms$). When the transition is thermalized,
as a result of further increase of either optical depth or density, the
$\xi$-dependence disappears too, and only the temperature dependence remains.

\subsection{Maser Pumping Scheme}

The populations of the 45 lowest rotational levels of ortho-H$_2$O were solved
for steady state from the set of rate equations in Equation\
(\ref{eq:steady-state}). Collision cross sections are not well known. Recent
calculations of \hto\ rotational excitations explored the low temperature
regime in cloud interiors (see Dubernet et al 2006 and references therein), but
the latest available tabulation at maser temperatures ($T \ga 200$ K) is from
Green et al (1993). We employ here these cross sections, assuming for
simplicity that hydrogen is purely molecular since separate atomic and
molecular hydrogen coefficients are not available.

From the solution for the normalized sublevel populations, $y_i$, we determine
the input properties of the standard maser pump model (see Appendix
\ref{App:Basics}). Denote by $m$ (= 1, 2) the two maser levels. All other
levels constitute the maser `reservoir', and interactions with the reservoir
levels populate each maser level at a pump rate per unit volume and sub-level,
$p_m$, and deplete it at the loss rate, $\Gamma_m$. These pump terms can be
read directly off Equation (\ref{eq:steady-state}),
\eq{
   \Gamma_m = n \sum_{j \ne m} R_{mj}, \qquad \qquad
   p_m = n\,n(\hto) \sum_{j \ne m} (g_j/g_m)\,y_jR_{jm};
}
note that the sums do not include transitions between the two maser levels,
which are handled separately (see Equations \ref{eq:m1} and \ref{eq:m2},
Appendix \ref{App:Basics}). Here we replace the loss rates of the two maser
levels, which our numerical results show to be slightly different, with the
common rate $\Gamma = (g_2\Gamma_2 + g_1\Gamma_1)/(g_2 + g_1)$. The three
quantities $p_1$, $p_2$ and $\Gamma$ fully describe the maser behavior under
all circumstances; in particular, the steady-state population of each maser
sub-level in the unsaturated regime is simply $p_m/\Gamma$. It is convenient to
replace the individual pump rates $p_2$ and $p_1$ with the rate coefficient for
their mean, $q$ (not to be confused with the collisional rate coefficients
$q_{ij}$ in Equation \ref{eq:rates}), and the inversion efficiency of the
pumping scheme, $\eta$, defined from
\eq{\label{eq:q_eta}
             \half (p_2 + p_1) \equiv n^2 x(\hto) q, \qquad
              \eta \equiv {p_2 - p_1\over p_2 + p_1}.
}
Note that this definition of $q$, which measures the pump rate per sublevel,
differs from that in EHM, which was per level; that is, $q_{\rm EHM} \propto
\half(g_2p_2 + g_1p_1)$. The functions $q$, $\eta$ and $\Gamma/n\ (= \sum_j
R_{mj})$ are expected to display the scaling property first noted in EHM: they
depend only on $\xi$ (and $T$) when the relevant rotational transitions become
optically thick at large \hto\ columns, with the density dependence totally
incorporated into $\xi$.

Figure 1 shows our results for the three pump parameters as functions of $\xi$
for a range of $n$ and $T$ relevant to observable masers. The scaling behavior
of $\Gamma/n$, $q$ and $\eta$ is evident from the left column panels, which
show their variation with $\xi$ at a fixed temperature. Even though the plots
span five orders of magnitude in density, all three quantities are largely
independent of $n$ whenever $\xi \ga 0.1$; scaling breaks down only for $\eta$
when $n \ga \E{10}$ cm$^{-3}$. Moreover, both $\Gamma/n$ and $q$ are further
independent of $\xi$ when $\xi \ga 0.1$, indicating that all level populations
are close to thermal equilibrium. The temperature variation of these pump
parameters, shown in the right-column panels, is well described by the simple
analytic approximations
\eq{\label{eq:q-G}
                    \Gamma_{-1}  \simeq 2.6n_9\, e^{-400/T}, \qquad
                    q_{-13} \simeq 3.2\, e^{-460/T},
}
where $\Gamma_{-1} = \Gamma/(\E{-1}\ \rm s^{-1})$ and $q_{-13} = q/(\E{-13}\
\rate)$. The accuracy of both expressions is within a few percent at all $T \ga
300$ K and $\xi \ga 0.1$; at $T$ = 200 K, the deviations reach only \about\
25\%.

As first noted by de Jong (1973), rotation levels on the ``backbone'' ladder
(the lowest level for each $J$) carry the bulk of the \hto\ population and
establish a thermal equilibrium among themselves. Levels off the backbone,
including the $J_{K-K+} = 6_{16}$ and 5$_{23}$ maser levels, are populated
predominantly by decays from higher backbone levels. This pattern leads to a
number of inverted transitions, with the 22 GHz having the longest wavelength
(1.35 cm) among them. Located 644~K above ground, the off-backbone 22 GHz maser
system contains such a tiny fraction of the \hto\ molecules (\about 1\% even at
the highest temperatures considered here) that it can be inverted with little
impact on the overall thermal distribution of level populations. The inversion
occurs because small contributions from radiative decays provide sufficient
competition with the collisions to maintain $p_2> p_1$ over a wide range of
parameters; we find that inversion is produced up to a density $n$ = 2\x\E{12}
\cc, although significant suppression of the maser line occurs for $n \ga
10^{10}$ cm$^{-3}$. The bottom panels of Figure 1 show the inversion efficiency
$\eta$. The right panel shows that $\eta$ is largely temperature independent
for $T > 200$ K, while the left panel displays the scaling property first noted
in EHM: when expressed as a function of $\xi$, $\eta$ is independent of density
as long as $n \la \E{10}$ cm$^{-3}$. Indeed, $\eta$ is well described over the
entire displayed range by the analytical approximation
\eq{\label{eq:eta}
         \eta_{-2} \simeq {4.5\over\xi^{0.5}}\, c_\eta
}
where $\eta_{-2} = \eta/\E{-2}$ and where the correction factor
\eq{\label{eq:ceta}
    c_\eta = \frac{1}{1 + 0.01\, n_9^{1.15}\xi^{-0.5}}\x
             \frac{1}{1 + 0.015\,n_9^{0.2}\xi^{1.5}}
}
displays explicitly the deviations from scaling and the thermalizing,
inversion-quenching effects of high densities and large optical depths. This
approximation reproduces the numerical results to within \about\ 20\% over the
entire phase space volume displayed in Figure 1, except for its very edge at
large $\xi$.

\subsection{Maser Geometry}
\label{sec:geometry}

The quantities $\eta$, $q$ and $\Gamma$ fully determine the pumping scheme,
enabling a complete solution of any maser model once its geometry is specified.
The planar geometry of the slab is the key to strong maser action. It allows
easy escape for the thermal photons through the slab thickness $d$, enabling
inversion everywhere. Simultaneously, maser amplification in the plane can
proceed along distance $ad$, where in principle the aspect ratio $a$ is
arbitrarily large but in practice is limited either by the curvature of the
shock or by the pathlength in the shock plane where velocity gradients shift
the component in the plane by the thermal width (see \S 5). The resulting
radiation is strongly beamed in the plane of the slab, and the strongest masers
will be seen from edge-on orientations. Indeed, Marvel et al (2008) find that
the outflows of the water maser associated with IRAS 4A/B in the star-forming
region NGC 1333 are nearly in the plane of the sky, with inclination of only
2\deg\ for IRAS 4A and about 13\deg\ for IRAS 4B.

The general solution of planar masers is presented in EHM92. Its essentials are
reproduced in Appendix \ref{App:planar}, together with a glossary of key
dimensions in Table \ref{Glossary}. The aspect ratio $a$ is the most important
geometrical property of such masers; for a circular disk with radius $\ell$ and
thickness $d$ (Figure 2) it is
\eq{\label{eq:a}
   a = {2\ell\over d}.
}
The shape of the masing material in the plane is largely irrelevant once the
maser saturates. For a circular \hto\ maser disk, the aspect ratio required to
bring about maser saturation is
\eq{\label{eq:asat}
   \asat \simeq 3.6{n_9\over\xi^{1/2}c_\eta}\,e^{60/T}
         \left[1 - {21\over T} + 0.12\ln\,{n_9\over\xi^{1/2}}\right],
}
obtained by inserting into the general expression for this geometry, reproduced
in Equation (\ref{eq:asat0}), the results of the \hto\ pumping scheme
(Equations \ref{eq:gammam} and \ref{eq:kod}). For comparison, the analogous
expression for a cylindrical maser with diameter $d$ is listed in Equation
(\ref{eq:asat-cyl}) in Appendix \ref{App:planar}. The two saturation aspect
ratios are nearly identical in \hto\ masers, the differences mostly involving
small logarithmic corrections. Since the saturation condition is the same for
the two extremes of planar geometrical shape, this ensures that maser
saturation is controlled solely by the length of the velocity coherent region.

With \ko\ the unsaturated absorption coefficient, the quantity $\asat\,\ko d$
is the maser optical depth along the disk diameter at saturation (cf Equation
\ref{eq:kod}); it is a measure of the amplification required along the maser
longest path in order to bring saturation. As is evident from Equation
(\ref{eq:asat0}), this quantity has similar values for all pumping conditions,
varying only logarithmically with the pumping parameters; some general
arguments show that the intrinsic properties of the \hto\ molecule imply
$\asat\,\ko d \sim$ 15 (E92; see also Eq.\ \ref{eq:asat0}). Strong masers can
be expected when saturation is reached at realistic elongations, i.e., moderate
aspect ratios \asat\ (\la\ 10). As we show below (see \S4, in particular Figure
11), J shocks produce \asat\ \la\ 5 over a large volume of parameter space,
ensuring strong maser action for a wide range of conditions. The near constancy
of \asat\ over such a large parameter region implies that $\ko d$ (roughly
proportional to 1/\asat; see Eq.\ \ref{eq:asat0}) too has only moderate
variation there.

As noted in EHM92, saturated masers can be distinguished by two types of
beaming. For {\it amplification-bounded} masers, whose prototype is the
spherical maser, the beaming angle depends on the amplification. These masers
are characterized by observed sizes significantly smaller than their projected
physical size. Furthermore, the observed size {\it increases} with frequency
shift from line center (Elitzur 1990). Such increases have been reported in a
recent study of \hto\ masers around evolved stars (Richards et al 2011). For
{\it matter-bounded} masers, whose prototype is the filamentary maser, the
beaming angle depends only on the geometry of the maser. They are characterized
by observed sizes that are equal to their projected physical size and constant
across the line profile. In principle, saturated planar masers produced by
shocks can display both types of behavior. For a shock moving across the line
of sight (shock velocity vector $\mathbf v_s$ in the plane of the sky), denote
by $\|$ the direction parallel to $\mathbf v_s$ and by $\perp$ the direction
orthogonal to both $\mathbf v_s$ and the line of sight. The dimension of the
masing medium along the $\|$-direction is the slab thickness, $d$, and the
dimension along the line of sight is $2\ell=ad$. Whereas $d$ is determined by
the structure of the shock, the dimensions of the masing medium in the two
directions in the slab plane are controlled by other factors, such as velocity
coherence and shock curvature. We term planar masers that are matter bounded in
the $\|$-direction ``thin," and those that are amplification bounded in that
direction ``thick."

Appendix \ref{App:planar} presents a detailed description of both thin and
thick disk masers, and Table \ref{Glossary} provides a glossary of maser
dimensions relevant for the two cases. As we shall see below, most interstellar
shocks are ``thin," with the maser structure as depicted in Figure 2. Let
$\dpar$ denote the observed size of the maser parallel to the shock velocity
and $d_\perp$ the observed size in the plane of the sky normal to the shock
velocity. A thin maser is matter bounded in the $\|$-direction and
amplification bounded in the $\perp$-direction, therefore $\dpar= d$ but
$\dperp$ is smaller than $ad$, the physical size in the
\hbox{$\perp$-direction}. Inserting the results of the \hto\ pumping scheme
(Eq.\ \ref{eq:kod}) into Equation (\ref{eq:dperp}) and utilizing Equation
(\ref{eq:asat}), the ratio $d/d_\perp$ is given by
\eq{\label{eq:shape1}
   {d\over\dperp} \simeq {3.3\over\asat}\x\frac
   {\DS 1 - {21\over T} + 0.12\ln\,{n_9\over\xi^{1/2}}}
   {\DS\left[1 - {14\over T} + 0.26\ln\,{n_9\over(\xi c_\eta)^{1/2}}
                                    - 0.13\ln a\right]^{1/2}}.
}
The maser will appear elongated either in the plane of the shock or along the
shock propagation, depending on the value of \asat\ that the pumping scheme
generates.

\subsection{Maser Brightness and Flux}

We now discuss the predictions of the \hto\ pumping scheme for observable
radiative quantities. These results are applicable only for the emission from
resolved individual maser spots. The brightness temperature at line center is
given by Equations (\ref{eq:unsat}) and (\ref{eq:Tb_sat}), respectively, for
the unsaturated and saturated regimes. While the pump properties can be
specified in terms of density-independent scaling quantities, the onset of
saturation does involve the density (Equation \ref{eq:asat}). Figure 3 shows
the variation of brightness temperature with $\xi$ for a wide range of
densities for disk masers with $a = 10$, chosen for illustration; the behavior
for other aspect ratios can be deduced from the explicit expressions for $T_b$
shown below. Each curve shows a steep exponential rise at the low-$\xi$ end,
corresponding to unsaturated maser growth. The break in the slope marks the
onset of saturation, and the behavior of $T_b$ at higher values of $\xi$ is
controlled by the variation of the maser pump properties. Saturation is reached
for all the displayed densities except for $n = 4\times\E9\, \cc$, which falls
just short of saturation---in that case \asat\ = 13 around the peak of the
$T_b$ curve. Therefore, $n \simeq 3\times\E9\, \cc$ is the highest density that
produces saturated disk masers with $a = 10$ at $T$ = 400 K. The curves for $n
= \E7$ and \E8\ \cc\ show an additional break at $\xi$ = 0.01 and 0.3,
respectively. This break marks the transition to a thick disk regime, a
transition that occurs only at lower densities (see Equation \ref{eq:thin}).
Masers with $n \ge 6\times\E8\,\cc$ are in the thin-disk domain for all values
of $\xi$.

On each curve in the figure, an X marks the value of $\xi$ generated by a J
shock with $v_{A\perp}=1$ km s$^{-1}$ and $\Delta v_D=1$ km s$^{-1}$ that
produces a maser density corresponding to that curve, as described below (see
\S 3).  This shows that J shocks with $a=10$ produce saturated thin-disk masers
in the entire range $\E7\,\cc\ \le n \le 3\times\E9\,\cc$. For thin-disk masers
with $a>a_{\rm sat}$, the brightness temperature at line center can be written
as
\eq{\label{eq:Tb}
  T_b \simeq 4.7\x\E{11}\,\xi^{1/2}c_\eta (d/\dperp) e^{-460/T}a_1^3,
}
where $a_1 = a/10$ (see Equation \ref{eq:Tb0}). This result amplifies our
earlier findings (EHM, EHM92): Apart from the gas temperature, the brightness
temperature of J-shock produced masers is determined almost exclusively by
$\xi$ and the maser aspect ratio $a$.  Further discussion of this effect
appears in \S 4.1.
The relevant quantity for maser detectability from a distance $D_{\rm kpc}$ in
kpc is its observed monochromatic flux. From Equation (\ref{eq:F0}), the
observed flux at line center is
\eq{\label{eq:Fobs}
    F_{\nu_0} \simeq 74\,\xi^{1/2}c_\eta e^{-460/T}
        {d_{13}^2\over D_{\rm kpc}^2}a_1^3        \quad \rm Jy.
}
Measured flux is frequently expressed in terms of the equivalent isotropic
luminosity $L_{\rm iso} = 4\pi D^2F$, where $F$ is the flux integrated over the
spectral range of the maser feature. Then
\eq{\label{eq:Liso}
    L_{\rm iso} = 4\pi D^2F_{\nu 0}\cdot\frac{\pi^{1/2}\Dv}{\lambda}
   \simeq 3\x\E{-6}\xi^{1/2}c_\eta e^{-460/T}\dvf d_{13}^2 a_1^3\quad \Lo .
}
Because of the beaming, the actual luminosity of a planar maser is only $L_m =
L_{\rm iso}/2a$ (EHM92).

\section{J-SHOCK STRUCTURE IN VERY DENSE GAS }

\subsection{Review of J-Shock Structure and Analytic Results}

A number of authors, including Hollenbach \& McKee (1979, 1989; hereafter HM79, HM89), Neufeld \& Dalgarno
(1989), Neufeld \& Hollenbach (1994),  Smith \& Rosen (2003), Guillet,  Jones,  \& Pineau Des For{\^e}ts (2009),
and Flower \& Pineau Des For{\^e}ts  (2010),
have discussed J-shock structure in
dense molecular gas.  EHM discussed the particular structure found in
the very dense J shocks that may give rise to 22 GHz water masers.  In fast J
shocks, the molecules are first completely dissociated by the extremely high
postshock temperatures ($\sim 10^5$ K) immediately behind the shock front.  In
this very hot region, dust may be partially or totally destroyed by thermal
sublimation, sputtering, and grain-grain collisions. Further downstream, where
the material cools down, H$_2$ molecules reform on the surviving dust grains
and are ejected to the gas phase with sizable internal energies, which provides
a source of heating for the gas if the postshock densities are sufficiently
high ($\ga 10^6$ cm$^{-3}$) to convert this internal energy into heat. In other
words, the rovibrationally excited H$_2$ molecule needs to be collisionally
de-excited, rather than suffering radiative decay, for the energy to be
converted to heat. This heating produces an ``H$_2$ re-formation plateau," a
nearly isothermal column of gas at a temperature $T_p\sim 300-400$ K. The
plateau gas is warm enough to drive all oxygen not locked in CO to form H$_2$O
and to collisionally populate the 22 GHz maser levels, which lie $ 644$ K above
ground. The dust temperature in the masing region is typically 50-100 K. The
hydrogen column density, $N_p$, of the heated plateau region can be as large as
$\sim10^{22-23}$ cm$^{-2}$, and the H$_2$O column as high as 3$\times 10^{19}$
cm$^{-2}$.\footnote{We note that Neufeld \& Dalgarno (1989) also found the same
plateau for fast, dense J shocks.} The H$_2$ re-formation plateau is an ideal
site for relatively low-lying H$_2$O masers like the 22 GHz masers; the
temperature $T_p$ may be too low to significantly excite higher excitation
H$_2$O maser levels, and C shocks have been proposed as sites of those masers
(Melnick et al 1993, Kaufman \& Neufeld 1996).

In both C and J shocks, the component of the magnetic field normal to the shock
velocity, $B_{0\perp}$, serves to limit the compression of the postshock gas.
This component of the magnetic field is related to the corresponding preshock
Alfven speed $v_{A\perp}$ by
\eq{
    B_{0\perp} = 1.7 v_{A\perp,5}n_{0,7}^{1/2}\ {\rm mG},
}
where $v_{A\perp,5}=v_{A\perp}/(1$ km s$^{-1}$) and where $n_{0,7}=n_0/(10^7$
cm$^{-3}$)  is the density of hydrogen nuclei in the preshock gas [i.e.,
$n_0=n_0$(H)$+2n_0$(H$_2)$]. If the shock velocity and the orientation of the
magnetic field are uncorrelated, the median value of $B_{0\perp}$ equals
$(\surd3/2)B_0$, so the distinction between $B_0$ and $B_{0\perp}$ is not
numerically important. Nonetheless, we shall retain this distinction here since
future observations might be able to determine the relative orientations of the
shock velocity and the upstream magnetic field. Typical preshock magnetic
fields in molecular clouds of widely varying density are characterized by
preshock Alfven speeds $v_A \sim 1-2$ km s$^{-1}$ (Heiles et al 1993). Fields
at high densities have recently been measured by Falgarone et al (2008) who
observed the Zeeman effect in CN. For the 8 measurements with positive
detections, the median value of $v_{A5}$ is 1. Including the 6 measurements
with no detections, but using the quoted error as the value, we find a median
of 0.6. The dispersion is large, however: a factor 6. Correcting for
inclination, we estimate $v_{A5} \simeq 1 \pm 0.8$ dex. This is quite crude,
however, since our treatment of the upper limits is very approximate and since
the Zeeman technique averages over fluctuations in the line-of-sight field. As
noted above, typically $v_{A\perp}\simeq v_A$ if the orientations of of the
shock velocity and the magnetic field are uncorrelated, so we shall adopt
$v_{A\perp,5}=1$ as a fiducial value.

The density $n_p$ in the masing (``plateau") region of a J shock\footnote{ The
warm region of a C shock occurs where the preshock gas is hardly compressed, so
that C shock masers are produced in gas with a density roughly equal to the
preshock density.   Although the final compression in a C shock is also limited
by magnetic fields, these compressed regions, unlike the case in the J shocks
we consider, are too cold to excite maser action.} is usually limited by the
value of $B_{0\perp}$ (HM79), and can be written as $n_p=\surd 2 n_0
v_s/v_{A\perp}$. In terms of the flux of H nuclei through the shock, $j\equiv
n_0v_s$, we have
\eq{
n_{p9}\equiv \frac{n_p}{10^9\  {\rm cm^{-3}}} = 1.4
\left(\frac{j_{14}}{v_{A\perp,5}}\right), \label{eq:np9}
}
where
\eq{
j_{14}\equiv n_{0,7}v_{s7},
}
$v_{s7}\equiv v_s/(100$ km s$^{-1}$) is the shock speed in units of 10$^7$ cm
s$^{-1}$, and $j_{14}= j/(10^{14}$~cm\ee~s\e). We shall find that many of
masing parameters mainly depend on $j$. The magnetic field in the masing region
of a J shock balances the ram pressure of the shock and is therefore
independent of the preshock field,
\eq{
B_p\simeq 0.24n_{0,7}^{1/2}v_{s7}\ \mbox{G} = 0.24j_{14}^{1/2}v_{s7}^{1/2}\
\mbox{G} = 0.24j_{14}n_{0,7}^{-1/2} \ \mbox{G}.
}
The analytic formulae for the density $n_p$ and the magnetic field $B_p$ in the
maser region apply when the magnetic pressure dominates there, or when
$v_{A\perp,5}\ga 2\times 10^{-2} v_{s7}^{-1}$, assuming the plateau temperature
is 300-400 K (HM79, EHM). Since $v_{A\perp,5}$ is usually $\ga 0.2$, this
condition is readily met.

Table 2 summarizes analytic solutions and approximations previously obtained in
HM79, HM89, and EHM, or, in the case of $T_b$ and $L_{\rm iso}$, taken from
Section 2 . We define $T_{b,11}\equiv T_b/10^{11}$ K and $L_{\rm iso,-6}\equiv
L_{\rm iso}/(10^{-6}\;L_\odot)$. The factor  $\gamma= 10^{-17}\gamma _{-17}$
cm$^3$ s$^{-1}$ is the average rate coefficient for H$_2$ formation on grains
in the temperature plateau region. In our numerical shock computations we use
the formulation for $\gamma$ from HM79. Note that because of the gas and {\it
dust} temperature sensitivity of $\gamma$ and because partial destruction of
dust near the shock front reduces the area of grains and therefore $\gamma$,
there is a ``hidden" additional dependence on $n_0$ and $v_s$, or on $d$ and
$v_{A\perp}$, when $\gamma$ appears in Table 2.  The same holds true for
$x_{-4}({\rm H_2O})$.   However, Table 2 is general to {\it any} formulation
for $\gamma$ and for {\it any} postshock H$_2$O abundance, in contrast to the
tables presented later in \S 4.2 that are specific to our particular
formulation of $\gamma$ and to our shock chemistry, which derives $x_{-4}({\rm
H_2O})$ at each point in the postshock gas.
 The column density in the masing region (or the
H$_2$  re-formation plateau region), $N_p$, is analytically determined by
finding the timescale $t_{\rm H_2}$ for H$_2$ formation in the plateau and then
taking $N_p=n_0v_s t_{\rm H_2}$ (EHM).  The timescale  $t_{\rm H_2}$ is given by
\eq{
 t_{\rm H_2}={1\over {n\gamma}}}
 leading then to
 \eq{
 N_p= 7\times 10^{21} \left({v_{A\perp,5} \over \gamma_{-17}}\right) \ {\rm cm^{-2}}}
The thickness $d$ of the postshock masing region (the spot size parallel to
the shock velocity) is given by 
\eq{
d=\frac{N_p}{n_p}=5.0\times 10^{12}\;\frac{v_{A\perp,5}^2}{\gamma_{-17} j_{14}}}
The formulae for $N_p$ and $d$ are
quite accurate when applied to the numerical results, but require a knowledge
of  $\gamma_{-17}$. The value of $\gamma_{-17}$ is of order 0.1-3.0 (HM79) for
the gas and dust temperatures typical of the masing plateau.  Note that the
column density in the plateau is independent of the preshock density and shock
velocity or of $j$, for {\it fixed} $\gamma$. The analytic formulae for $\xi$,
$T_b$, and $L_{\rm iso}$ in the masing region come from Equations (2-4),
(2-15), and (2-17), respectively. The expressions for $n_p$, $T_p$, $d$, $\xi$,
$T_b$ and $L_{\rm iso}$ in Table 2 depend only on $j$ and not separately on
$n_0$ and $v_s$.    The parameter $B_p$ is the sole parameter dependent on the
additional parameter $v_s$, but only as the square root.

Table 2 tabulates quantities  in terms of the preshock variables $j$ and
$v_{A\perp}$ in column 1 and in terms of the observable quantities $d$,
$d/d_\perp$ (the "shape" of the maser spot) and $v_{A\perp}$ in column 2.
Essentially, we have eliminated $j=n_0v_s$ from column 1 in favor of $d$ in
column 2. Column 2 is added to aid observers in estimating the shock
parameters. However, it must be noted that the average values of
$\gamma_{-17}$, $x_{-4}({\rm H_2O})$, and $\Delta v_{D5}$ in the masing plateau
appear in these equations. As we will see in the subsections below, these all
have values near unity for most cases, enabling an estimate to be made of the
shock parameters. The numerical results presented in this paper can be
understood, interpolated and extrapolated by applying these formulae.

\subsection{Physical Processes in the J-shock Model}

HM79, HM89 and  Neufeld \& Hollenbach  (1994) describe in detail the physical
processes included in the 1D steady state shock code we have used in this
paper.  The fundamental input parameters to this code are the preshock hydrogen
nucleus density $n_0$, the shock velocity $v_s$, the Alfven speed $v_{A\perp}$
in the preshock gas, the velocity dispersion $\Delta v_D$ in the line-emitting
gas, and the gas phase abundances of the elements. In our standard runs we take
$v_{A\perp}= 1$ km s$^{-1}$, $\Delta v_D= 1$ km s$^{-1}$, and gas phase
abundances listed in HM89 (the main number abundances relative to hydrogen
nuclei that are relevant here are those for carbon, 2.3$\times 10^{-4}$, and
oxygen, 5.4$\times10^{-4}$).

The code uses the Rankine-Hugoniot jump conditions to set the physical
parameters immediately behind the shock front, and the various continuity
equations to numerically solve for the temperature, density and chemical
structure in the cooling postshock gas.  The chemistry includes 35 species and
about 300 reactions. For the fast ($v_s \ga 20$ km s$^{-1}$), dense ($n_0 \ga
10^5$ cm$^{-3}$) shocks considered in this paper, important chemical processes
include collisional dissociation and ionization, photodissociation and
photoionization by the UV photons produced in the (upstream) hot postshock gas,
neutral-neutral reactions with activation barriers, and the re-formation of
H$_2$ molecules on warm ($T_{\rm gr} \simeq 50-150$ K) dust grains.  All but
the last are either well determined experimentally or well understood
theoretically. 

   We discuss here the formation rate coefficient of H$_2$ on warm dust grains in some detail, as
 this process is critical to forming the high temperature plateau where the H$_2$O maser
 is produced.   We use the theoretical model of HM79 for the formation of
H$_2$ on warm grains. In this formulation the formation rate coefficient
$\gamma$ is a function of both the gas temperature $T$ and the dust temperature $T_{\rm gr}$.   At 
relatively low $T$ ($\la 100$ K) and $T_{\rm gr}$ ($\la 30$ K), the rate coefficient has been
inferred observationally in diffuse clouds and molecular cloud surfaces to be 
$3\times 10^{-17}$ cm$^{3}$ s$^{-1}$.   At the somewhat higher gas temperatures $T\sim 400 $ K
in the plateau, the coefficient drops by about a factor of $\sim$ 2 due to the decreased sticking 
probability of incoming H atoms (e.g., HM79, Cuppen et al 2010).  However, a more important
effect in the plateau is caused by the increased $T_{\rm gr}\sim 100$ K, which causes
$\gamma$  to drop even more because of the evaporation of
H atoms from grain surfaces prior to H$_2$ formation.  The exact amount of this drop cannot
be well determined for realistic interstellar dust.   However, HM79, Cuppen et al (2010), and
Cazaux et al (2011) have used theoretical modeling to try to estimate the effect.   All three of
these studies are in quite good agreement, given the inherent uncertainties.   Including both
the sticking probability and the probability of H$_2$ formation on the grain surface, and normalizing
to obtain the above standard rate at low $T$ and $T_{\ gr}$,  Cuppen 
et al get a H$_2$ rate coefficient for 400 K gas and 100 K dust of $2.2\times 10^{-18}$ cm$^3$ s$^{-1}$, whereas
HM79 find $3.8\times 10^{-18}$ cm$^3$ s$^{-1}$.  In addition, Cazaux et al  find
a rate coefficient for 400 K gas and 125 K dust of $1.5\times 10^{-18}$ cm$^3$ s$^{-1}$, whereas
HM79 find $1.4\times 10^{-18}$ cm$^3$ s$^{-1}$.    Therefore, the HM79 formation rate coefficient agrees well
with the more recently obtained values {\it in the region of parameter space ($T\sim 400$ K, $T_{gr} \sim 100$ K)
where the H$_2$ re-forms in the postshock gas, and where the H$_2$O maser is produced}.
This agreement is far better than the uncertainties in these models, and therefore there could be
fairly large differences between these values and the values  for real shocked grains at high dust temperatures.
Because of these uncertainties, we  consider the
sensitivity of our results to the H$_2$ formation rate coefficient  in the next subsection.

Our model also includes the partial destruction of dust grains in the shock, which
reduces the grain surface area per H nucleus, and which therefore also reduces the rate 
coefficient for H$_2$ formation on grains.
 Our J-shock maser model relies on at least some grains
surviving the shock, since the H$_2$ re-formation plateau is caused by H$_2$
re-formation on grain surfaces.  However, shocks with $v_s \ga 200$ km s$^{-1}$
will completely destroy dust grains by sputtering and grain-grain collisions
(Jones et al 1996).  We therefore only consider shocks with $v_s < 200$ km
s$^{-1}$.  In addition, shocks with $n_{0,7} v_{s7}^3 \ga 100$ will sublimate
all grains with sublimation temperatures $T_{\rm sub} \la 1500$ K, which is the
maximum sublimation temperature of a likely interstellar grain material.  Therefore,
no dust exists above this constraint as well.   For $n_0$ and $v_s$ values that
are low enough to provide at least some dust survival, we
adopt the grain composition mixture of Pollack et al (1994), and allow for the
sublimation of the less refractory material at lower values of $n_{0,7}
v_{s7}^3$ as each sublimation temperature is exceeded. 

In summary, the
conditions $n_{0,7} v_{s7}^3 \la 100$ and $v_{s7} \la 2$ provide upper limits
for J-shock masers produced by the H$_2$ re-formation plateau. We find below,
however, somewhat more stringent conditions on preshock density occur due to
the quenching of the H$_2$O maser by high postshock densities and H$_2$O line
optical depths. An example of this is shown in Figure 3, which shows the
quenching that occurs in the case of slabs with $a=10$ and
$v_{A\perp,5}=\dvf=1$; the maser is quenched for $n_p \geq 4\times 10^{9}$
cm$^{-3}$ since then $\asat>a$. The upper limit on the preshock density for
effective maser emission in this particular case is therefore $n_0\la 3\times
10^7\left(v_{A\perp,5}/v_{s7}\right)~{\rm cm^{-3}}$ from Equation
(\ref{eq:np9}).

The cooling of the postshock gas is treated with the escape probability
formalism
(including the effects of dust absorption),
since a number of important cooling transitions become optically thick in the lines.  For
this study, we focus mainly on the postshock temperature region bounded by 3000
K $> T >$ 50 K, where molecular formation occurs and H$_2$O masers may be
produced. In the masing region, the gas cooling is dominated by optically thick
rotational transitions of H$_2$O and by gas collisions with the cooler dust
grains.
The gas heating in the masing region is dominated by the re-formation of H$_2$.
There are two main contributions to this heating process. The newly formed
molecules can be ejected from the grain surfaces with kinetic energies greater
than $kT$, thereby heating the gas. In addition, a newly formed and ejected
molecule may carry with it rovibrational energy that can be transferred to heat
by collisional de-excitation in the gas. These processes are not well
determined.  We adopt the theoretical formulation of HM79,  in which the newly
formed molecule is ejected with 0.2 eV of kinetic energy and 4.2 eV of
rovibrational energy and use the de-excitation rate coefficients for H and
H$_2$ collisions quoted in HM79. However, Tielens and Allamandola (1987)
speculate that the H$_2$ molecule may lose a significant portion of its
formation energy (the rovibrational energy) to the grain, before leaving the
grain surface. Since the formation heating is proportional to the H$_2$
formation rate times the energy delivered per H$_2$ to the gas, we test the
sensitivity of our results to the energy partition when we test the sensitivity
of the plateau temperature to the uncertain formation rate coefficient.

\subsection{Numerical Results for J-shock Structure}
\label{sec:numjshock}

Figure 4 presents the shock profile for our standard model: $n_0=10^7$
cm$^{-3}$, $v_s=100$ km s$^{-1}$, $v_{A\perp}=1$ km s$^{-1}$, and $\Delta
v_D=1$ km s$^{-1}$. The column density of hydrogen nuclei, $N$, and the
position, $z$, are measured from the shock front. The ultraviolet radiation
from the shock has processed the preshock gas before it enters the shock front.
As a result, the molecular preshock gas is photodissociated and partially
photoionized prior to being shocked. Using the results of HM89, we take the
initial abundances at the shock front for the standard case to be
$x(H^+)=0.47$, $x(H)=0.36$, $x(H_2)=0.087$; the trace species are largely
atomic and singly ionized as well.  For slower shocks, the precursor field is
less important and the shock front abundances are initially largely molecular.
The gas then collisionally dissociates and partially ionizes in the hot
postshock gas just downstream of the shock front.  Figure 4 shows that the 100
km s$^{-1}$ shock heats the plasma to about $2\times 10^5$ K, and the gas cools
by collisional ionization and by UV and optical emission to 10$^4$ K in a
column $N\simeq 4\times 10^{17}$ cm$^{-2}$.  The Lyman continuum photons from
the $\sim 10^5$ K gas maintain a Str\"omgren region at $T\sim 10,000$ K to a
column $N\simeq 10^{19.5}$ cm$^{-2}$. Once the Lyman continuum photons are
absorbed, the electrons and protons recombine and the gas cools until the
heating due to H$_2$ re-formation maintains the temperature at $T_p\simeq
300-400 $ K. This is the \lq\lq H$_2$ re-formation plateau". Note that the size
scale of this plateau, shown at the top of the figure, is $d\sim 10^{13}$ cm
for $v_{A\perp} = 1$ km$^{-1}$ and for our assumed formulation for
$\gamma_{-17}$.  After the molecular hydrogen has nearly completely reformed,
at a column of about $2\times 10^{22}$ cm$^{-2}$, the heating rate drops and
the gas temperature drops to $\la 100$ K.

The H$_2$O number abundance relative to hydrogen nuclei, $x$(H$_2$O), is also
plotted in Figure 4.  The abundance is negligible for $N\la 10^{21}$ cm$^{-2}$,
but the H$_2$O abundance rapidly climbs once the H$_2$ abundance rises in the
re-formation plateau. CO re-forms even more rapidly; typically all the gas
phase carbon is incorporated into CO once $x$(H$_2$)$\ga 10^{-3}$.  Therefore,
the abundance of H$_2$O is limited to the abundance of oxygen that remains once
an oxygen atom has combined with every gas phase carbon atom.  We have taken
for elemental gas phase abundances $x_{\rm O}$=$5.4\times 10^{-4}$ and $x_{\rm
C}$=$2.3\times 10^{-4}$; thus, $x_{\rm max}$(H$_2$O)$\simeq 3\times 10^{-4}$.
Typically, $x$(H$_2$O)$\simeq x_{\rm max}$(H$_2$O) once $x$(H$_2$) $\ga 0.25$.
However, the reactions that lead to H$_2$O have large activation energies
($\Delta E/k \sim 4000$ K) and proceed slowly in the plateau region. In many
cases the timescales spent in the plateau are insufficient to reach chemical
equilibrium; as a result, the H$_2$O abundance varies somewhat with $T_p$ and
$N_p$ and consequently with $n_0$ and $v_s$, as shall be demonstrated below.

We have also plotted the grain temperature, $T_{\rm gr}$, in Figure 4 to
emphasize the fact that the grain temperature is significantly below the gas
temperature in the postshock gas if $n_{0,7} v_{s7}^3 \la 100$  (HM79).  The
grains are only weakly coupled to the gas through gas collisions and through
the line radiation from the gas.  At the same time, radiative grain cooling is
very efficient; the result is that the grains are considerably cooler than the
gas.   Because the dust is optically thin in the H$_2$O rotational transitions
and the lines themselves have finite opacity, the effective temperature of the
radiation field is cooler than the gas kinetic temperature.  Consequently,
collisions with H atoms and H$_2$ molecules excite the H$_2$O and the escaping
IR photons from H$_2$O rotational transitions create non-LTE populations and
the population inversion of the maser levels. Absorption by dust competes with
escape of the IR photons when the dust optical depth reaches unity for the IR
photons; for $\hto$ IR photons with typical wavelengths of 50 $\mu$m, this
occurs at a column density $N_p\simeq 3\times 10^{23}$ cm$^{-2}$ if we account
for some reduction in dust abundance in the shock (HM79).  Using the expression
for $N_p$ in Table 2, we estimate that dust absorption is not important
for
\eq{ {\gamma_{-17} \over v_{A\perp,5}} \ga 0.02,
}
which is generally the case.  If dust absorption is important, and the dust is cooler than the gas, the
presence of dust  enhances the effective escape probability of the
H$_2$O IR photons (Collison \& Watson 1995).

Figure 5 plots contours of $T_p$, the temperature of the H$_2$ re-formation
plateau, as a function of the shock parameters $n_0$ and $v_s$.  The gas
temperature declines slightly with $N$ in the H$_2$ re-formation plateau; we
have defined $T_p$ as the temperature of the gas when $x$(H$_2$)=0.375 (i.e.,
when 75\% of the hydrogen is molecular). Figure 5 and the subsequent 3 figures
are the results of a grid of shock models ($n_0= 10^5,\ 10^6,\ 10^7,\ 10^8$,
and $10^9$ cm$^{-3}$; $v_s=20,\ 40,\ 80,\ 100$, and 160 km s$^{-1}$;
$v_{A\perp}=1$ km s$^{-1}$; $\Delta v_D= 1$ km s$^{-1}$). This grid is
sufficiently coarse that the contours are somewhat approximate.  As noted
above, we have taken $v_s=160$ km s$^{-1}$ as our upper limit because shocks
with $v_s \ga 200$ km s$^{-1}$ destroy essentially all of the dust grains. Once
there are no grain surfaces upon which to form H$_2$, molecular re-formation in
the postshock gas effectively ceases, no warm H$_2$O is produced, and postshock
H$_2$O masing action is destroyed. In addition, we have carried out
calculations up to $n_0 = 10^9$ cm$^{-3}$ since dust grains sublimate at higher
preshock densities, but in fact maser emission is generally quenched at
considerably lower pre-shock densities as discussed above.

The main results from Figure 5 are that $T_p\simeq 300-400$ K and that $T_p$ is
very insensitive to $n_0$ and $v_s$ as long as $n_0\ga 10^5$ cm$^{-3}$ and $v_s
\ga 30 $ km s$^{-1}$.  EHM discussed this insensitivity as due to the balance
between the gas heating by H$_2$ formation being balanced by H$_2$O and grain
cooling of the gas. An analytic fit to the numerical results gives:
\eq{
    T_p \simeq 350n_{0,7}^{0.12}v_{s7}^{-0.12}\Delta v_{D5}^{-0.22}\ {\rm K},
}
accurate to a factor of 1.3 for 10$^6$ cm$^{-3}\la n_0 \la 10^8$ cm$^{-3}$ and
30 km s$^{-1} \la v_s \la 160$ km s$^{-1}$. For preshock densities $n_0 \la
10^5$ cm$^{-3}$, the H$_2$ formation heating rapidly drops because the
newly-formed, vibrationally-excited H$_2$ molecules radiate away their
vibrational energy before collisions can transform this excitation energy into
heat. However, for $n_0\ga 10^5$ cm$^{-3}$, the H$_2$ re-formation plateau
provides a temperature environment where the chemical production of H$_2$O is
efficient and where collisional excitation of the 22 GHz H$_2$O maser, which
lies $ 644 $ K above ground, is possible.

Figure 6 provides contour plots of the H$_2$O abundance at the point where
$x$(H$_2$)=0.375 and $x$(H)=0.25, the same position in the re-formation plateau
where we measure $T_p$.  The main result is that an appreciable fraction of the
available oxygen is converted to water for $n_0\ga 10^6$ cm$^{-3}$.  At
$n_0=10^5$ cm$^{-3}$, the water abundance is only $\sim 10^{-5}$, caused by a
combination of lower plateau temperature (see Figure 4) and lower plateau
column density $N_p$. The former suppresses the rate of H$_2$O formation
because of the activation barriers present in this process.  The latter reduces
the protective shielding by the dust of the dissociating UV photons, and
reduces the time available for H$_2$O to form.  However, for $n_0\ga 10^{6}$
cm$^{-3}$, a simple analytic fit to the numerical results gives:
\eq{
x_{-4}({\rm H_2O}) \simeq 1.6n_{0,7}^{0.2}v_{s7}^{-0.3},
}
accurate to a factor of 1.2 for 10$^6$ cm$^{-3}\la n_0 \la 10^8$ cm$^{-3}$ and
30 km s$^{-1} \la v_s \la 160$ km s$^{-1}$.

The postshock density, $n_p$, as a function of $n_0$ and $v_s$ in the masing
plateau is accurately given by the equation in Table 2, and we therefore do not
present a contour plot for it. However, $d$ depends on $\gamma _{-17}$ and
$\xi$ depends on $\gamma _{-17}$ and $x$(H$_2$O), and therefore these
parameters are accurately determined only by numerical solutions of shock
structure. Figure 7 plots the thickness $d_{13}=d/(10^{13}$ cm) of the masing
plateau as a function of $n_0$ and $v_s$. Comparing the numerical results with
the equation in Table 2, we see that $\gamma _{-17}$ declines as $n_0$ and
$v_s$ increase, because denser and faster shocks have higher grain
temperatures, which reduces the rate of H$_2$ formation on grain surfaces.  In
addition, there is reduction in grain area at high values of $n_{0,7} v_{s7}^3$
due to partial sublimation of grains. A simple fit to the numerical results
gives
\eq{
    d_{13}\simeq 1.3n_{0,7}^{-0.7}v_{s7}^{-0.2}v_{A\perp,5}^2\ ,
}
\eq{\label{eq:gamma}
    \gamma_{-17}\simeq 0.38n_{0,7}^{-0.3}v_{s7}^{-0.8},
}
accurate to a factor of 1.5 for 10$^6$ cm$^{-3}\la n_0 \la 10^8$ cm$^{-3}$ and
30 km s$^{-1} \la v_s \la 160$ km s$^{-1}$. We note (see also Table 2) that $d
\propto v_{A\perp}^2$, so that the maser spot size may be the observable
parameter that is most sensitive to the strength of the preshock magnetic
field.  Typical preshock densities of $n_{0,7} \sim 0.1-1$ and observed maser
spot sizes of $\sim 10^{13}-10^{14}$ cm imply that $v_{A\perp} \sim 1$ km
s$^{-1}$, in line with the observations discussed above in Section 3.1.

Figure 8 presents the contours of the maser emission measure $\xi$ [measured
from the shock front to the point in the postshock plateau where
$x($H$_2)$=0.375] as a function of $n_0$ and $v_s$ for $v_{A\perp,5}$=1 and
$\Delta v_{D5}$=1.  The main result is that $\xi$ increases monotonically with
increasing $n_0$, as would be expected from its dependence on the relevant
parameters seen in Table 2.  From this one might predict that denser masers
will be brighter; however, the pump efficiency $\eta$ decreases with increasing
$n_0$ and ultimately the maser quenches as collisions and line trapping create
LTE conditions (see \S2).  A simple  fit to the numerical results gives
\eq{\label{eq:xi3}
    \xi\simeq 4.0(n_{0,7}v_{s7})^{1.5}\Delta v_{D5}^{-1}=4.0j_{14}^{1.5} \Delta v_{D5}^{-1},
}
 accurate to a factor of 1.4
for 10$^6$ cm$^{-3}\la n_0 \la 10^8$ cm$^{-3}$ and 30 km s$^{-1} \la v_s \la
160$ km s$^{-1}$.

Utilizing Equations (2-8), (2-9), (\ref{eq:np9}), and (\ref{eq:xi3}) we find
good fits\footnote  {We use these equations to give a rough fit, and then
slightly adjust the normalization coefficients to better fit the numerical
results.  In the case of $c_\eta$, we assume the second term in Eq. 2-9
dominates at high $j$, but then adjust slightly the power law dependence to
take into account the first term.} to the results of the numerical shock runs
for $c_\eta$ and $\eta$:
\eq{\label{eq:ceta3}
 c_\eta \simeq \left({1+0.045{j_{14}^{2.8} \over
    {v_{A\perp,5}^{0.2} \Delta v_{D5}^{1.5}}}}\right)^{-1},
}
\eq{\label{eq:eta3}
\eta \simeq 0.032{{\Delta v_{D5}^{0.5} c_\eta} \over {j_{14}^{0.75}}},
}
which are accurate to within a factor of 1.2 for $c_\eta$ and 1.4 for $\eta$ in
the parameter range 10$^6$ cm$^{-3}\la n_0 \la 10^8$ cm$^{-3}$ and 30 km
s$^{-1} \la v_s \la 160$ km s$^{-1}$.  The fit to $c_\eta$ will be useful in
subsequent analytic fits to $a_{\rm sat}$, $d/d_\perp$, $T_b$ and $L_{\rm
iso}$.

Figures 4-8 are valid for $v_{A\perp,5}=1$, which corresponds with measured
values (to within a factor $\sim 6$) for a wide range of cloud densities in the
Galaxy. However, as discussed earlier, there may be environments (such as very
dense gas or the nuclei of galaxies) where $v_{A\perp,5}$ deviates
substantially from unity.  Therefore, in Figure 9 we have plotted the variation
of $d$, $T_p$, and $\xi$ as functions of $v_{A\perp}$ for the standard case
$n_0=10^7$ cm$^{-3}$, $v_s=100$ km s$^{-1}$ and $\Delta v_{D5}=1$.  For
convenience, we have plotted the ratios of these parameters to their values
$d_{\rm std}$, $T_{p,\, \rm std}$, and $\xi_{\rm std}$ at the standard
$v_{A\perp,5}$=1.  The results follow the predictions from Table 2: $d$ varies
as $v_{A\perp}^2$ whereas $T_p$ and $\xi$ are relatively insensitive to
$v_{A\perp}$.

Figures 4-9 assume the HM79 model for the formation rate $\gamma$ of H$_2$
molecules on grains and for the kinetic and rovibrational energy delivered to
the gas per H$_2$ formation.  Because the formation process is uncertain, we
test the sensitivity of the results to variations in these parameters.  Since
the heating rate is proportional to the formation rate, we vary only $\gamma$
in this test. In the HM79 model $\gamma$ is a complicated function of gas and
grain temperature.  Figure 10 presents the results of models with constant
$\gamma$ and shows the sensitivity of $T_p$ and $\xi$ to variations in
$\gamma$.  In this figure we test only the standard case ($n_{0,7}=1$,
$v_{s7}=1$, $v_{A\perp,5}=1$, and $\Delta v_{D5}=1$). The plateau temperature
varies slowly with $\gamma$, changing from 180 K to 550 K as $\gamma$ increases
from $3\times 10^{-19}$ cm$^{3}$ s$^{-1}$ to $3\times 10^{-17}$ cm$^{3}$
s$^{-1}$, where the latter corresponds to the maximum H$_2$ formation
efficiency on grains.  The emission measure $\xi$ drops rapidly for decreasing
$\gamma \la 3\times 10^{-18}$ cm$^3$ s$^{-1}$, because of the inefficient
production of H$_2$O when the plateau temperature drops below $\sim 250$ K.
The water abundance drops from $\sim 3\times10^{-4}$ for $\gamma\sim 3\times
10^{-18}$ cm$^3$ s$^{-1}$ to $\sim 3\times 10^{-7}$ for $\gamma \sim 3\times
10^{-19}$ cm$^3$ s$^{-1}$.  Therefore, referring to the equation for $\xi$ in
Table 2, $\xi \propto x({\rm H_2O})/\gamma$ decreases by a factor of $\sim$100
over this range.

%%%%%%%%%%%%%%%%%%%%%%%%%%%%%%%%%%%%%%%%%%%%%%%%%%%%%%%%%%%%%%%%%%%%%%%%%%%%%%%

\section{H$_2$O MASER SLAB MODELS APPLIED TO SHOCK RESULTS}

In \S2 we performed a detailed calculation of the H$_2$O level populations and
the radiative transfer in a uniform slab characterized by $\xi$, $n$, and $T$.
In \S3 we found the values of $\xi$, $n_p$ and $T_p$ in the H$_2$ re-formation
plateau behind an interstellar J shock as functions of the H nucleus flux into
the shock $j=n_0v_s$,  the Alfven speed $v_{A\perp}$, and velocity dispersion
in the line-emitting gas $\Delta v_D$.  In this section we merge the results
from these two numerical computations to produce useful predictions concerning
the H$_2$O maser properties of astrophysical J shocks.

\subsection{Numerical Results for $a_{\rm sat}$, $d/d_\perp$, $T_b$, and $L_{\rm iso}$}

Perhaps the most important parameter for the application to J-shock models is
the aspect ratio of the maser (the ratio of the length along the line of sight
to the thickness) required for saturation, \asat, which is shown in Figure 11.
Shock-produced astrophysical H$_2$O masers are weak and unobservable as long as
they remain unsaturated, so that we shall require $a > \asat$.  However, the
coherence length in the shock plane is finite and $a$ cannot exceed \about\
30--100 because of the curvature of the shock and because of velocity gradients
in the plane (e.g., see \S 5). Therefore, $\asat \la 30-100$ is a constraint on
observable H$_2$O masers produced in J shocks. Figure 11 shows the dependence
of $a_{\rm sat}$ on $n_0$ and $v_s$ for $v_{A\perp,5}$=1. When $n_0 \la 10^6$
cm$^{-3}$, $a_{\rm sat}$ rapidly increases to $\ga 100$ because of the low
value of $\xi$ in the shock due to low densities and low values of $x_{-4}({\rm
H_2O})$.  Since such large aspect ratios are extremely unlikely in
astrophysical shocks, we conclude that low-density shocks cannot produce
saturated masers beamed in the shock plane and that, therefore, such shocks
will produce weak, unsaturated masers that are difficult to detect. They are
``starved" for sufficient collisions to the highly excited states that feed the
maser.   At the other extreme, when $n_0 \ga10^8 $ cm$^{-3}$, $a_{\rm sat}$
again rapidly increases to $\ga 100$ because the maser quenches (levels
approach LTE and the inversion is weak) and large coherence paths are needed to
reach saturation. At such high preshock densities the plateau density $n_p \ga
4\times 10^9$ cm$^{-3}$ and $\xi \ga 10$ (see Table 2 and Figure 8), and
therefore quenching is significant (e.g., see $\eta$ in Figure 1 or the
quenching of $T_b$ at high $\xi$ and $n$ in Figure 3). When $10^6$ cm$^{-3} \la
n_0 \la 3\times10^7(v_{A\perp,5}/v_{s7}) $ cm$^{-3}$ and $0.3 \la v_{s7} \la
1.6$, $a_{\rm sat} \simeq 1-10$. An analytic fit to \asat \ can be obtained
using Equations (2-11), (\ref{eq:np9}), and (\ref{eq:xi3}) as guides:
\eq{\label{eq:asat4}
a_{\rm sat} \simeq 2.5 \;{j_{14}^{0.25} \dvf^{0.5}\over {v_{A\perp,5} c_\eta}},
}
where the expression for $c_\eta$ given in Equation (\ref{eq:ceta3}) completes
the analytic fit. This expression is good to a factor of 1.3 over the main
maser parameter space 10$^6$ cm$^{-3}\la n_0 \la 10^8$ cm$^{-3}$ and 30 km
s$^{-1} \la v_s \la 160$ km s$^{-1}$.

The ratio of the maser spot diameter in the parallel direction to that in the
$\perp$ direction, $d/d_\perp$, behind J shocks is another observational
diagnostic of the shock conditions. The shape $d/d_\perp$ of the maser is given
in Equations (2-11) and (2-12) in terms of the general parameters $n$, $T$,
$\xi$, and $a$.   Figure 12 plots $d/d_\perp$ for our numerical shock results
over the shock parameter space $n_0$ and $v_s$, assuming $v_{A\perp,5}=1$,
$\Delta v_{D5} =1$, and $a=10$.
The dashed lines demarcate the zone where $a_{\rm sat}<10$; above the top
dashed line and below the lower dashed line $a_{\rm sat}>10$.  Our expressions
for $d/d_\perp$ are no longer valid if the maser is not saturated and therefore
we do not plot $d/d_\perp$ outside the dashed lines since $a=10< a_{\rm sat}$
there.
We see that $d_\perp \sim d$  in the strongly masing region of parameter space.
Maser spots will be approximately circular and shocked masers can be
approximated by equivalent cylinders of diameter $d$ and length $ad$. However,
we predict some variation in the shape of the maser spot. An analytic fit can
be obtained using Equations (2-12) and (\ref{eq:asat4}),
\eq{\label{eq:ddperp}
\frac{d}{d_\perp} \simeq {3.3 \over a_{\rm sat}} \simeq 1.3\; {{v_{A\perp,5}
c_\eta} \over j_{14}^{0.25}\dvf^{0.5}},
}
good to a factor of 1.5 over  the main maser parameter space 10$^6$ cm$^{-3}\la
n_0 \la 10^8$ cm$^{-3}$ and 30 km s$^{-1} \la v_s \la 160$ km s$^{-1}$
 as long as $a>a_{\rm sat}$.   Note that if $a>10$, then the dashed lines move
 to accompany the slightly more allowed $n_0$, $v_s$ parameter space (see Figure 11).
The equation shows that the elongation of the maser in the direction of the
shock velocity is directly proportional to the Alfven speed in the ambient
medium.

Figure 13 shows the dependence of $T_{b,11}=T_b/(10^{11}$ K) on $n_0$ and $v_s$
for $v_{A\perp,5}=1$, $\Delta v_{D5} =1$, and $a=10$.
The dashed lines are the same as in Figure 12.    Since this figure applies to
$a=10$, regions outside the dashed lines with $a_{\rm sat}>10$ will have
exponentially reduced $T_b$ (see Eq. B13) since they will be unsaturated.   The
effect of entering the unsaturated regime is seen in Figure 3, where very small
changes in $\xi$ lead to extremely large changes in $T_b$.    For $n_0 \la
10^6$ cm$^{-3}$ the maser is unsaturated and ``starved" for exciting collisions
as discussed above. For $n_0 \ga 3\times10^7(v_{A\perp,5}/v_{s7}) $ cm$^{-3}$
in the case of $a=10$, the maser is rapidly quenched and $T_b$ precipitously
drops.
For intermediate densities, where the maser is saturated ($a>a_{\rm sat}$),
an approximate analytic fit (using Eqs. 2-13, 3-13, and
\ref{eq:ddperp}) to these numerical results is
\eq{
    T_{b,11} \simeq 2.5 j_{14}^{0.5} \left({v_{A\perp,5}\over \Delta v_{D5}}\right)  c_\eta^2 a_1^3 \ {\rm K},
}
which is accurate to a factor of 1.4 for $10^6$ cm$^{-3} \la n_0 \la 10^8$
cm$^{-3}$ and 30 km s$^{-1} \la v_s \la 160$ km s$^{-1}$.  Note that as long as
$c_\eta$ is of order unity, that is, as long as $\xi$ and $n$ are not so large
that the maser quenches, $T_b$ increases with increasing $j$ and/or increasing
$v_{A\perp}$.   However, if $j$ becomes too large, the maser quenches, $c_\eta$
plummets, and $T_b$ drops.   Although raising $v_{A\perp}$ for fixed $a$ raises
$T_b$, increasing $v_{A\perp}$ also has the effect of increasing $d$; observed
maser spot sizes limit the size of $d$, thereby limiting the possible
$v_{A\perp}$ and therefore $T_b$ in the region.  In addition, since
$a=2\ell/d$, increasing $d$ can lower $a$; this then can lead to lower $T_b$
even as $v_{A\perp}$ and $d$ increase.

Ever since the detailed study of W51 by Genzel et al.\ (1981), maser spot sizes
have been shown to be uncorrelated with brightness temperature, a finding
reaffirmed by the thorough investigation of W49 by Gwinn, Moran \& Reid (1992)
and Gwinn (1994b).  Figure 13 indicates this lack of correlation for fixed
$v_A$. The brightness temperature does not vary much in the strong masing
region, even though (see Figure 7) $d$ varies by a factor of roughly 20 from
$\sim 5\times 10^{12}$ cm at the upper boundary to $\sim 10^{14}$ cm at the
lower boundary. In addition, note that Figure 13 is for fixed $a=10$.    $T_b$ varies
with $a^3$ and therefore the $T_b$ spread in a given source arises mostly from
variations in $a$.   In summary, at fixed $v_A$
the brightness is practically independent of the observed dimensions and dependent
almost entirely on $a$, in
agreement with observations.    One can increase $T_b$ and $d$ by holding 
$j$, $a$ and $\Delta v_{D5}$
fixed, but increasing $v_A$.  In this case, $T_b \propto v_A
\propto d^{1/2}$. Here, there is a weak dependence of $T_b$ on $d$, but the
very strong dependence on the aspect ratio ($T_b \propto a^3$) likely washes
this out. 

Figure 14 plots the contours of $L_{\rm iso,-6}\equiv L_{\rm iso}/(10^{-6}$
$L_\odot)$ as a function of $n_0$ and $v_s$ for our standard values of
$v_{A\perp,5}=1$, $\Delta v_{D5} =1$, and $a=10$.  The dashed lines are the
same as in Figures 12 and 13 (i.e., they demarcate $a_{\rm sat}=10$). For a
fixed aspect ratio, the luminosity peaks at somewhat lower $n_0$ compared with
$T_b$ because $L_{\rm iso}/T_b$  is proportional to $d^2$, and $d$ increases as
$n_0$ decreases (see Table 2 or Fig. 7). Using Equation (2-15) and the analytic
fits (\ref{eq:xi3}) for $\xi$ and (\ref{eq:ddperp}) for $d/d_\perp$, we find a
fit for $a>a_{\rm sat}$:
\eq{\label{eq:liso1}
 L_{\rm iso,-6} \simeq 2.2\left( {{v_{s7} v_{A\perp,5}^4 \Delta v_{D5}^{0.5}c_\eta}\over {j_{14}^{0.65} }}\right) a_1^3 ,
}
which is accurate to a factor of 2 for $10^6$ cm$^{-3} \la n_0 \la 10^8$
cm$^{-3}$ and 30 km s$^{-1} \la v_s \la 160$ km s$^{-1}$.
Recalling that $j=n_0v_s$, we see that $L_{\rm iso}$ is proportional to
$n_0^{-0.65}v_{s7}^{0.35}a_1^3$ as long as $c_\eta$ is of order unity.
Therefore, for fixed aspect ratio $a$, the luminosity increases with decreasing
preshock density due to the increase in $d$, as discussed above.     In
addition, regions with high preshock magnetic fields (i.e., high $v_{A\perp}$)
will produce much more luminous maser spots, because the maser spot size $d
\propto v_{A\perp}^2$.  In both cases the masers will not be much brighter, but
bigger and more luminous.  However, there is a very important caveat to this
discussion. Recall that the aspect ratio $a= 2\ell/d$, the coherence path
length divided by the shock thickness.   Therefore, as $d$ gets larger, it is
likely that $a$ gets smaller.    In both Figures 13 for  $T_b$ and Figure 14
for $L_{\rm iso}$, $a$ is held constant, even as $d$
increases
from roughly $3\times 10^{12}$ cm at $n_0 \sim 10^8$ cm$^{-3}$ to $10^{14}$ cm
at $n_0 \sim 10^6$ cm$^{-3}$.   This implies that the coherence length is
assumed to increase by a factor of roughly 30 as $n_0$ decreases over this
range.    This is likely not physically plausible; $\ell$ will not exactly
scale as $d$, and, in fact, $a$ will likely decrease as $d$ increases.   Both
$T_b$ and $L_{\rm iso}$ scale with $a^3$.   Therefore, the regions of higher
preshock density may be more likely to give larger $T_b$ and $L_{\rm iso}$.

All the equations derived in this subsection assume that the shocked slab is
geometrically thin, that is, $d<\dthin$ (see \S  2.3, Table 1, and Eqs. B7 and
B8).  In this case, the maser spot size observed in the parallel direction is
$d$, that is, it is limited by the thickness of the masing region--it is
"matter" bounded. In Appendix C we justify this assumption, using the
approximate analytic formulae we have derived.
The equations in this section also assume the \htwo\ formation rate
coefficient, $\gamma$, determined by HM79. As discussed at the end of Section
\ref{sec:numjshock}, the value of $\gamma$ has a significant effect on the
shock structure. For example, if $\gamma$ drops from $3\times 10^{-18}\;\rate$
to $3\times 10^{-19}\;\rate$, Table 2 shows that this decrease in $\gamma$
leads to a decrease in the brightness temperature by a factor of  $\sim 30$
when we include the $T_p$ dependence.  On the other hand, the maser spot size
$d$ increases by a factor of 10; as a result, for a fixed aspect ratio $a$, the
isotropic luminosity $L_{\rm iso}$, which is proportional to $d^2 T_b$,
actually increases by a factor of $\sim 3$ for this decrease in $\gamma$. This
increase in the maser luminosity is primarily due to the increased area of the
maser spot and the increased length of the coherence path $2\ell$ associated
with the assumption of a constant aspect ratio.

\subsection{Summary of Results for H$_2$O Masers Produced By Fast J Shocks}

We summarize the approximate analytic fits to the numerical results of \S2, \S3
and \S4 in two tables.  Table 3 presents the fits to the parameters $d$,
$\gamma$, $\xi$, $T_p$, $x_{-4}$(H$_2$O), $c_\eta$, $\eta$, $a_{\rm sat}$,
$d/d_\perp$, $T_b$ and $L_{\rm iso}$ as functions of $j=n_0v_s$, $v_s$,
$v_{A\perp}$, $\Delta v_D$ and $a$. The fits are good to better than a factor
of 2 (see Section 4.1 for the individual error estimates) over the range 10$^6$
cm$^{-3}\la n_0 \la 10^8$ cm$^{-3}$ and 30 km s$^{-1} \la v_s \la 160$ km
s$^{-1}$, which is the range that produces strong J-shock H$_2$O masers.

Table 4 inverts the equations in Table 3 so that the shock parameters $n_0$,
$v_s$, $j$, $B_{0\perp}$ or $v_{A\perp}$ and $a$ are derived in terms of the
potential
observables $v_{A\perp}$
or $B_{0\perp}$, $d$, $d/d_\perp$, $B_p$, and either $T_b$ or $L_{\rm
iso}$.\footnote{
For shocks propagating in the plane of the sky, which give the brightest
masers, $B_{0\perp}$ can be inferred via the Chandrasekhar-Fermi method. The
corresponding Alfv\'en velocity, $v_{A\perp}$, can be inferred only if the
ambient density, $n_0$, can be measured also. We have included $v_{A\perp}$ as
a ``potential observable" and have given most of the parameters in Table 4 in
terms of it because $B_0$ scales as a moderate power of the density (Crutcher
et al. 2010 find $B_0\propto n_0^{0.65}$) so that $v_A$ varies with the ambient
conditions much less than $B_0$.}  
We also give an expression for $c_\eta$
since it is needed in the expression for $a$.  It may
 be the case that $v_{A\perp}$ is not directly observable, but that a
rough estimate of $n_0$ can be obtained. In this case we can use
\eq{
v_{A\perp,5}\simeq  n_{0,7}^{0.45} d_{13}^{0.5} B_p^{0.1},
}
which is obtained by  inverting the expression for $n_{0,7}$ given in the top
line of Table 4. Recall $B_p$ is measured in Gauss. Note the weak dependence on
$n_0$ and especially $B_p$, which enables an estimate of $v_{A\perp}$ even when
$n_0$ is only roughly estimated and $B_p$ is even more uncertain.

Figure 15 graphically plots the J-shock parameter space that produces strong 22
GHz H$_2$O masers, and indicates the physical mechanisms that intercede to
reduce maser activity in J shocks. Above $n_0 \ga 10^8$ cm$^{-3}$, the maser
inversion is quenched in J shocks by the high densities and high optical depths
in the H$_2$O infrared transitions, which drive the H$_2$O rotational levels to
LTE and reduce the inversion in the maser levels. Below about $n_0\sim 10^6$
cm$^{-3}$, masers with $a\la 10-100$ are weak and unsaturated (``starved").

Above $v_s \ga 200$ km s$^{-1}$, the J shocks destroy most of the dust grains,
leaving no grain surface upon which H$_2$ can reform.  As a result,
insufficient columns of warm H$_2$O are produced in the postshock gas, and no
observable H$_2$O masers are excited.

For $v_s \la 40$ km s$^{-1}$, C shocks rather than J shocks may form in dense
molecular gas (cf., Draine \& McKee 1993, and references therein).  We have
marked the boundary of C shocks with J shocks with a dashed  vertical line in
Figure 15 at $v_s= 40$ km s$^{-1}$. This is appropriate if the gas is weakly
ionized (ionization fractions of $\la 10^{-7}$), so that charged grains mediate
the C shock.  However, if the dense preshock gas is more highly ionized,
perhaps by the UV photons from nearby faster J shocks, this boundary between C
and J shocks moves to lower values of $v_s$, and the J-shock maser parameter
space is extended to lower $v_s$.    For example, for ionization fractions of
about $10^{-5}$ the boundary between J and C shocks occurs near $v_s\sim 10$ km
s$^{-1}$ (Smith \& Brand 1990).  To the left of the solid line ($v_s \la 15$ km s$^{-1}$, marked ``too
cold"), no H$_2$ re-formation plateau is produced in J shocks because too few
preshock H$_2$ molecules are dissociated in  J shocks.

\subsection{J-Shock Masers Versus C-Shock Masers}

Figure 15 also roughly indicates the region of parameter space where C shocks
may produce water masers. Kaufman \& Neufeld (1996) model H$_2$O maser emission
from such C shocks. Here, the H$_2$ is not dissociated, but is kept warm over a
large column by the ambipolar heating of the neutrals as they drift through the
ions.  In non-dissociating C shocks, the low density boundary (marked by solid
horizontal line at $n_0= 10^7$ cm$^{-3}$) is at a higher density than in J
shocks because of less compression of the gas in the warm region. Similarly,
the upper boundary marked by quenching is raised in C shocks to $n_0\sim 10^9$
cm$^{-3}$.

Several constraints bound the velocity range of C shock masers.
For C shocks with low ionization fraction, the postshock peak
temperatures are too cold to excite the maser for $v_s \la 15$ km s$^{-1}$.
However, for higher ionization fractions, C shocks are warm enough to excite
maser emission at shock velocities as low as $v_s\sim 5$ km s$^{-1}$.  
Another factor
affecting the low-velocity boundary of C-shock masers is the velocity required to
sputter water-ice mantles off the grains.   The C-shock masers occur in regions of high
density, $n_0 \ga 10^7$ cm$^{-3}$, and the freeze-out times for gas phase molecules
is very short, $\la 100$ years.   The grains are likely warm enough to thermally desorb
CO, but may not be warm enough ($T_{gr} \la 100$ K) to prevent the formation of water-ice mantles.
In addition, the FUV radiation fields may sufficiently attenuated to prevent photodesorption of the
ice.  Therefore, for C shocks to produce strong water maser emission, they must sputter the
ice mantles off the grains in these regions, and Draine (1995)  estimates that 
only 10\% of the water ice is sputtered off by C shocks with $v_s=20$ km s$^{-1}$.
Hence, unless the radiation field in the C-shock maser region is high enough
to warm the grains to $T_{gr} \ga 100$ K, C shocks must have
$v_s \ga 20$ km s$^{-1}$ to produce strong water-maser emission.
As
noted above in our discussion of J shocks, the high-velocity boundary for C
shocks also depends on the ionization fraction in the gas, and is likely of
order $v_s \la 30-50$ km s$^{-1}$.

Are most water masers produced in J shocks or in C shocks?   This is a
difficult question to answer with certainty.  One measure might be the ram
pressure $\propto n_0v_s^2$ needed to drive masing shocks.   C shocks require
lower shock velocities but higher preshock densities than J shocks.  As seen in
Figure 15, these two effects roughly cancel each other, suggesting J shocks and
C shocks require roughly the same driving pressure and, thus, from this point
of view, could be equally likely.  However, there may be more gas in the
density range that can produce J-shock masers than in the higher density range
required for C-shock masers, which would favor J shocks.  Water masers require
densities of $\sim 10^8-10^9$ cm$^{-3}$.
Regions of this density are rare, especially at distances $\ga 100$ AU from a
central protostar along the jet axis, where many masers are observed. The maser
emission from C shocks must come from gas that is close to this density,
whereas the emission from J shocks comes from gas that has been compressed from
a density (only)  $\sim 10^6-10^7$ cm$^{-3}$.
Another factor to consider is the relative values of the key maser parameter $\xi$, which
is proportional to the product of warm postshock column $N_p$ times postshock density $n_p$.
J shocks not only have the advantage in producing higher postshock densities for a given
preshock density, as discussed above, but J shocks also produce larger columns of warm
gas.   In J shocks, the column is determined by the time to reform H$_2$ in the postshock
gas, and $N_p \sim 10^{22}$ cm$^{-2}$, as we have shown.    In C shocks the warm column
is determined by the column needed for ions to collide with neutrals and drag the neutrals up
to the shock speed.   In dense regions, the ionization fraction is low and small charged grains
mediate the C shock.   The warm coupling column here is only $N_p \sim 10^{21}$ cm$^{-2}$
(e.g., Kaufman \& Neufeld 1996).  If gas ions dominate rather than charged grains, the column
is even smaller.
The smaller value of $N_p$ results in a smaller value of $\xi$ and therefore of $T_b$ and
$L_{\rm iso}$, both of which vary as $\xi^{1/2}$ (Eqs. \ref{eq:Tb} and \ref{eq:Liso}).

Another method of distinguishing between the two types of shocks would be to
infer the shock velocity, with the idea that slower masers might be C-shock
masers. However, in shock masers the maser is beamed perpendicular to the shock
velocity (in the plane of the shock).  Therefore, even if the shock velocity is
high, the Doppler velocity observed will be low. Proper motion studies are
needed to try to estimate the shock velocity. Unfortunately, these studies
determine the velocity of the shocked gas, not of the shock itself. For
example, if high-velocity gas containing dust grains impacts a stationary dense
clump or a protoplanetary disk and produces a J shock, the postshock gas would
be {\it decelerated} to a speed similar to that of the dense gas, resulting in
a small proper-motion velocity.  The contrary could also occur: if the observed
masing gas had a high velocity, one cannot be sure that the maser was induced
by J shocks since the emission could originate in fast moving clumps with slow
C shocks moving through them.   In short, it is difficult to distinguish
J-shock masers from C-shock masers by velocity information alone.

Liljestrom \& Gwinn (2000) observed 146 maser spots in W49N in the 22 GHz water
maser line.   Although no attempt was made to compare their observations with
C-shock models, they found good agreement with our J-shock models, inferring
shock velocities of order 30-100 km s$^{-1}$ and aspect ratios of 30-50. In
addition, their inferred values of $T_p$ and $d$ matched the predictions of
J-shock maser models.  Many of their masers features had Doppler  velocities in
excess of 30 km s$^{-1}$ and up to $\pm 200$ km s$^{-1}$, again suggesting, but
not proving, that the masers were produced by J shocks.

J-shock masers might be distinguished by their atomic and ionic infrared line emission.
The shocks producing the maser spots likely have typical sizes of order the
masing region, or $\ga 100$ AU.   This is a lower limit; in massive star-forming regions
like W49 the size is likely of order $10^{17}$ cm.   The shock area $A_{\rm shk}$ therefore at
least $\ga 10^{30}$ cm$^{-2}$ in low mass star-forming regions and could be as high
as $10^{34}$ cm$^{-2}$ in high mass star-forming regions. 
The shock is very embedded, so that the
emergent cooling lines must lie in the mid to far infrared, so that they can
penetrate the high dust extinction.   J shocks differ from C shocks in that
they create singly ionized and atomic species, which are strong coolants.   C
shocks are molecular and mainly cool via molecular rotational lines.    One
observational test of a J-shock origin is therefore to look for strong infrared
cooling transitions from atomic or singly ionized species.    For example, in
our standard model of a J shock with $n_0= 10^7$ cm$^{-3}$ and $v_s= 100$ km
s$^{-1}$, we find that the luminosities in the [NeII] 12.8 $\mu$m, [FeII] 26
$\mu$m, and [OI] 63 $\mu$m lines are $3.1\times 10^{-5}
(A_{\rm shk}/10^{30} {\rm cm^{-2}})$ L$_\odot$, $2.1\times 10^{-4} (A_{\rm
shk}/10^{30} {\rm cm^{-2}})$ L$_\odot$,   
 and $1.7\times 10^{-3}(A_{\rm
shk}/10^{30} {\rm cm^{-2}})$ L$_\odot$, respectively.   Even with an area of $A_{\rm
shk}= 10^{30} {\rm cm^{-2}}$, these lines can be detected by the Stratospheric Observatory for
Infrared Astronomy (SOFIA) in nearby ($\la 1$ kpc) maser
regions.   The angular resolution of SOFIA for these lines
may not be sufficient to spatially resolve these lines,
but SOFIA has the sensitivity to detect the fluxes from these
lines.  We note that although the maser lines originate from portions of the shock
nearly edge on, the full shock will likely have considerable portions directed along the line
of sight, and therefore the shock IR emission lines could be distinguished from background
photodissociation (PDR) regions or HII regions by their width or velocity shifts, which will be of
order 30-200 km s$^{-1}$.    The [NeII] 12.8 $\mu$m line can be also observed with high
spectral resolution by 8 meter
class telescopes from the ground, at least in the nearest
($\la 1$ kpc) masing regions.   Here, the spatial resolution is roughly 3 times better than
the obtained on SOFIA, and this will also 
help to disentangle the J shock masing region from background PDRs or HII regions.
We note, however, that the [NeII] line is very sensitive to the J shock velocity, and is strong
only for $v_s \ga 100$ km s$^{-1}$.

Finally, one might appeal to ratios of different masing lines to determine the
temperature of the masing gas.   As discussed in this paper, J-shock masers
likely cannot heat the masing gas to temperatures greater than about $T_p\sim
400$ K.   However, C-shock masers can heat the masing regions to $T_p\ga 1000$
K.  As discussed in the Introduction, maser regions as hot as 1000 K will
excite not only the 22 GHz maser, but also a number of submillimeter masers
(Kaufman \& Neufeld 1996).  These authors have applied their results to
observations of submillimeter masers, which almost certainly are produced by C
shocks.   However, there are many more regions where only 22 GHz masers are
seen (see Introduction), and these masers are likely produced in cooler $T_p
\sim 400$ K gas.   Although this is suggestive of J-shock masers, again the
proof is not definitive since C shocks can also produce dense molecular gas
with maximum temperatures of about 400 K.

\section{Global Luminosity of a Masing Region}
\label{sec:global}

Up to this point we have been discussing the maser emission from a single spot.
We have often used a planar disk maser as a model that provides a single maser
spot for an observer in the plane. The total maser luminosity from the disk,
$L_m$, includes the emission seen by observers at all orientations with respect
to the maser; it is less than the isotropic luminosity, $\liso$, since the
emission is confined to solid angle near the plane of the disk. However, it is
unlikely that the maser emission is confined to a single region associated with
a given maser spot. Astrophysical shock waves generally cover a significant
solid angle as measured from the source of the shock, and as a result they are
likely to produce many maser spots, as is often observed. It is therefore
instructive to adopt a global viewpoint: What is the total maser luminosity,
$L_{m,G}$, emanating from a shock that is produced by a given astronomical
phenomenon, such as a wind, an accretion flow, a density wave, or an explosion?
For a given shock geometry, which in principle can be inferred from the
geometry and kinematics of the maser spots, it is possible to predict the
global isotropic luminosity of the maser emission, $\lisog$. Provided we do not
have a special location with respect to the maser, this global isotropic
luminosity will be about the same as the total isotropic luminosity of all the
observed maser spots.

Consider a shock with an area $A_{\rm shk}$ that produces masing gas with a
thickness $d$. The shape of the shock, such as part of a spherical shell, is
determined by the mechanism that produced the shock. The total volume of the
masing gas is $V_{m,G}=A_{\rm shck} d$. If a fraction $f_m$  of this volume is
saturated, then the total luminosity of the maser---the global sum of maser
spots radiating in all directions permitted by the shock geometry---is
 \beq
   L_{m,G}=\Phi_m h\nu_0 f_m V_{m,G}
 \eeq
where $\Phi_m$ is the volume production rate of maser photons (see Equation
\ref{eq:Phi_m}). Using the maser photon production rate per Hz from Equation
(\ref{eq:Phi}) and integrating over the line profile, the maser emission per
unit area is then
 \beq
    \call_{m}\equiv\frac{L_{m,G}}{A_{\rm shk}} = 0.075 f_m\dvf \xi^{1/2} c_\eta
         e^{-460/T}~~~\mbox{erg cm\ee\ s\e}, \label{eq:call}
 \eeq
which improves upon the result given by Maoz \& McKee (1998). This expression
is quite general, and applies to masers excited by X-rays (Neufeld, Maloney \&
Conger 1994) as well, provided the appropriate value of $\xi$ for the X-ray
heated gas is used. If the medium is turbulent on scales larger than the shock
thickness, then $f_m$ in this expression should be interpreted as the areal
covering factor of the saturated emission. We do not expect significant
turbulence on scales smaller than the shock thickness; however, if there were
significant density fluctuations on such small scales, our results would not
apply.  The total maser luminosity is proportional to the area, and is
naturally much greater for observable extragalactic masers than for galactic
ones.

The maser luminosity is not an observable quantity, however; rather, it is the
global isotropic luminosity, $L_{{\rm iso,G}} \equiv 4 \pi D^2 F_{{\rm
obs},G}$,
that is measurable, where $F_{{obs},G}$ is the total flux measured by an
observer from all the spots in a masing region. 
If the maser emission covers a fraction $C$ of the sky---i.e., if the masing region radiates
into a solid angle $\oem=4\pi C$, which means that $C$ is also the fraction of random observers
who will see the masers from the region---then the average isotropic luminosity in that solid
angle is
 \beq
 L_{{\rm iso,G}} = \frac{1}{C}\; L_{m,G} = \frac{1}{C}\; \call_{m} A_{\rm shk} .
 \label{eq:liso}
 \eeq
In general,  emission from the masing region will vary with direction inside
$\oem$, so that the isotropic luminosity inferred by a given observer might
differ from the average somewhat. It should be noted that $\oem$ differs from
the maser beaming angle
of a single maser spot,
$\Omega=F_m/I$, which relates the flux emitted at the maser surface to the
intensity of the maser radiation. For example, a sphere has $\oem=4\pi$,
corresponding to a covering factor of unity,
whereas its maser emission can be tightly beamed, with $\Omega\ll 4\pi$. In
Appendix \ref{App:planar}, we show that
for a single maser spot
$\oem/\Omega=A_m/\aobs$, where $A_m$ is the area over which the maser radiation
is emitted and $\aobs$ is the observed size of the maser. Furthermore, both
$\oem$ and $\Omega$ differ from the observed angular size of the maser,
$\oobs=\fobs/I=\aobs/D^2$.

Disks and cylinders are idealized models for maser emission on the micro-scale.
Such structures can be produced by large scale flows associated with accretion
disks or with shocks driven by winds or explosions. In accretion disks, maser
emission can be produced in density-wave shocks (Maoz \& McKee 1998) or by
X-ray illumination (Neufeld et al 1994). In both cases, the emission is from a
ring of gas, and it is generally beamed close to the plane of the disk. If the
maser emits into an angle $2\theta_{\rm em}$ above and below the plane, then
the maser emission from a ring is concentrated in a solid angle $\oem =
2\pi\times 2\sin\theta_{\rm em} = 4\pi\sin\theta_{\rm em}$,
corresponding to a covering factor $C=\sin\them$. 
In the case of a
density-wave shock, the emission comes from a ring of vertical thickness $h$;
at a radius $R$, the area of the shock is then $A_{\rm shk}=2\pi Rh$. The
average isotropic luminosity of such a ring is then
 \beqa
 L_{{\rm iso,G}} &=& \left(\frac{2\pi Rh}{\sin\theta_{\rm em}}\right) \call_{m},\\
  &=&1.2 \dvf\xi^{1/2} c_\eta e^{-460/T}\left(\frac{f_mR_{18}h_{16}}
  {\sin\theta_{\rm em}}\right)~~~L_\odot,
 \label{eq:lisoring}
 \eeqa
where $R_{18}=R/(10^{18}$~cm), etc.; the normalizations have been chosen in
conformity with Maoz \& McKee (1998). This is the total isotropic luminosity of
the ring, including emission from both sides of the disk; the isotropic
luminosity corresponding to just one side of the disk (i.e., to either the blue
or the red emission) is half this. A given observer may see the emission from
the ring as arising from a number of individual spots, which may result from
the alignment of different filamentary masers (Kartje, K\"onigl, \& Elitzur
1999). However, the time-averaged emission of all the spots at a given velocity
should correspond to half the average isotropic luminosity in Equation
(\ref{eq:lisoring}).

Outflows and explosions drive shocks that can give rise to maser emission. For
a complete spherical shell of radius $R$, the average isotropic luminosity is
simply $L_{{\rm iso,G}}=4\pi R^2\call_{m}$. Outflows from protostars and AGN
are more likely to produce shocks that extend over only a part of the sky. We
approximate such a shock as being part of a spherical shell that subtends a
solid angle $\osh=2\pi(1-\cos\thsh)$ as seen from the center of the sphere; the
area of the shock is then $A_{\rm shk}=R^2\osh$. One can then show that the
maser emits into a solid angle 
$\oem=4\pi\sin\thsh$ for $\thsh\leq \pi/2$,
provided $\thsh$ is not too small. Observe that for $\thsh=\pi/2$, the emission
fills $4\pi$ sr; thus, a hemisphere emits in all directions. For $\thsh>\pi/2$,
the emission solid angle remains $\oem=4\pi$. However, if $\thsh$ is too small,
the beaming is determined by the thickness of the shell rather than its
curvature. In this case, the  partial shell approaches a disk of radius
$\ell=R\thsh$. If we define the emission angle $\them$ through 
\beq
C= \frac{\oem}{4\pi} =\sin\them,
 \label{eq:oem}
 \eeq
which is consistent with the above discussion of emission from a ring, then
(for $\them=1/(2a)\ll 1$) $\them=C=
L_{m,G}/L_{{\rm iso,G}}=d/4\ell$ (EHM92).
There is a critical value of $\thsh$ for which $\them$ for the disk equals
$\them=\thsh$ for the shell; this value is \beq \thsd\equiv\frac 12
\left(\frac{d}{R}\right)^{1/2}. \eeq The beaming is like that due to a disk for
$\thsh<\thsd$. The maximum size of a disk that can fit into a shell is then
$\ell_{\rm max}=R\thsd=\frac 12(Rd)^{1/2}$, and the corresponding maximum
aspect ratio is $a_{\rm max}=(R/d)^{1/2}$. From Equations (\ref{eq:liso}) and
(\ref{eq:oem}), the average isotropic luminosity of a shell is then
 \beqa
 L_{{\rm iso,G}}&=&2\pi R^2 \left(\frac{1-\cos\thsh}{\sin\them}\right) \call_{m},
 \label{eq:liso5}\\
  &=& 1.2\times 10^{-4} R_{15}^2\left(\frac{1-\cos\thsh}{\sin\them}\right)
    f_m \dvf\xi^{1/2} c_\eta e^{-460/T} ~~~L_\odot
 \eeqa
 where
 \beq
 \them=\left\{\begin{array}{l} \dis\frac{d}{4\ell}~~~~~~~~\thsh<\thsd=\frac12 \left(\frac{d}{R}\right)^{1/2}\\
 	\dis\thsh~~~~~\thsd\leq\thsh\leq\frac{\pi}{2}\\
	\dis\frac{\pi}{2}~~~~~~~~\frac{\pi}{2}\leq\thsh
	\end{array} \right.
\label{eq:them} \eeq
 For $\thsh\leq\thsd$ and $f_m=1$,
this reduces to Equation (\ref{eq:Liso}) for a disk, as it should. At high
resolution, the shell will break up into individual spots with a spatial
distribution that reflects the overall geometry of the shell.

As discussed in \S 4.1, Equation (\ref{eq:liso1}) for a maser spot can be
misleading since the aspect ratio is $2\ell/d$, and $d$  depends on $n_0$,
$v_s$, and $B_{0\perp}$, whereas $\ell$ can depend on other quantities such as
the shock curvature $R$.  Therefore, $a$ can decrease with increasing $d$, and
should not be treated as a constant independent of $d$.   The dependence of the
isotropic luminosity of an entire masing region on the 
geometric
properties of the source
is more clearly shown by the expression for $L_{\rm iso,G}$ in Equation
(\ref{eq:liso5}), which is based on the assumption that the masing gas is part
of a spherical shell of radius $R$ that subtends an angle $2\theta_{\rm shell}$
as seen from the center of the sphere:
 \beq 
 L_{\rm iso,G}\simeq0.50\times 10^{-4} R_{15}^2\left(\frac{1-\cos\thsh}{\sin\them}\right)
    f_m\dvf^{0.5} j_{14}^{0.75} c_\eta~~~L_\odot,
 \eeq
where $\them$ is given by Equation (\ref{eq:them}); recall that $f_m$ is the
filling factor of the masing gas. If the masing segment of the shell is small
[$\thsh<0.5(d/R)^{1/2}$], then 
%cfmnew I have simplified the discussion
$L_{\rm iso,G} = \liso$ as noted above; for fixed $\ell$, both vary as $1/d$.
%cf
%djhnew  It is also proportional to xi^{1/2}
%cfmnew Since the whole paragraph is about d, I don't think we need more than what I said above about the geometric properties.
For the more typical case in which the shell is not small, the global isotropic
luminosity is independent of the shell thickness for fixed $j$; it represents the isotropic luminosity of all the maser spots in the source and no longer scales in the same way as the isotropic luminosity of a single maser spot.
%cf
%djhold  I do not exactly understand the above sentence.   (cfm--which is not different) As you noted, the isotropic luminosity
%does depend on the shock area times d, that is, the volume of the masing gas.   It seems like
%what you have done is bury that d dependence in your parameters above for j.
%cfm: Here is my explanation:
%This is at first somewhat surprising, since the pathlength through the shell increases as $d^{1/2}$, so one might expect an increase in the global isotropic luminosity as $d$ increases. We can see why the global isotropic luminosity is constant by considering a simple model in which the shell is made up of a number of disks. In such a model, an observer would see a number of spots arrayed along the rim of the  shell with a spacing equal to $2\ell$. The global isotropic luminosity would then be the luminosity of each disk $\propto \ell^3/d$ times the number of maser spots $\propto 1/\ell$, so that the total global isotropic luminosity is constant: $ (\ell^3/d)\times 1/\ell\propto \ell^2=$~const. Of course, in reality the maser spots do not form a regular array due to turbulence, butthe overall energetic argument is unchanged.
%
%cf

%djhnew  I added this final paragraph (from Moshe slightly rewritten by me, to the end of this subsection
%However, the a_max that you get from the geometry of the curved slab is a factor of 8^{1/2} bigger than
%what Chris found in his analysis.  I am still working on understanding that, and we can discuss.
We conclude this section with a final point on the aspect ratios that can be achieved
in a shocked shell, expanding outwards at velocity $v_s$, and with a radius of curvature $R$.
There is a geometric limit on $a$ as noted 
%cfm  in EHM:  
above: 
%cf
If $d$ is the thickness of the shell,
then the maximum physical length of the coherence path 2$\ell$ is $(
%cfm   8
Rd)^{1/2}$ and the maximum
velocity coherence length is $R\Delta v_D/v_s$.  As a result, the maximum aspect
ratio is
\eq{
   a_{\rm max} = \min\left[(\Delta v_D/v_s)(R/d), (
%cfm 8
   R/d)^{1/2}\right]
}
For example, if $\Delta v_D=1$ km s$^{-1}$ and $v_s= 50$ km s$^{-1}$, then $\Delta v_D/v_s = 2\x\E{-2}$, an aspect ratio of 50 can be
achieved in principle if $d= 10^{13}$ cm and $R\simeq 2.5\x\E{16}$ cm, comparable with the
observed sizes of clusters of H$_2$O 22 GHz maser spots.
%cfm It is important to have added this. I think we should also say:
Note that this limit on $a$ also applies to the case in which two masing filaments are aligned: the total aspect
ratio of the combined masing regions cannot exceed this limit if the masers are part of an expanding spherical shell.
%cf
%cfmnew: We shall work on this further while the referee is working on the paper. My task is to show that this is a valid limit provided the maser emits uniformly over the covering factor. For a single maser spot, the global isotropic maser luminosity is the same as that from a spot; hence, the expression for LisoG is an upper limit on equation 2-15 for Liso from a spot. I have to see if that gives the limit I quote. However, it is possible that the maser does not emit uniformly and that some observers will see larger values of Tb and Liso. It occurred to me that we could approximate this by dividing the observed part of the shell into disks of thickness d/2, d/4, etc. For the d/2 disk, the effective ell is 2^1/4 root Rd; from eq. 2-15, I find that Liso is reduced by only a factor 1/2^{1/4} from the value for a thickness d (note that xi does vary as d here, since we are integrating only over part of the maser). If you place a maser of thickness d/4 on top of this, then the amplification will be dominated by rays from the larger disk, so an observer in the plane of the larger disk will not see much from this one. Approximating 2^1/4 as unity, this gives an aspect ratio of 2 ell/d = (8Rd)^1/2. Hence, we are both right, but we need to explain why. I think that my result gives the average Liso, whereas the other approach gives the maximum. Now that I think of it, Dave, if you want to explain this in the text before submitting, please feel free to do so.

\section{CONCLUSIONS AND COMPARISON TO OBSERVATIONS}

Using a grid of numerical shock models coupled with a grid of slab models for
H$_2$O maser production, we have shown that J shocks in the range $10^6$
cm$^{-3} \la n_0 \la 10^8$ cm$^{-3}$ and 30 km s$^{-1} \la v_s \la 200$ km
s$^{-1}$ produce strong, saturated, beamed 22 GHz H$_2$O masers.  The masers
are generally beamed because the velocity coherence pathlength $2\ell$ in the
shock plane is usually greater than the masing pathlength $d$ in the direction
of the shock velocity; this beaming is therefore characterized by the aspect
ratio $a \equiv 2\ell /d\ga 1$. The numerical results of the combined shock and
maser models are shown in Figures 11-15.

We have also presented useful analytic formulae (Tables 2-4) that show how the
observed maser spot size $d$, shape ($d/d_\perp$), flux (or isotropic
luminosity, $L_{\rm iso}$), and brightness temperature $T_b$ scale with the
physical parameters in the shock regions ($n_0$, $v_s$, $v_{A\perp}$, $\Delta v_D$, and $a$).
In addition we invert these equations so that the observed quantities can be
used to derive the physical parameters in the shock region.  Table 2 presents
analytic equations derived from shock and maser theory, whereas Tables 3 and 4
provide analytic fits to the numerical shock models. We note in Table 3 and \S 5 that a number
of the key parameters ($\xi$, $c_\eta$, $\eta$, $a_{sat}$, $d/d_\perp$, $L_{{\rm iso,G}}$, and $T_b$)
depend only on the combination $j=n_0v_s$, rather than on $n_0$ and $v_s$
separately.  The maser results in all
three tables assume that the maser is saturated. A key difference among the
tables is that in Table 2 the average values in the maser plateau of $x({\rm
H_2O})$ and of $\gamma$, the rate coefficient for H$_2$ formation on grain
surfaces, appear in the equations; in Tables 3 and 4 the numerical results for
these parameters provided by the shock models are incorporated into the
resulting equations, so that these parameters do not appear.

We conclude with a summary of how 22 GHz water maser observations  of $T_b$,
$L_{\rm iso}$, $d$, $d/d_\perp$, $B_p$, $T_p$, $n_p$, beaming, maser
velocity, and maser transience correspond to the theoretical models described in this paper.

{\it $T_b$}.   Observed 22 GHz maser brightness temperatures range from $T_b
\sim 10^{11}$ to $10^{14}$ K (e.g., Genzel 1986, Gwinn 1994b).   Figure 3 shows
that it is impossible for any 400 K slab of gas with an aspect ratio $a=10$ in
the plane of the slab to produce brightness temperatures in excess of about
$10^{12}$ K.    This is independent of whether the slab was produced by shocks,
or by some other mechanism.   Therefore, to reach brightness temperatures of
$10^{14}$ K, either high aspect ratios, $a\ga50$ are required, or there must be
two masing regions lined up along the line of sight such that their
``effective" $a$ is of this order (see Elitzur, McKee \& Hollenbach 1991).
Figure 13 shows that J shocks characterized by preshock densities roughly in
the range $10^6$ cm$^{-3} \la n_0 \la 10^8$ cm$^{-3}$ and $30$ km s$^{-1} \la
v_s \la 160$ km s$^{-1}$ produce $T_b \sim 1-2 \times 10^{11}$ K if $a=10$, and
require $a\sim 100$
(or again two regions lined up along the line of sight to simulate $a \sim
100$) to produce $T_b \sim 10^{14}$ K.    Since $d \sim 10^{13}-10^{14}$ cm, an
aspect ratio $a=100$ corresponds to a coherence path length of about
$10^{15}-10^{16}$ cm.   Maser brightness temperatures $T_b$ have been observed
to be uncorrelated with maser spot size $d$ (Genzel et al 1981;  Gwinn, Moran,
\& Reid (1992), Gwinn (1994b).  We have also shown that shock models do not
produce a significant correlation of $T_b$ with $d$.  The dependence of $T_b$
on $d$ is very weak, and variations in $T_b$ are primarily controlled by the
aspect ratio $a$, since  $T_b \propto a^3$.

{\it $L_{\rm iso}$}.   Observed isotropic 22 GHz maser luminosities  range from
$\sim 10^{-7}- 10^{-1}$ $L_\odot$ from individual maser spots in the Galaxy (Genzel
\& Downes 1977; Walker, Matsakis \& Garcia-Barreto 1982; Genzel 1986; Gwinn
1994a). In spatially unresolved maser regions, the global isotropic luminosity $L_{\rm iso,G}$ is
higher, since all the spot luminosities are added together. In particular,
Genzel \& Downes (1977) find that sources with maser spectra classified as
``singles" have a mean value of $L_{\rm iso,G}$ of \E{-5}\Lo. Assuming such spectra to
be dominated by one bright maser spot, this would imply $a$ = 18 for the mean
aspect ratio of a single feature if we use $d$ = 1AU in Equation (\ref{eq:Liso}). The
mean value of $L_{\rm iso,G}$ increases with the complexity of the source spectrum;
this can be attributed to an increase in the number of spots contributing to
the overall emission;
physically, it is due to an increase in the overall size of the masing region (Eq. \ref{eq:liso5}). 
The exceptionally luminous maser region W49N is a
Galactic outlier. With $L_{\rm iso,G} = 1.3 L_\odot$, its brightest spot has
$L_{\rm iso} = 0.08 L_\odot$ which is about 8 times the {\em total} isotropic
luminosity of the most luminous maser sources outside the W49 complex. This
outlier status can be attributed to the short lifetime ($\la$ 1,000 years) of
the bright maser phase in high-mass star forming regions (Mac Low et al 1994;
Elitzur 1995). Starburst galaxies can be expected to contain more W49-class
maser sources, and indeed Brogan et al (2010) find three maser regions in the Antennae interacting
galaxies with $L_{\rm iso,G}$ ranging from 1--6 times the
W49 luminosity. 
For the H$_2$O masers associated with star formation and
outflows, the H$_2$O luminosity is correlated with the mechanical luminosity
seen in the CO outflow (Felli, Palagi \& Tofani 1992, Claussen et al 1996,
Furuya et al 2001). Such a correlation is expected in a shock model; the mass
loss produces the shocks that, in turn, produce the masers.
Figure 14 shows that in the range of saturated masers with $10^6$ cm$^{-3}
\la n_0 \la 10^8$ cm$^{-3}$ and $30$ km s$^{-1} \la v_s \la 160$ km s$^{-1}$,
the isotropic luminosity $L_{\rm iso}$ ranges from about $3\times 10^{-7}-10^{-5}$
$L_\odot$ for $a=10$. To achieve isotropic luminosities as high as 0.08
$L_\odot$ would require $a\ga 200$.  Since the $0.08 L_\odot$ upper limit arose
from the extreme case of a maser spot in W49, this spot could correspond to two
coherent regions lining up to give an exceptionally high effective $a$.
For both $T_b$ and $L_{\rm iso}$, we note that the predictions of the model are
dependent on the collisional rate coefficients to excite H$_2$O.   These rate
coefficients, especially for collisions with atomic H (the H$_2$ rates are
often scaled from theoretical results of He collisions, but H is more reactive
than He or H$_2$), are somewhat uncertain, and larger rate coefficients might
also give higher $T_b$ and $L_{\rm iso}$ without requiring 
such
exceptionally high values of the aspect ratio, $a$.

{\it $d$}.   Observed 22 GHz maser spot sizes are of order $10^{13}-10^{14}$ cm
(Genzel 1986; Gwinn 1994a; Torrelles et al 2001a,b; Lekht et al 2007; Marvel et al 2008) when the
maser is spatially resolved by very long baseline interferometry.   Figure 7
shows that this size range falls right in the middle of our optimum J-shock
maser range. Relatively bright and luminous maser spots (Figures 13 and 14) are
predicted in our standard model to occur when $n_0\sim 10^6$ cm$^{-3}$ if the
aspect ratio remains high even as $d$ increases with decreasing $n_0$. Since
shocks are driven by high pressure (ram or thermal) and the frequency of
occurrence of high pressures in the ISM  is a decreasing function of pressure,
one would expect more masers to be found with $n_0= 10^6$ cm$^{-3}$ than with
higher preshock densities.    However, our standard model predicts these masers
to have size $d\sim 10^{14}$ cm, at the upper end of the typical maser size.
If most luminous maser spots are smaller, one explanation for this apparent
discrepancy is that the aspect ratio $a$ decreases with increasing $d$.   This
would greatly lower $T_b$ and $L_{\rm iso}$ (proportional to $a^3$) for masers
with preshock density $n_0\sim 10^6$ cm$^{-3}$. The other possibility is that
our models have overestimated $d$.  Referring to Table 2, we see that $d$ is
proportional to $v_{A\perp}^2/\gamma$.   For our standard shock models, we use
$v_{A\perp}=1$ km s$^{-1}$ and the H$_2$ rate coefficient $\gamma$ is taken
from the $T$ and $T_{\rm gr}$-dependent formulation given in HM79. If
$v_{A\perp}^2/\gamma$ is of order 3 times smaller than we have assumed, the
typical maser spot size will come more closely in alignment with the
observations.  In fact, the equation for $d$ in Table 2 predicts that
$v_{A\perp,5}^2/(\gamma_{-17} j_{14}) \sim 6$ for $d\sim 3\times 10^{13}$ cm.

{\it $d/d_\perp$}.  Figure 12 shows that in the main region of saturated 22 GHz
masers, $d/d_\perp$ in J-shock models lies between 1 and 3.   We therefore
predict that maser spot shapes are often fairly circular and that
masers usually extend in the direction of the shock velocity.
Equation (4-2) shows that the shape variation comes mostly from the Alfven
speed $v_{A}$, since the dependence of $d/\dperp$ on $j$ is rather weak. If the
ambient magnetic field is roughly uniform over the maser region, then the
shapes of the maser spots will vary primarily due to the variation in the
relative orientation of the magnetic field and the shock velocities.
Observationally, maser spot sizes are determined from circular-Gaussian fits to
spectral features in maps of the correlated flux (e.g.,  Gwinn 94, Richards et
al 2011). Discerning the elongations predicted here would require a more
detailed analysis of maser maps that employs elliptical Gaussian fits to
individual maser features, and could provide a new method of inferring the
properties of regions with \hto\ masers.

{\it $B_p$}.    Fiebig \& G\"usten (1989) have observed the Zeeman splitting of
the 22 GHz H$_2$O maser in W49 and estimated the component of the B field along
the line of sight to be about 100 mG.   Sarma et al (2008) and Alves et al (2012) report Zeeman 
splittings in a high mass star-forming region (OH 43.8-0.1) and a low mass star-forming region (IRAS 1693-2422), respectively, 
and estimate B fields of 10-20 mG and 100 mG, respectively.  The shock model predicts  similar values
for $B_p$ in the masing region (Table 2).

{\it $T_p$}.   The 22 GHz water maser lies 644 K above the ground state of
water. Collisional excitation of this maser therefore requires $T\ga 300$ K. On
the other hand, their observed linewidths ($\la 1$ km s$^{-1}$) suggest thermal
temperatures generally $\la 1000$ K (Liljestrom \& Gwinn 2000).
 Millimeter observations, the observed 321 GHz H$_2$O maser (Neufeld \&
Melnick 1990), and the observation that there are not enough external photons
to pump the maser all point to warm gas in the range $300-1000$ K. J shocks
produce a large column of H$_2$O in the lower part of this temperature range,
at 300-400 K (see Figure 5), where collisions can pump the 22 GHz maser.
However, in some regions, observations of other maser transitions of H$_2$O
indicate higher temperatures than J shocks seem to be able to provide in the
H$_2$ re-formation plateau, and C shocks may be implicated since these shocks
can produce higher gas temperatures in an extended molecular column. Kaufman \&
Neufeld (1996) have modeled such shocks and applied their results to
multitransition observations of water masers (Menten, Melnick, \& Phillips
1990; Chernicharo et al 1990; Menten et al 1990; Melnick et al 1993).

{\it $n_p$}.   Genzel (1986) reviews observational evidence that the density in
the 22 GHz masing region is $\ga 10^9$ cm$^{-3}$ (for a recent study, see 
Alves et al 2012).   We see from Figure 3 that
any $T=400$ K slab, whether produced by a shock or not, has much lower $T_b$
once $n\gg 10^9$ cm$^{-3}$ because of the quenching of the maser.     The
beauty of the shock model is that regions of density $n\sim 10^9$ cm$^{-3}$ are
rare, especially $\ga 100$ AU from a central protostar along the jet axis,
where many masers are observed, and  the J shock compresses preshock gas of
density (only)  $\sim 10^7$ cm$^{-3}$ to this density.  We note that C shocks
produce much of the maser emission in their warmer regions, which are not
nearly as compressed as J shocks.   Therefore, C shocks require substantially
higher preshock densities (see Figure 15), which may be rarer.

{\it Beaming}. As expected from the planar shock geometry, maser radiation is
preferentially beamed perpendicular to the motion of the emitting material.
This can be inferred indirectly from the inverse correlation between measured
Doppler velocity and maser brightness (e.g., Genzel 1986) or from the increased
numbers of masers observed with low los velocities compared with high los velocities
(e.g., Walsh et al 2011) and directly from
spatially resolved observations which show that the maser velocity vectors lie
nearly in the plane of the sky (Marvel et al 2008). Beaming angles themselves
are immeasurable and can only be inferred indirectly. Gwinn (1994c) has modeled
the diffuse H$_2$O maser ``haloes" around maser spots in W49N as arising from
scattering by nearby ionized plasma and concluded that the masers are indeed
beamed. His analysis suggests that the beaming angles lie in the range 0.002
$\la \theta _m \la 0.02$ radians, corresponding to aspect ratios $500 \ga a \ga
50$. Unfortunately, in the absence of any followup work to verify the
assumptions entering into the chain of analysis in this pioneering study, these
bounds can only be considered as order-of-magnitude estimates. Aspect ratios of
order $\la 50$, similar to Gwinn's lower bound, are expected in our model for single
maser features while values of \about 200--500 could arise from the alignments
of two maser regions (Deguchi \& Watson 1989; Elitzur, McKee \& Hollenbach
1991), needed to explain the high-end of brightness temperatures (\about \E{14}
K). Another indirect determination of the beaming angle is through time
monitoring. Using spectroscopic results of 146 water maser outbursts in W49N,
Liljestr{\"o}m \& Gwinn (2000) derive aspect ratios of 16--28 if the masers are
filamentary and 29--52 if they are disks. One big flare stands out with aspect
of either 70 or 126, depending on the geometry. All results are consistent with
aspect ratios of order a few tens for single maser regions.

{\it Observed maser velocity.}      As noted above in the discussion on
beaming, the masers with low los velocities are brighter.   However, when time
lapse images show motions in the plane of the sky, we find that, for example in
W49 and IRAS 05413-0104, the space velocities are almost always in excess of 25
km s$^{-1}$ (Gwinn 1994a, Claussen et al 1998, Liljestr{\"o}m \& Gwinn 2000,
Marvel et al, 2008). In W75N, 20 years of monitoring show maser speeds of
$\sim 75$ km s$^{-1}$ (Lekht et al 2007).  These speeds are in line with the velocities needed to
excite masers in J shocks (see Figure 15), although we note that shocks in the
range $v_s\sim 20-40$ km s$^{-1}$ are likely to be C shocks and not J shocks,
unless the ionization fraction is higher than $\sim 10^{-7}$. However, In
applying these models to observations, some thought must be given concerning
the dependency of  the observed line of sight and proper motion velocities of
interstellar H$_2$O masers on the shock velocity $v_s$, the wind/jet velocities
$v_w$ observed in the masing region, and the velocity $v_a$ of the ambient gas.
In strong J shocks, the flow velocity  of the masing gas with respect to the
shock front is very slow, $\sim (n_0/n_p)v_s \sim 10^{-2}v_s$, so that the
masing gas moves at $\sim v_s$ {\it with respect to the preshock gas}. We
envisage the shocks that produce H$_2$O masers as arising when high speed (jet,
wind, or a shell or clump driven by the wind) material moving at $v_w$ from a
protostar interacts with slower \lq\lq ambient" material moving at $v_a$.  The
ambient material might be either a clump in the ambient molecular gas, a
circumstellar disk, or a slowly moving shell of already shocked material
(sometimes, this is identified as ``outflow" material). In this picture the
shock velocity is then $v_s\simeq v_w-v_a$. If the high speed material is
denser than the ambient material, then the ambient gas is shocked up to the
speed of the wind, and the H$_2$O maser is shocked ambient gas observed as a
high velocity clump moving at $\sim v_w$.  On the other hand, if the high speed
material is less dense than the ambient gas (for example, when wind hits the
accretion disk around protostars; an observation of one such case is presented
in Moscadelli et al 2006), then the wind shocks {\it down} to the ambient
speed, and the H$_2$O maser is shocked wind material moving at $\sim v_a$.  In
other words, masers with small space velocities less than 10 km s$^{-1}$ can
still be produced by J shocks if high speed material is shocked {\it down} to
low speed.  In this case, it is important to note that the wind material must
contain dust grains, for the masing plateau to be formed due to the heat of
H$_2$ re-formation. In any of these cases, the maser is beamed perpendicular to
the shock velocity and in the shock plane (i.e., $a$ is higher in the plane of
the shock), so that brighter masers will have higher proper motions speeds than
line-of-sight speeds.

{\it Transience of maser regions.} The \hto\ maser phase of star formation is both widespread and
selective. In the low-mass case, while all Class 0 protostars likely have water
masers, none are found in Class II (Furuya et al 2001, 2003). In the case of
high-mass star formation, the core of the W49 region contains at least ten
distinct ultra-compact HII regions arranged in a ring-like structure with a
diameter of 2 pc (Welch et al.\ 1987).  Only one of these objects is also a
water maser, by far the most powerful in the Galaxy. 
The large number and spatial extent of the maser spots in W49 imply that the
covering factor of the maser emission is large---we are not in a special place
from which to observe the maser, but instead must be at a special time.
This conclusion is reinforced by
Figure 15, which
summarizes our shock and pumping detailed calculations and
shows that the phase-space for \hto\ maser action is
rather large, spanning a substantial range of densities and shock velocities.
Strong \hto\ maser emission is a robust phenomenon, generated for a wide range
of physical conditions. Since there is no need for fine tuning of parameters,
maser action could be expected to occur at some stage of the star formation
process, perhaps in all sources. However, although the phase space for maser
action is large, the conditions are somewhat extreme; in particular, the
preshock density ($\ga$ \E6\,\cc) is rather high. The dimensions of a region
containing such high densities probably cannot exceed \about\,\E{17} cm or so.
As discussed above, maser spots are observed to have space velocities 
that are generally in excess of 25 \kms, and will therefore
move across such a small
region in less than 2000 years. 
In observing a maser in any particular source
we are witnessing a transient phenomenon, which may help explain its
selectiveness --- although it is easy to generate an \hto\ maser, that maser
does not last very long.   In summary, in a given source water maser spots might be
observed for an extended period of time ($\la 2000$ yr), but a given maser spot
tends to have a much shorter life, depending on the time for the aligned
coherent region to point elsewhere.    Similarly, maser emission due to density-wave shocks in accretion disks, as
in the model of Maoz \& McKee (1998), could persist for long periods of time, although
each maser spot would be transient. Maser emission from accretion shocks at the surfaces 
of disks could also
persist, but there is no definitive evidence for such masers at present.

All these comparisons make a clear case that J shocks provide a natural
explanation for many observed characteristics of 22 GHz water masers.
C shocks likely also produce
water masers, and several of the above features of J-shock models apply
equally well to any shock model, in particular that the brightest spots should have
low Doppler velocities and that they are likely to be transient phenomena. J-shock maser models
are distinguished from C-shock models by their high shock velocities and lower required ambient densities.
We have given explicit predictions for the maser spot sizes and shapes. J shocks produce strong
emission in atomic IR lines, which are absent in C shocks. C shocks can produce
strong submiillimeter water masers because the temperatures of their masing regions can exceed $\sim 1000$ K, 
whereas J shocks cannot because their masing regions never exceed 400-500 K.
For masers that are identified as being due to
J shocks, the results of this paper provide powerful diagnostics for determining the
physical conditions in the region of maser emission.

Acknowledgments.  The research of DJH and CFM during the early portion of this
work was supported in part by a NASA grant (RTOP 344-04-10-02) to the Center
for Star Formation Studies, a consortium of theoretical researchers from NASA
Ames, the University of California at Berkeley, and the University of
California at Santa Cruz. CFM's research is also supported by NSF grants
AST-0098365 and AST-1211729.

\newpage

\appendix

\section{MASER BASICS}
\label{App:Basics}

Here we briefly recapitulate the basics of astrophysical masers from E92 and
relate the results described in the text to the treatment in E92. Let $n_i'$ be
the density of molecules in level $i$ and let $n_i\equiv n_i'/g_i$ be the
density per sublevel. For the 22 GHz maser levels, the nuclear spin contributes
a factor 3 to the statistical weights, so that $g_1=33$ and $g_2=39$. Let
$n_{i\nu}'$ be the density in level $i$ per unit frequency of the molecules
that can interact with the maser radiation at frequency $\nu$. Then the maser
level populations are determined by
\eq{\label{eq:m1}
 \non'[\Gamma_1+(g_2/g_1)B_{21} J_\nu]=P_1\phi_\nu +\ntn'(A_{21}+B_{21} J_\nu),
\label{eq:pop1}
}
\eq{\label{eq:m2}
 \ntn'(\Gamma_2+A_{21}+B_{21} J_\nu)=P_2\phi_\nu+\non' (g_2/g_1)B_{21} J_\nu,
\label{eq:pop2}
}
where $J_\nu$ is the angle-averaged maser intensity, $\Gamma_i$ is the loss
rate from level $i$ to non-maser levels, $P_i$ is the pump rate from non-maser
levels into level $i$ and $\phi_\nu$ is the Doppler profile describing the
molecular motions. The standard form $B_{21}J_\nu n_\nu$ for the interaction
rate with maser radiation at frequency $\nu$ is strictly correct only for
linear masers, where both photons and molecules move along a single line so
that there is a unique relation between velocity and frequency in the masing
gas. In realistic geometries with higher dimensions, this expression provides
an adequate approximation within a frequency core with width $x_s\Delta\nu_D$
around line center (Elitzur 1994). The dimensionless, geometry-dependent $x_s$
is 1.7\asat\ for filamentary masers and \about\ \asat\ for disk masers, where
\asat\ is the aspect ratio at the onset of maser saturation (see below). Since
the \hto\ pumping scheme has $\asat \ga 1$ (fig.\ 11),
deviations from the standard radiative rates generally occur sufficiently far
from line center that they can be ignored in most practical applications.

For simplicity, we henceforth take for both levels the same loss rate $\Gamma =
(13\Gamma_2 + 11\Gamma_1)/24$, where the $\Gamma_i$ are the actual results of
the numerical calculations for the 45 ortho-\hto\ rotation levels. With the
conventions we have adopted, $A_{21}$ and $B_{21}$ are related by $A_{21} =
(2h\nu_{21}/\lambda_{21}^2) B_{21}$. The spontaneous transition probability
$A_{21}=1.9\times 10^{-9}$ s\e\ is negligible and may be ignored in Equations
\ref{eq:pop1} and \ref{eq:pop2}. Let $p_i\equiv P_i/g_i$ be the pump rate per
sublevel and define $\Delta p \equiv p_2-p_1$. Then the unsaturated populations
(i.e., the populations evaluated at $J_\nu=0$) are $\noo = p_1/\Gamma$ and
$\nto = p_2/\Gamma$, and the population difference at any maser intensity is
\eq{\label{eq:Dn}
   \Delta n_\nu\equiv \ntn-\non
        = \frac{\Delta p\,\phi_\nu}{\Gamma+(g_1+g_2)B_{21}J_\nu/g_1}
        = \frac{\Delta p\,\phi_\nu}{\Gamma(1+ J_\nu/\Js)},
}
where
\eq{\label{eq:Js}
    \Js \equiv \left(\frac{g_1}{g_1+g_2}\right)\frac{\Gamma}{B_{21}}
}
is the intensity at which the maser saturates.

Before proceeding to specific results for planar masers we first list a number
of important geometry-independent general properties of the \hto\ pump. For a
maser that amplifies its own spontaneous emission, the intensity starts as the
(absolute value of the) unsaturated source function
\eq{
   S_0 = {A_{21}\over B_{21}}\,{n_{2,0}\over n_{2,0} - n_{1,0}}
}
From the definition of the pump efficiency $\eta$ (Equation \ref{eq:q_eta}),
\eq{
    {n_{2,0}\over n_{1,0}} = {p_2 \over p_1} = {1 + \eta \over 1 - \eta}
}
therefore
\eq{\label{eq:S0}
   S_0 = {A_{21}\over B_{21}}\,{1 + \eta \over 2\eta}
}
Saturation occurs when the maser intensity that starts as $S_0$ grows to the
saturation level \Js, and the required degree of amplification is controlled by
$\gamma_m = \Js/S_0$. From Equations \ref{eq:Js} and \ref{eq:S0},
\eq{\label{eq:gammam}
    \gamma_m = {2g_1\over g_2 + g_1}\,{\eta\Gamma\over (1 + \eta)A_{21}}
        \simeq 5.6\x\E6\,{n_9c_\eta\over\xi^{1/2}}\,e^{-400/T}.
}
The second equality, which holds when $\eta \ll 1$, is our specific result for
the \hto\ pumping scheme with the analytic approximations derived for $\Gamma$
and $\eta$ in Equations \ref{eq:q-G} and \ref{eq:eta}, respectively. In this
and following equations, the analytic approximations are valid for $\xi > 0.1$
and $T > 200$ K.

The net production rate of maser photons per unit volume and unit frequency is
$\Phi_{m,\nu} = g_2 B_{21}J_\nu\Delta n_\nu$. For a saturated maser ($J_\nu >
\Js$), Equation \ref{eq:Dn} shows that $\Phi_{m,\nu} = \Phi_m\phi(\nu)$ where
\eq{\label{eq:Phi_m}
        \Phi_m  = g_2 B_{21}\Js{\Delta p\over\Gamma}
                = {g_2g_1\over g_2 + g_1}\,(p_2 - p_1).
}
From the definitions of $q$ and $\eta$ in Equation \ref{eq:q_eta},
\eq{
   \Delta p = \eta(p_2 + p_1) = 2n^2x(\hto) \eta q
}
Replacing the product $n^2x(\hto)$ with the scaling parameter $\xi$ (Equation
\ref{eq:xi}), the photon production rate at line center of a saturated maser is
\eq{\label{eq:Phi}
    \Phi_{m,\nu_0} = 2.7\x\E{-5}\,{\xi\eta_{-2}q_{-13}\over d_{13}}
         \simeq 3.9\x\E{-4}\,{\xi^{1/2}c_\eta\over d_{13}}\,e^{-460/T}
         \quad   {\rm phot\,cm^{-3}\,s^{-1}\,Hz^{-1}}.
}
where the second equality, again, utilizes the analytic approximations for the
\hto\ pumping scheme.

The final relevant quantity is the unsaturated absorption coefficient at
line-center, \ko; independent of the saturation degree, it provides the natural
length scale for the maser. The corresponding optical depth across the slab
thickness, $\tau_{21} = \ko d$, is readily obtained from Equation \ref{eq:tau}.
Expressing similarly the unsaturated maser populations with the parameters of
the \hto\ pumping scheme, the result is
\eq{\label{eq:kod}
    \ko d = 0.82\,{\xi\eta_{-2}q_{-13}\over \Gamma_{-1}}
          \simeq 4.5\,{\xi^{1/2}c_\eta\over n_9}\,e^{-60/T}.
}

The expressions for $\gamma_m$, $\ko d$ and $\Phi_{m,\nu_0}$ in the \hto\
pumping scheme determine the maser properties in any geometry.

\section{PLANAR MASERS}
\label{App:planar}

Here we apply the results of the previous section to planar masers, whose
general solution is presented in EHM92. In this discussion, we shall need to
solve equations of the form
 \beq
    e^x=bx^n
 \eeq
for $n>0$. This equation has solutions only if $b\geq (e/n)^n$. (EHM92
incorrectly stated that solutions exist only for $b\geq e$.) For $x \neq n$,
there are two such solutions, one with $x> n$ and one with $x<n$; since in our
equations $x$ is proportional to $\kappa_0 d$ to some power, we shall assume
that the former solution, with the higher opacity, is the physically relevant
one. An approximate solution of this equation for $x\geq n$ that is accurate to
within about 3\% for $n=1$ and 25\% for $n=3$ is
 \beq
    x\simeq \ln(n^n b)+n\left(2\ln\ln nb^{1/n}\right)^{1/2}.
    \label{eq:expsoln}
 \eeq
For simplicity, we shall generally keep only the first term, which is accurate
to within a factor 1.6 for both $n=1$ and $n=3$.

With the aid of this result, one finds that the saturation condition for a
circular disk maser (see Figure 2) given in EHM92 corresponds to an aspect
ratio (Eq.\ \ref{eq:a}) given by
\eq{\label{eq:asat0}
   \asat =  {1\over\ko d}\ln\left[3(3\pi)^{1/2}{\gamma_m\over\ko d} \right].
}
Equation (\ref{eq:asat}) in the text is obtained by inserting the results of
the \hto\ pumping scheme (Equations \ref{eq:gammam} and \ref{eq:kod}) into this
expression. It is instructive to compare the disk with a cylindrical maser with
diameter $d$ and length $ad$. Such a filamentary maser saturates at the aspect
ratio (Elitzur, McKee \& Hollenbach 1991)
\eq{\label{eq:asat-cyl}
   a_{\rm sat}^{\rm cyl} =
   {1\over\ko d}\ln\left[64{\gamma_m\over(\ko d)^2} \right]
    \simeq 3.7{n_9\over\xi^{1/2}c_\eta}\,e^{60/T}
           \left[1 - {17\over T} + 0.18\ln\,{n_9\over\xi^{1/2}}
           - 0.06\ln\,c_\eta\right];
}
this can also be obtained from EHM92 with the approximation in Equation
(\ref{eq:expsoln}).

Consider now an edge-on planar maser. Denote by $\|$ the direction parallel to
the slab thickness (i.e., parallel to the shock velocity in shock models for
masers) and by $\perp$ the direction orthogonal to it in the plane of the sky.
The EHM92 disk maser solution assumes matter-bounded beaming in the
$\|$-direction, so that the observed size in that direction is the physical
size, $\dpar = d$. In the $\perp$-direction, beaming in the disk plane limits
the observed maser size to \dperp, which is less than $ad$, the physical
dimension in that direction. The size \dperp\ is related to the radius of the
core, $r_{\rm s}$ (Fig.\ 2), through $\ko\dperp = (\pi\ko r_s)^{1/2}$ (note
that EHM92 give the expression for the maser observed area $A_{\rm obs}$, which
is equal to $d\dperp$). When the core is unsaturated and small compared to the
disk radius, \dperp\ is determined from the equation
\eq{\label{eq:dperp0}
   \frac{e^{\frac{2}{\pi}(\ko\dperp)^2}}{(\ko\dperp)^3} =
   \frac{64\gamma_m}{\pi a^2(\ko d)^3}
}
whose approximate solution is
\eq{\label{eq:dperp}
    \ko\dperp \simeq  \left[
    {\pi\over2}\ln\left\{24(3\pi)^{1/2}{\gamma_m\over a^2(\ko d)^3} \right\}
    \right]^{1/2}
}
The observed size, \dperp, is slowly decreasing when the disk size, $ad$, is
increasing. At sufficiently large aspect ratio, the core saturates and \dperp\
begins to grow. This limit is of little interest in the thin disk regime as it
generally requires excessive values of $a$ for water masers.

These results hold so long as beaming in the $\|$-direction remains matter
bounded. This condition breaks down when regions along the disk axis become
saturated, since then the size of the maser spot in the $\|$-direction,
$\dpar$, becomes less than the slab thickness, $d$---i.e., the maser becomes
amplification bounded along all lines of sight. In EHM92 we estimated the disk
thickness at this transition by treating its unsaturated core not as a disk but
as a cylinder with radius $r_s$ and length $d$, and demanding that this
cylinder not develop saturated caps. Here we take a slightly different
approach. Consider instead a saturated spherical maser and imagine removing
material from its caps, producing a structure whose flat top and bottom are
parallel to the line-of sight, separated by distance $d$. Initially, the core
of this structure retains a roughly spherical shape, producing an amplification
bounded ``thick disk". Removing more material and decreasing $d$, rays
propagating along the short axis of the shaved structure are less intense and
need stronger amplification across the core to induce saturation upon exit from
the core. To provide this additional amplification the core begins to expand
along the axis, changing from a spherical to ellipsoidal shape elongated in the
$\|$-direction. Eventually, when $d$ is sufficiently small, the core becomes
unsaturated along the disk axis, reaching the pillbox shape depicted in Figure
2. Denote by $\dthin$ the diameter of a sphere just reaching saturation. This
diameter is given by the relation (EHM92)
\eq{\label{eq:dthin}
    \frac{e^{\ko\dthin}}{\ko\dthin} = 2\gamma_m.
}
Combining the approximate solution of this equation (see Equation
\ref{eq:expsoln}) with the expression for $\ko d$ (Equation \ref{eq:kod})
yields
\eq{\label{eq:thin}
   \frac{d}{\dthin} \simeq 0.28\,\frac{\xi^{1/2}c_\eta}{n_9}\x\frac
   {\DS e^{-60/T}}
   {\DS 1 - {25\over T} + 0.06\ln\,{n_9c_{\eta}\over\xi^{1/2}}}
}
Disks with $d < \dthin$ are certain to be matter bounded in the $\|$-direction
because their thickness is smaller than the smallest sphere that can produce
saturated regions; we term these ``thin disks." In contrast, disks with $d >
\dthin$ will develop saturated regions along the axis, becoming amplification
bounded; we term these ``thick disks."

We describe the matter-bounded behavior of thin disks ($d<\dthin$) with the
EHM92 disk solution. Note that the validity of this solution requires as an
additional constraint the {\it filamentary condition}
\eq{\label{eq:filament}
    a \gg \max[1, \ko d/8]
}
to ensure that the amplification along all rays between the two caps of the
observed filamentary volume (see Figure 2) is roughly the same. Very thick
disks ($ d \gg \dthin$), with amplification-bounded behavior in the
$\|$-direction, can be approximated with the solution of a spherical maser
whose radius $\ell\ (= \half ad)$ is equal to the disk radius. The sphere's
observed radiation is effectively confined to a cylinder aligned with the line
of sight, whose diameter \dl\ is determined exclusively by $\ell$ and the
pumping scheme; it is related to the radius $r_s$ of the sphere's core via
$\ko\dl = 2(\ko r_s)^{1/2}$ (EHM92). When the core is unsaturated, \dl\ is
determined from the equation
\eq{\label{eq:dl0}
   \frac{e^{\frac12(\ko\dl)^2}}{(\ko\dl)^6} =
   \frac{12\gamma_m}{a^4(\ko d)^4}
}
with the approximate solution
\eq{\label{eq:dl}
    \ko\dl \simeq
    \left\{2\ln \left[{2592\gamma_m\over a^4(\ko d)^4}\right] \right\}^{1/2}\,.
}
The core size decreases slowly with $a$ and eventually the core saturates; in
contrast to the thin-disk case, core saturation is relevant for thick disks.
Core saturation occurs when $\ko\ell = 1.6\gamma_m^{1/4}$, corresponding to the
aspect ratio (Elitzur 1990)
\eq{\label{eq:ac}
   \ac = 3.2\; \frac{\gamma_m^{1/4}}{\ko d}
}
In this fully saturated domain, where the sphere is saturated throughout, the
core diameter is
\eq{\label{eq:dls}
    a \ge \ac: \qquad  \qquad   \dl = \frac{ad}{(6\gamma_m)^{1/4}}\,;
    \qquad \qquad \phantom{a \ge \ac:}
}
that is, the core size now {\em increases} linearly with $a$ so that $\dl/d$
remains constant. Thus the behavior of the maser observed shape is as follows:
\begin{itemize}
  \item $d < \dthin$: Beaming is matter bounded in the $\|$-direction, and
      the EHM92 disk solution is applicable for all masers that obey the
      filamentary condition. The observed maser size is $d$ in the
      $\|$-direction and \dperp\ in the $\perp$-direction.

  \item $d \gg \dthin$: Beaming is amplification bounded in both $\|$- and
      $\perp$-directions. The maser shape is a circle with diameter \dl,
      given by Equation (\ref{eq:dl}) when the core is unsaturated ($a <
      \ac$; see Equation \ref{eq:ac}) and by Equation (\ref{eq:dls}) when it
      is saturated ($a > \ac$).
\end{itemize}
A description of the transition between the two regimes, from matter-bounded
($d < \dthin$) to amplification-bounded ($d > \dthin$) behavior in the
$\|$-direction, requires numerical studies because the angular integration of
the intensity cannot be performed in closed form. Also, the approximations in
Equations (\ref{eq:dperp}) and (\ref{eq:dl}) can become inadequate and require
replacement with numerical solutions of Equations (\ref{eq:dperp0}) and
(\ref{eq:dl0}). Nevertheless, the discussion here captures the essence of the
maser behavior as the disk thickness increases.

The brightness temperature of an unsaturated maser depends only on the
inversion efficiency, $\eta$, and the amplification along the propagation path,
$a\ko d$; it is independent of the geometry. Denote by $T_0$ the temperature
equivalent of the source function $S_0$ (Equation \ref{eq:S0}) in the
Rayleigh--Jeans limit, i.e., $kT_0 = \frac12\lambda^2S_0$. Then the brightness
temperature in the unsaturated domain is
\eq{\label{eq:unsat}
   a < \asat: \hspace{1.5in} T_b = T_0\,e^{\DS a\ko d}\,. \hspace{2in}
}

The intensity of a strongly saturated maser does depend on the geometry. The
overall photon production rate at line center of such a maser is
$\Phi_{m,\nu_0}V_m$, where $V_m$ is the volume of the maser. This luminosity is
radiated away through area $A_m$ with a line-center flux $F_{m,\nu_0}$ measured
at the surface of the maser. Following EHM92, we assume that the disk emits
primarily through its rim, neglecting maser emission from the two faces. Then
in steady state the line-center maser luminosity is
\eq{\label{eq:Lm}
  L_{m,\nu_0} = F_{m,\nu_0}A_m
              = F_{m,\nu_0}\cdot2\pi\ell d = h\nu_0 \Phi_{m,\nu_0}\,\pi\ell^2d.
}
The flux emitted at line center from the surface of the maser is thus
\eq{\label{eq:Fm}
    F_{m,\nu_0} = \half h\nu_0 \Phi_{m,\nu_0} \ell.
}
Because maser radiation is beamed, $F_{m,\nu_0} = I_{\nu_0}\Omega$, where
$I_{\nu_0}$ is the intensity and $\Omega$ the beaming angle at line center. An
observer at large distance $D$ will measure the line-center flux $F_{\nu_0} =
I_{\nu_0}\Aobs/D^2$, where \Aobs\ is the maser observed area.\footnote{
To relate this discussion to that in Section \ref{sec:global}, which is given
in terms of $\oem$, the solid angle of all the emission from the maser, instead
of in terms of $\Omega$, the solid angle into which the maser emission is
beamed at a given point on the surface of the maser, we first note that the two
equations in the sentence preceding this footnote imply that
$F/F_m=\Aobs/(D^2\Omega)$. We also have $F=L_{\rm iso}/(4\pi
D^2)=L_m/(D^2\oem)$ from Equation (\ref{eq:liso}). Since the maser luminosity
is $L_m=F_m A_m$, it follows that the emission solid angle and the beaming
angle are related by $\oem/\Omega=A_m/A_{\rm obs}$.
}
Therefore,
\eq{
   F_{\nu_0} = \half h\nu_0 \Phi_{m,\nu_0} \frac{\ell}{D^2}\frac{\Aobs}{\Omega}.
}
We take the beaming solid angle from the disk solution when $d < \dthin$ and
from the sphere solution when $d > \dthin$. Then from expressions in EHM92,
\begin{eqnarray}\label{eq:beam}
 d < \dthin:&      \Aobs = d\dperp, \quad &\Omega = {\Aobs\over2\ell^2};
 \nonumber \\
 d > \dthin:& \Aobs = \frac14\pi d_\ell^2, \quad &\Omega = {\Aobs\over\ell^2}.
\end{eqnarray}
Combining the last two equations for the case $d < \dthin$ yields
\eq{\label{eq:F0}
  F_{\nu_0} = h\nu_0 \Phi_{m,\nu_0} \frac{\ell^3}{D^2}
  \simeq 5.0\x\E{14}\,\xi\eta_{-2}q_{-13}\,a^3 {d^2\over D^2} \quad \rm Jy,
}
where the last equality utilizes Equation (\ref{eq:Phi}) for the photon
production rate. If $d > \dthin$, $F_{\nu_0}$ is reduced by a factor of 2.
Equation (\ref{eq:Fobs}) in the text follows directly.

Expressing the maser intensity at line center in terms of the equivalent
brightness temperature $T_b$, Equations (\ref{eq:Fm}) and (\ref{eq:beam}) show
that in all cases $T_b$ can be brought to the common form
\eq{\label{eq:Tb_sat}
   a > \asat: \hspace{1.25in}
   kT_b = {hc\lambda\over16}\Phi_{m,\nu_0}d\,a^3s, \hspace{2in}
}
where the ``shape factor" is
\eq{\label{eq:shape}
   s  = \cases{{\DS d\over\dperp}                   & $d < \dthin$,      \cr \cr
       \frac{2}{\pi}\left({\DS d\over\dl}\right)^2  & $d > \dthin$. \cr}
}
From Equation (\ref{eq:Phi}) for the photon production rate, the brightness
temperature becomes
\eq{\label{eq:Tb0}
    T_b = 3.3\x\E7\,\xi\eta_{-2}q_{-13}\,a^3\,s \quad \rm K,
}
which leads directly to Equation (\ref{eq:Tb}) in the text.

In deriving these estimates we neglected maser emission from the two faces of
the disk. The equivalent approximation in filamentary masers produces a beaming
solid angle smaller than the actual one by factor 11/16 (Elitzur, McKee \&
Hollenbach 1991). Based on this result, the expression for $\Omega$ in Equation
(\ref{eq:beam}) can be expected to produce $\sim \sqrt{11/16} = 0.83$ of the
actual beaming angle, for an error of order 20\% in our results for the maser
flux and brightness.

\section{J-SHOCK MASERS ARE GEOMETRICALLY THIN}

Here we check the assumption made in \S 4.1 that J-shock masers are
geometrically thin. Equation (B8) gives an expression for $d/d_{\rm thin}$ in
terms of $\xi$, $c_\eta$, and $n_9$.   Using our analytic expressions for these
(Eqs. \ref{eq:np9}, \ref{eq:xi3}, and \ref{eq:ceta3}), we find
\eq{
\frac{d}{d_{\rm thin}} \simeq 0.4 \left({{v_{A\perp,5}  c_\eta} \over
{j_{14}^{0.25}\Delta v_{D5}^{0.5}}}\right),
}
valid in the strong masing region 10$^6$ cm$^{-3}\la n_0 \la 10^8$ cm$^{-3}$
and 30 km s$^{-1} \la v_s \la 160$ km s$^{-1}$. The condition that the J-shock
maser slab be thin is then
\eq{
j_{14}=n_{0,7}v_{s7}\ga 2.6\times 10^{-2} v_{A\perp,5}^4 \Delta v_{D5}^{-2}
c_\eta^4,
}
again valid in the strong masing region.   For the standard values of
$v_{A\perp}$ and $\Delta v_D$, we see that the J-shock masers are thin in the
entire strong J-shock region above $n_0\ga 10^6$ cm$^{-3}$.

These equations are approximate fits to the numerical results.   We further
check our assumption of geometrical thinness by applying exact numerical
solution of two representative J-shock cases: a high-density model with $n_p$
=\E9\,\cc, $T$ = 400\,K and $\xi$ = 2.34, and a low-density one with $n_p =
3\x\E7\,\cc$, $T$ = 300 K and $\xi$ = 0.03. Note that the low-density model is
outside the scaling range for $\Gamma$ (Figure 1).   Note also that these
constant density models correspond to $n_{0,7}v_{s7} \simeq 0.7 v_{A\perp,5}$
and $n_{0,7}v_{s7} \simeq 0.02 v_{A\perp,5}$.   The high-density model lies in
the upper high density range of strong J-shock masers, while the low-density
model lies very close to the low density boundary of strong J-shock masers.  We
find that the high-density model has $d=0.28 \dthin$, firmly in the thin-disk
regime.  The low density case has $d=0.96\dthin$. Therefore, it lies at the
boundary of the thin/thick transition.   The strong maser region is therefore
almost entirely geometrically thin, save perhaps for a small region near the
low-density boundary.

The above estimate of $\dthin$ is based on the assumption that rays from the
saturated parts of the disk do not contribute significantly to the mean
intensity along the disk axis. This is true so long as the disk is far from
core saturation, $(\ko\ell)^2\ko d \ll 2.74\gamma_m$ (EHM92). The core
saturates only when $a \ge 668$ in our high-density case and $a \ge 30$ in the
low-density one. Core saturation is thus unlikely, although could be reached
for the low-density case in some exceptional situations.

\let\Ref=\item
{}

\newpage

\begin{table}[htbp]
\begin{center}
\caption{Glossary of maser dimensions \label{Glossary}}\centerline{}

\begin{tabular}{c@{$\cdots\,$} p{0.8\hsize}}
\tableline \tableline \noalign{\medskip}

$d$ & Thickness of the masing region; determined by the shock properties.
\\ \noalign{\medskip}
\dthin & Diameter of a spherical maser that has just reached saturation;
determined by the pump properties (equation \ref{eq:dthin}). Planar masers with
$d < \dthin$ are ``thin" and can be described by the EHM92 disk maser solution
when they also obey the filamentary condition $a \gg \max[1, \ko d/8]$.
``Thick" disk masers have $d > \dthin$ and can be described by the solution for
a saturated sphere with the same diameter as the disk.
\\ \noalign{\medskip}
\dpar & Maser observed size in the direction parallel to the shock propagation;
equal to $d$ for thin disk masers.
\\ \noalign{\medskip}
\dperp & Maser observed size in the direction perpendicular to the shock
propagation; given in Equation (\ref{eq:dperp}) for thin disk masers.
\\ \noalign{\medskip}
\dl & Diameter of the observed circular shape of a thick disk maser. Given by
equation (\ref{eq:dl}) when the core is unsaturated, and equation
(\ref{eq:dls}) when it is saturated.
\\ \noalign{\medskip}
\tableline

\end{tabular}
\end{center}
\end{table}

\newpage

\def\FRAC#1#2{{\displaystyle #1 \over \displaystyle #2}}% big letters fraction

%%%% Table 2

\begin{table}[htbp]
\begin{center}
\caption{Equations from Analytic Shock and Saturated Maser Slab Model}
\centerline{}

\begin{tabular}{lll}
\tableline \tableline \\
  Parameter
& Preshock variables
$j=n_0 v_s$
and $v_{A\perp}$
& Observable variables $d$, $d/d_\perp$ and $v_{A\perp}$
\\ \\
\tableline \noalign{\medskip}
 $n_p$
 & $1.4\times 10^9\left(\FRAC{j_{14}}{v_{A\perp,5}}\right) $\ cm$^{-3}$
 & $7\times 10^8\left(\FRAC{v_{A\perp,5}} {\gamma_{-17}d_{13}}\right)$\ cm$^{-3}$
 \\
 $B_p^*$
 &$0.24j_{14}^{1/2}v_{s7}^{1/2}$\ G
 &$0.17\left(\FRAC{v_{A\perp,5}v_{s7}^{1/2}} {\gamma_{-17}^{1/2}d_{13}^{1/2}} \right)$
 \ G
 \\
 $N_p$
  &$7\times 10^{21}\left(\FRAC{v_{A\perp,5}}  {\gamma_{-17}}\right) $\ cm$^{-2}$
 &$7\times 10^{21}\left(\FRAC{v_{A\perp,5}}  {\gamma_{-17}}\right)$\ cm$^{-2}$
 \\
 $d$
 &$5\times 10^{12}\left(\FRAC{v_{A\perp,5}^2} {\gamma_{-17}j_{14}}\right)$\ cm
 &$d$
 \\
 $j_{14}$
 &$j_{14}$ & $0.5 \left(\FRAC{v_{A\perp,5}^2} {\gamma_{-17}d_{13}}\right)$
 \\
 $\xi$
 &$1.0\left[\FRAC{x_{-4}({\rm H_2O})j_{14}}  {\gamma_{-17}\Delta
 v_{D5}}
\right]$
 &$0.5\left[\FRAC{x_{-4}({\rm H_2O})v_{A\perp,5}^2} {\gamma_{-17}^2d_{13}\Delta v_{D5}
}\right]$
 \\
  $T_{b,11}^+$
 &$4.7\left[\FRAC{x_{-4}({\rm H_2O})j_{14}}
  {\gamma_{-17}\Delta v_{D5}}\right]^{1/2}\left({d\over d_\perp}\right) c_\eta e^{-460 {\rm K}/T_p} a_1^3 $
 &$3.3 \left[\FRAC{x_{-4}({\rm H_2O})v_{A\perp,5}^2}
 {\gamma_{-17}^2 d_{13}\Delta v_{D5}}\right]^{1/2}\left({d\over d_\perp}\right) c_\eta e^{-460 {\rm K}/T_p} a_1^3 $
 \\
 $L_{\rm iso,-6}^+$
 &$0.75\left[\FRAC{x_{-4}({\rm H_2O})^{0.5} \Delta v_{D5}^{0.5}v_{A\perp,5}^4 }{\gamma_{-17}^{2.5}(j_{14})^{1.5}}\right] c_\eta e^{-460 {\rm K}/T_p} a_1^3$
 &$2.1\left[\FRAC{x_{-4}({\rm H_2O})\Delta v_{D5}v_{A\perp,5}^2 d_{13}^3}{\gamma_{-17}^2}\right]^{1/2} c_\eta
 e^{-460 {\rm K}/T_p} a_1^3$
  \\ \\
 \tableline
 \\

\end{tabular}
\end{center}
$^*$ Note that $B_p$ is the only one of these parameters to depend directly on
the shock velocity $v_s$ in addition to the indirect dependence through $j$

$^+$ These expressions for $T_b$ and $L_{\rm iso}$ are only valid if $a>a_{\rm
sat}$
(see Eq. \ref{eq:asat4}).
Recall that $a_1=a/10$.
\end{table}

%%%% Table 3

\begin{table}[htbp]
\begin{center}
\caption{Approximations Derived From Numerical Shock Results} \centerline{}

\begin{tabular}{ll}
\tableline \tableline \\
  Parameter
 & Approximation (preshock variables $j$, $v_s$, $v_{A\perp}$)$^*$
\\ \\
\tableline \noalign{\medskip}

 $d$
 &$1.3\times 10^{13}j_{14}^{-0.7} v_{s7}^{0.5}v_{A\perp,5}^2\ \ {\rm cm}$
 \\
 $\gamma $
 &$3.8\times 10^{-18}j_{14}^{-0.3}v_{s7}^{-0.5}$\ \ cm$^3$\ s$^{-1}$
 \\
 $\xi$
 &$4.0j_{14}^{1.5}\Delta v_{D5}^{-1}$
 \\
 $T_p$
 &$350j_{14}^{0.12}v_{s7}^{-0.24}\Delta v_{D5}^{-0.22}\ \ {\rm K}$
 \\
 $x_{-4}({\rm H_2O})$
 &$1.6j_{14}^{0.2}v_{s7}^{-0.5}$
 \\
 $c_\eta$
 &$\left(1+0.045 {j_{14}^{2.8}\over {v_{A\perp,5}^{0.2} \Delta v_{D5}^{1.5}}}\right)^{-1}$
 \\
 $\eta$
 &$0.03 j_{14}^{-0.75}\Delta v_{D5}^{0.5} c_\eta$
 \\
 $a_{\rm sat}$&$2.5j_{14}^{0.25} \Delta v_{D5}^{0.5}v_{A\perp,5}^{-1} c_\eta^{-1}$
 \\
 $d/d_\perp$
 &$1.3 j_{14}^{-0.25} \Delta v_{D5}^{-0.5}v_{A\perp,5} c_{\eta}$
 \\
 $T_b $
 &  $ 2.5 \times 10^{11} j_{14}^{0.5} \Delta v_{D5}^{-1} v_{A\perp,5} c_\eta^2 a_1^3 \ \ {\rm K}$
 \\
 $L_{\rm iso}$
 &  $2.2 \times 10^{-6} j_{14}^{-0.65} v_{s7} \Delta v_{D5}^{0.5} v_{A\perp,5}^4 c_\eta a_1^3 \ \ {L_\odot}$
   \\ \\
 \tableline
 \\

\end{tabular}
\end{center}
$^*$ Recall that $j=n_0v_s$ so that, equivalently, these expression show the
preshock density dependence. These expressions assume that the HM79 prescription for the
rate coefficient $\gamma$ of H$_2$ formation is correct.   These approximations are good to better than a
factor of 2
(see text for individual error estimates)  in the range $10^6$ cm$^{-3} \la n_0
\la 10^8$ cm$^{-3}$, 30 km s$^{-1} \la v_s \la 160$ km s$^{-1}$, 0.5 $\la
v_{A\perp,5} \la 5$, and $0.5\la \Delta v_{D5} \la 3$.  The approximations for
$d/d_\perp$, $T_b$ and $L_{\rm iso}$ assume $a>a_{\rm sat}$
(see Eq. \ref{eq:asat4}).
\end{table}

%%%% Table 4

\begin{table}%[htbp]
\begin{center}
\caption{Approximations Derived From Numerical
Saturated
Maser Slab and Shock Models}
\centerline{}

\begin{tabular}{ll}
\tableline \tableline \\
  Parameter
 & Approximation (observable variables $d$, $v_{A\perp}$ or $B_{0\perp}$, $T_b$, $L_{\rm iso}$, $B_p$)$^*$
\\ \\
\tableline \noalign{\medskip}

   $n_{0,7}$ &$1.0 d_{13}^{-1.11} v_{A\perp,5}^{2.22}B_p^{-0.22}$\\

   $v_{s7}$ &$4.2d_{13}^{0.56} v_{A\perp,5}^{-1.11}B_p^{1.11}$\\

   $j_{14}$  &  $4.2 d_{13}^{-0.56} v_{A\perp,5}^{1.11} B_p^{0.89}$\\

   $B_{0\perp}$ & $1.7 \times 10^{-3} d_{13}^{-0.56} v_{A\perp,5}^{2.11}B_p^{-0.11} \ {\rm Gauss}$\\

   $v_{A\perp,5}$ & $20.5 d_{13}^{0.27} B_{0\perp}^{0.47} B_p^{0.052}$\\

   $c_\eta$ & $\left(1 + 2.5 d_{13}^{-1.57} v_{A\perp,5}^{2.91} \Delta v_{D5}^{-1.5} B_p^{2.49} \right)^{-1}$\\

$a$ & $5.8 T_{b,11}^{0.33} d_{13}^{0.094} v_{A\perp,5}^{-0.52} \Delta v_{D5}^{0.33} B_p^{-0.15} c_\eta^{-0.67}$\\

$a$ & $6.5 L_{\rm iso,-6}^{0.33} d_{13}^{-0.31}  v_{A\perp,5}^{-0.72} \Delta v_{D5}^{-0.17}B_p^{-0.18} c_\eta^{-0.33}$\\

  \\ \\
 \tableline
 \\

\end{tabular}
\end{center}
$^*$ As in Table 3, these expressions assume that the HM79 formulation for the rate coefficient of H$_2$ formation
on grain surfaces is correct.   These expressions are good to better than a factor of 2 in the same
range quoted in Table 3.  In column 2, $B_{p}$ and $B_{0\perp}$ are measured in
Gauss.

\end{table}

\newpage

%%%%%%%%%%%%%%%%%%%%%%%%%%%%%%%%%%%%%%%%%%%%%%
\begin{figure}
  \centering
  \includegraphics[width=0.85\hsize,clip]{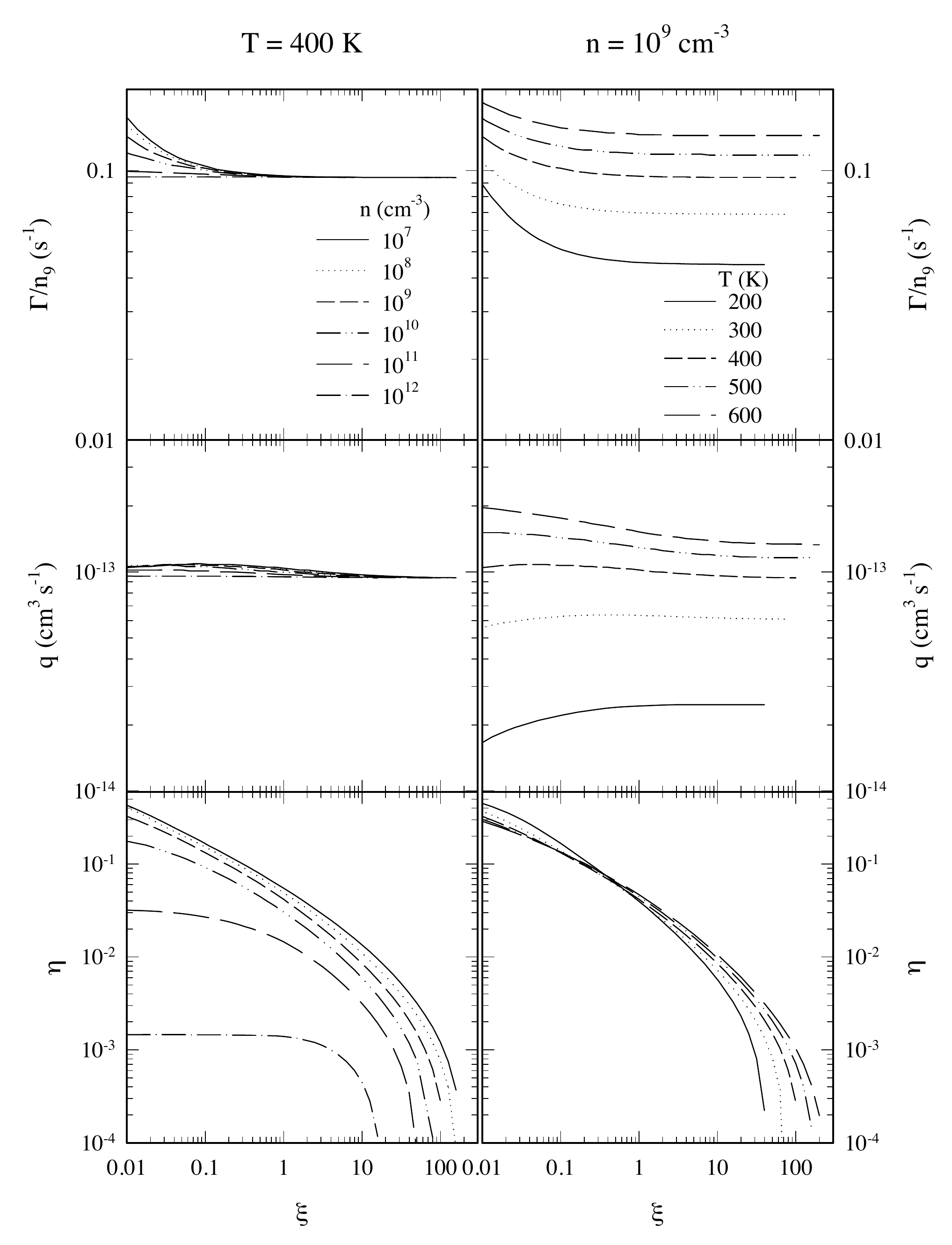}
\caption{Modeling results for \hto\ pumping at $T$ = 400 K and various
densities (left panels) and at $n$ = \E9\ \cc\ and various temperatures (right
panels). Plotted are the maser loss rate $\Gamma$, mean pump rate coefficient
$q$ and inversion efficiency $\eta$ (Equation \ref{eq:q_eta}) as functions of the
scaling parameter $\xi$ (Equation \ref{eq:xi}).
}
\end{figure}
%%%%%%%%%%%%%%%%%%%%%%%%%%%%%%%%%%%%%%%%%%%%%%
\newpage
%%%%%%%%%%%%%%%%%%%%%%%%%%%%%%%%%%%%%%%%%%%%%%
\begin{figure}
  \centering
  \includegraphics[width=\hsize,clip]{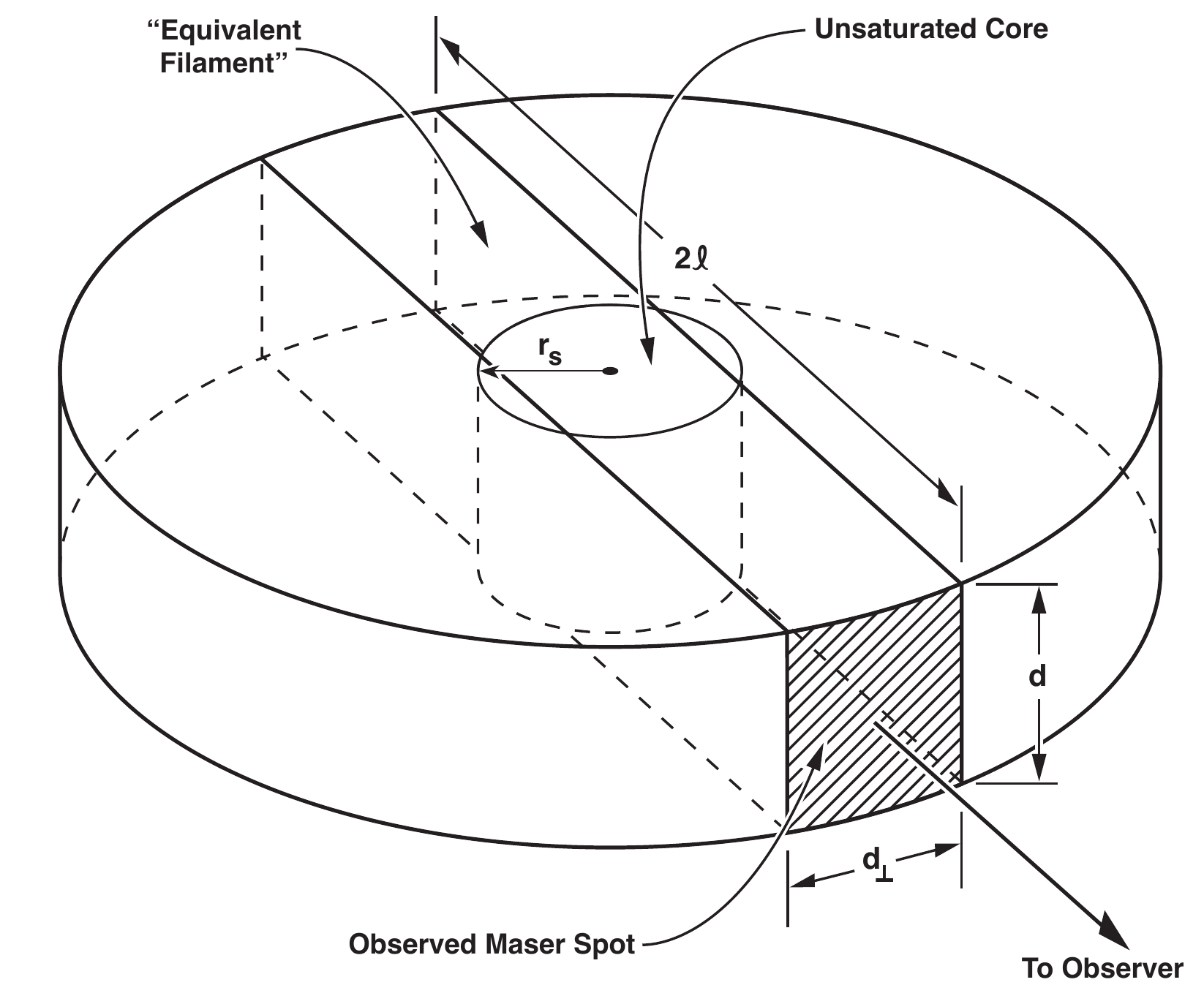}
\caption{Illustration of the geometry of the shock maser model.  The warm H$_2$O
forms a slab of thickness $d$ behind the shock front, in the plane of the
shock. A simple model for the region of velocity coherence is that of a planar
disk of diameter $2\ell$. The maser aspect ratio is defined as $a
\equiv 2\ell/d$. However, the results of this paper apply to
any velocity coherent region in a slab that has its longest dimension $2\ell
> d$.  An unsaturated central region of radius $r_s$ forms.  The
masing photons are beamed out in rays in the disk plane passing through the
unsaturated core.  The observer sees a maser spot size $d\times\dperp$, where
$\dperp$ is defined in Equation (\ref{eq:dperp}).  EHM92 show that $d_\perp /r_s =
(\pi /(\kappa_0 r_s))^{1/2}$ so that $d_\perp /r_s$ ranges from $\sim 0.5$ when the maser
is just saturated to $\sim 1$ when the maser approaches core saturation (see Appendix B).
}
\end{figure}
%%%%%%%%%%%%%%%%%%%%%%%%%%%%%%%%%%%%%%%%%%%%%%
%%%%%%%%%%%%%%%%%%%%%%%%%%%%%%%%%%%%%%%%%%%%%%
\begin{figure}[ht!]
\label{fig:maser}
  \centering
  \includegraphics[width=\hsize,clip]{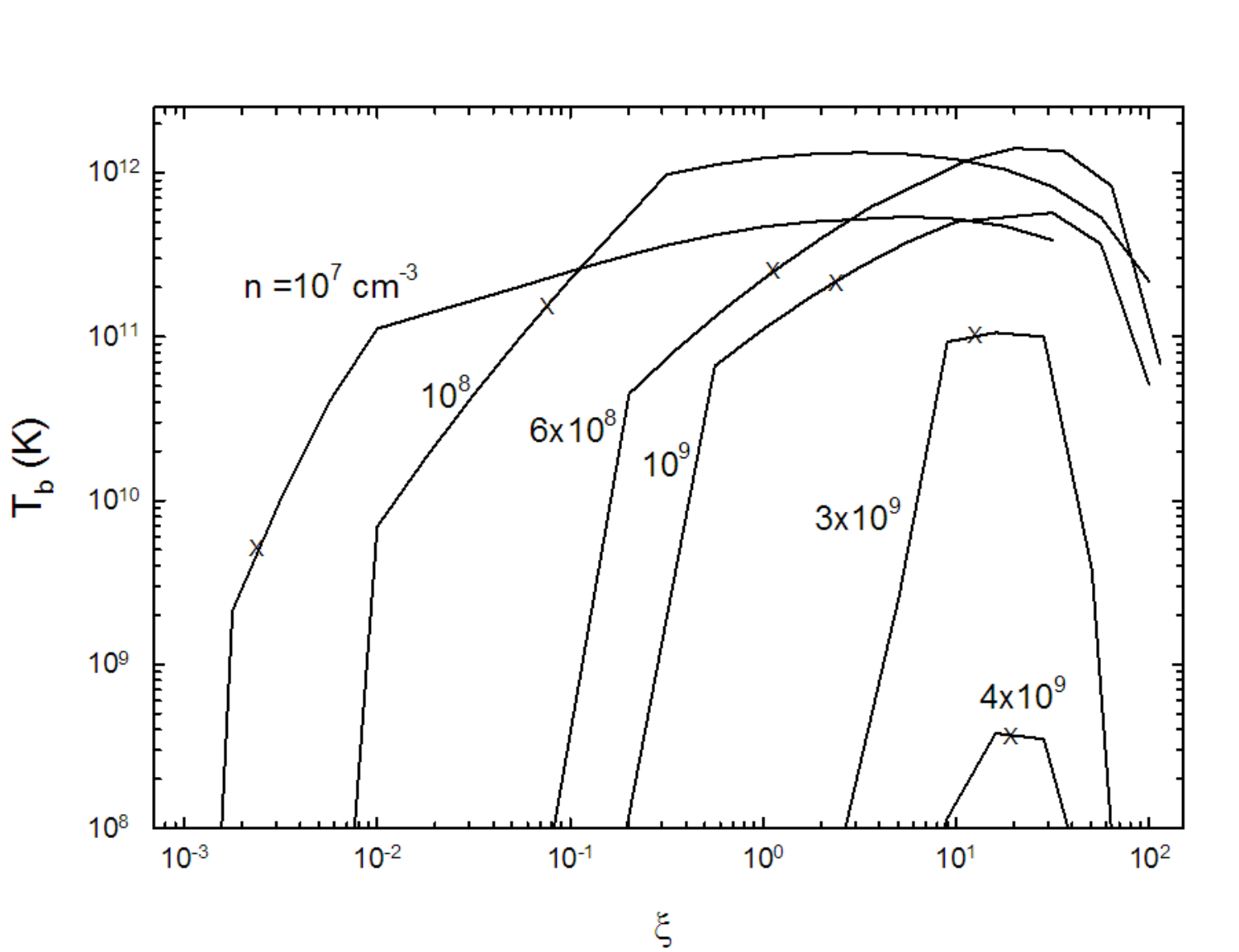}

\caption{Brightness temperatures of disk masers with $a = 10$ as functions of
$\xi$ at $T = 400$ K and different densities, as marked. On each curve, an X
marks the value of $\xi$ generated by a J shock that produces the corresponding
post-shock density for $v_{A\perp,5} = \dvf = 1$ (see text for details).
}
\end{figure}

\clearpage
\begin{figure}[ht!]
\label{fig:shock,ps}
\plotone{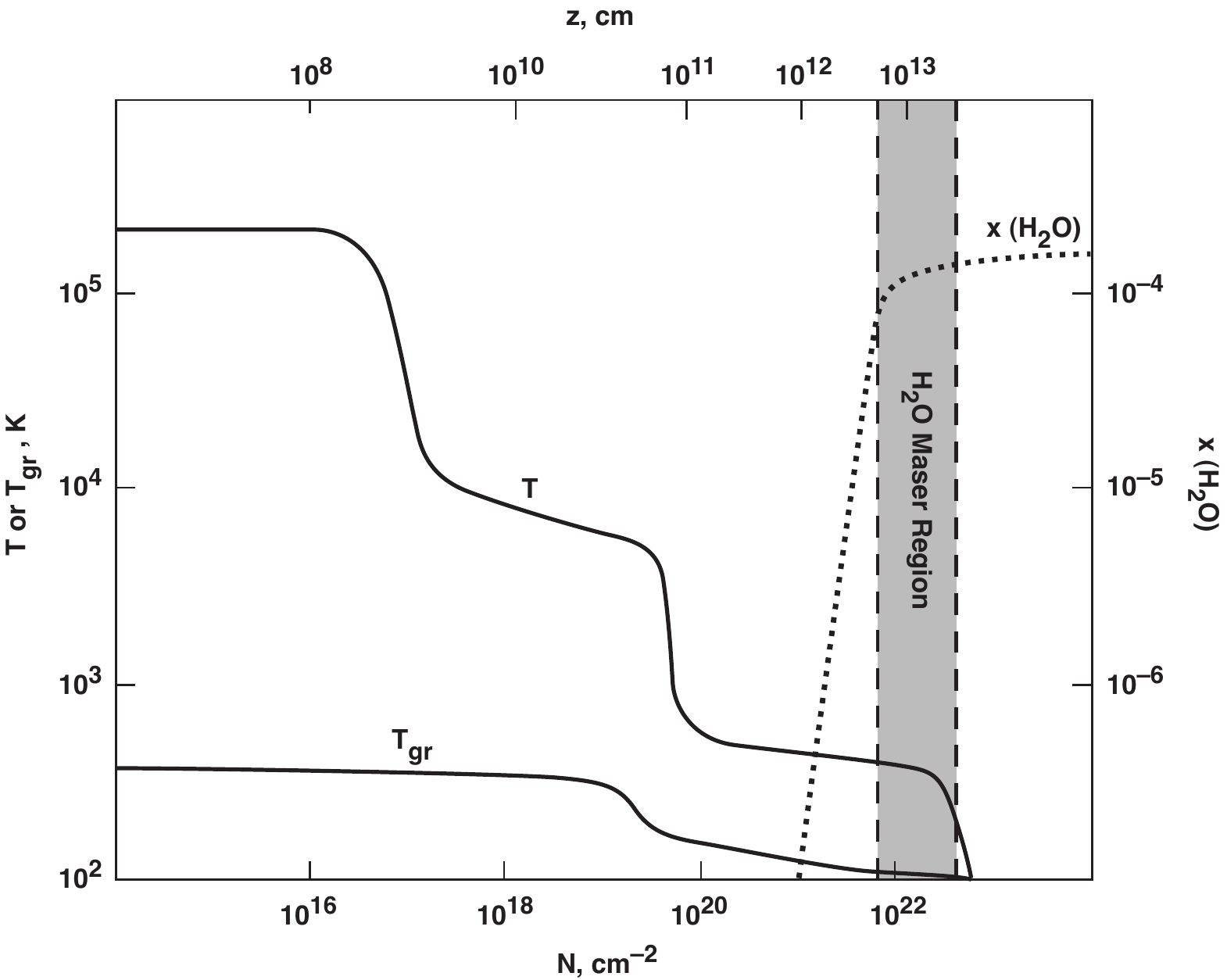}
\caption{Shock structure for the standard run with preshock
density $n_0=10^7$ cm$^{-3}$ ($n_{0,7}=1$), shock velocity $v_s=100$ km s$^{-1}$
($v_{s7}=1$), and preshock magnetic field given by the Alfven speed $v_{A\perp}= 1$ km
s$^{-1}$
($v_{A\perp,5}=1$).  The $x$ axis (bottom) is the column density $N$ of hydrogen
nuclei
downstream from the shock front; the $x$ axis (top) is the corresponding
distance $z$.  The $y$
axis (left) is the temperature ($T$, $T_{\rm gr}$) of the (gas, dust grains); the
$y$
axis (right) is the abundnace of H$_2$O.  Key features include $T_{\rm gr}\ll T$ and
the gas temperature plateau $T \simeq T_p \simeq 350$ K from $N \simeq 10^{20.5}
- 10^{22.5}$ cm$^{-2}$ caused by the heating due to H$_2$ re-formation.  This
plateau region includes a large column of warm H$_2$O molecules that are
collisionally-excited
to mase at 22 GHz.
}
\end{figure}

\clearpage
\begin{figure}[ht!]
\label{fig:Tp}
\plotone{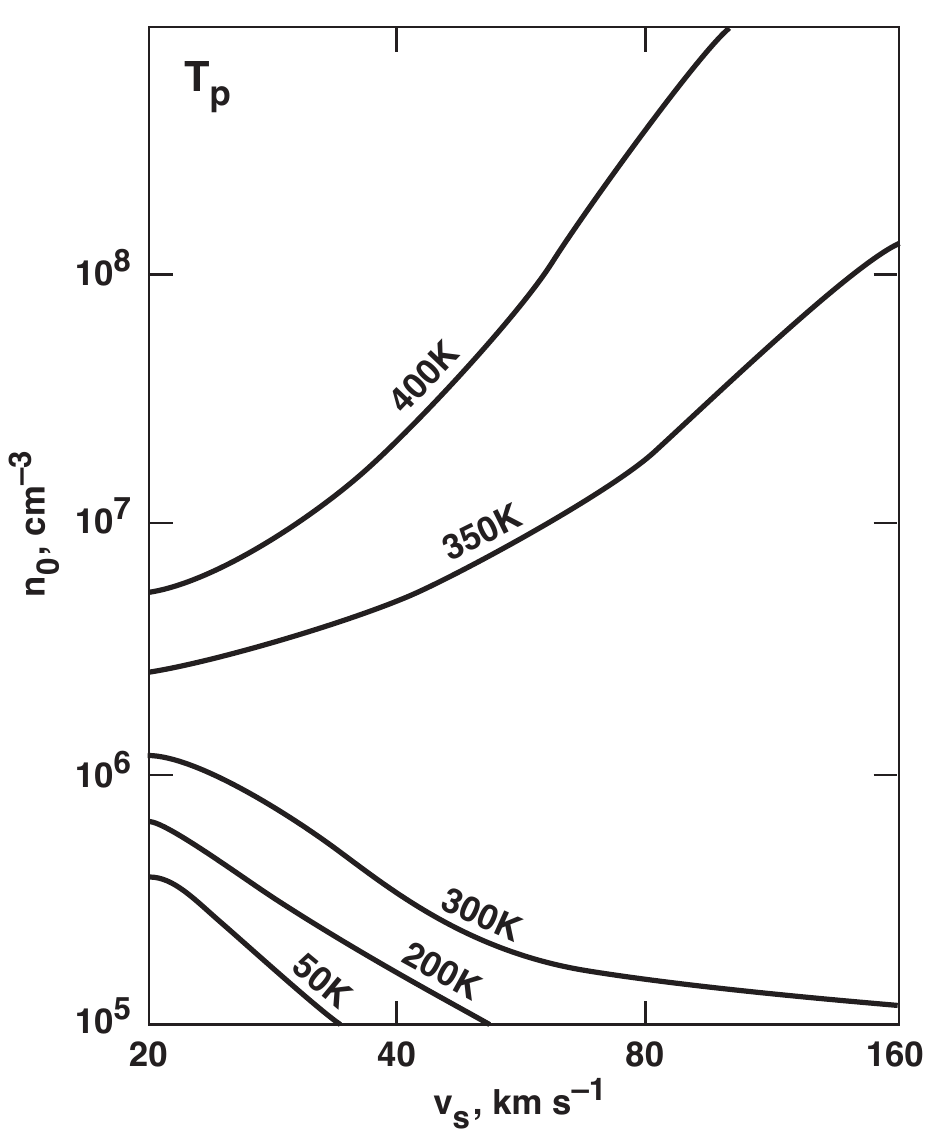}
\caption{Contours of the gas temperature $T_p$ of the plateau region
as a function
of $n_0$ and $v_s$.  $T_p$ is
defined as the temperature when 75\% of the hydrogen has been re-formed into
molecular H$_2$
[i.e., $x({\rm H_2})=0.375$].
For $n_0 \ga 10^5$ cm$^{-3}$, $T_p \simeq 350$ K, and is quite insensitive to
$n_0$ or $v_s$
(this is not sufficient for strong maser emission, however---see Fig. 12).
For $n_0 \la 10^5$ cm$^{-3}$, collisional de-excitation of re-forming H$_2$ in
vibrationally-excited states does not occur frequently, and the heating and
$T_p$ drop.
}
\end{figure}

\clearpage
\begin{figure}[ht!]
\label{fig:wa}
\plotone{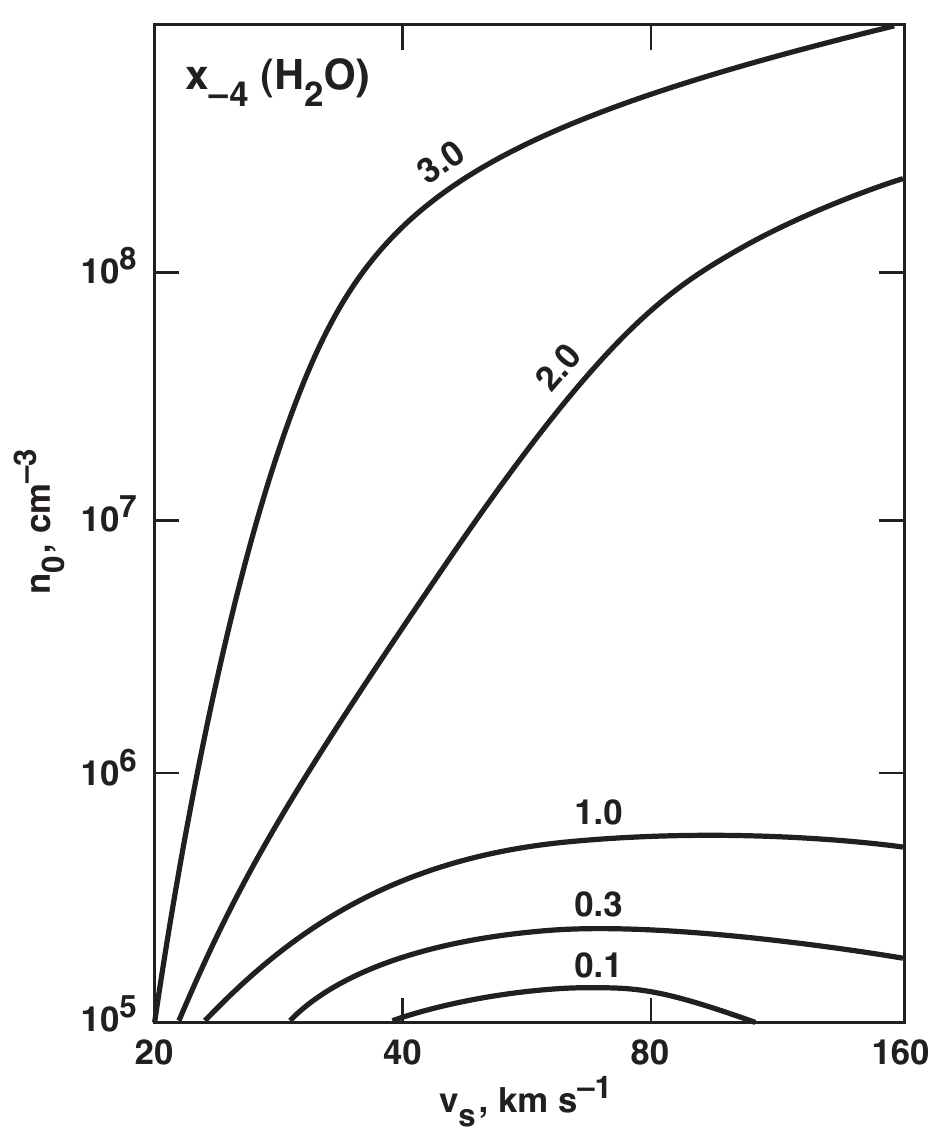}
\caption{The abundance $x_{-4}({\rm H_2O}) \equiv x({\rm
H_2O})/10^{-4}$ is
plotted versus $n_0$ and $v_s$ in the postshock gas at the point where $x({\rm
H_2})=0.375$,
a value near the end of the temperature plateau.
For $n_0 \ga 10^5$ cm$^{-3}$, the
gas
is sufficiently warm and H$_2$ sufficiently abundant to rapidly drive nearly all
the oxygen
not in CO into
H$_2$O
(this is not sufficient for strong maser emission, however---see Fig. 12).
}
\end{figure}

\clearpage
\begin{figure}[ht!]
\label{fig:d}
\plotone{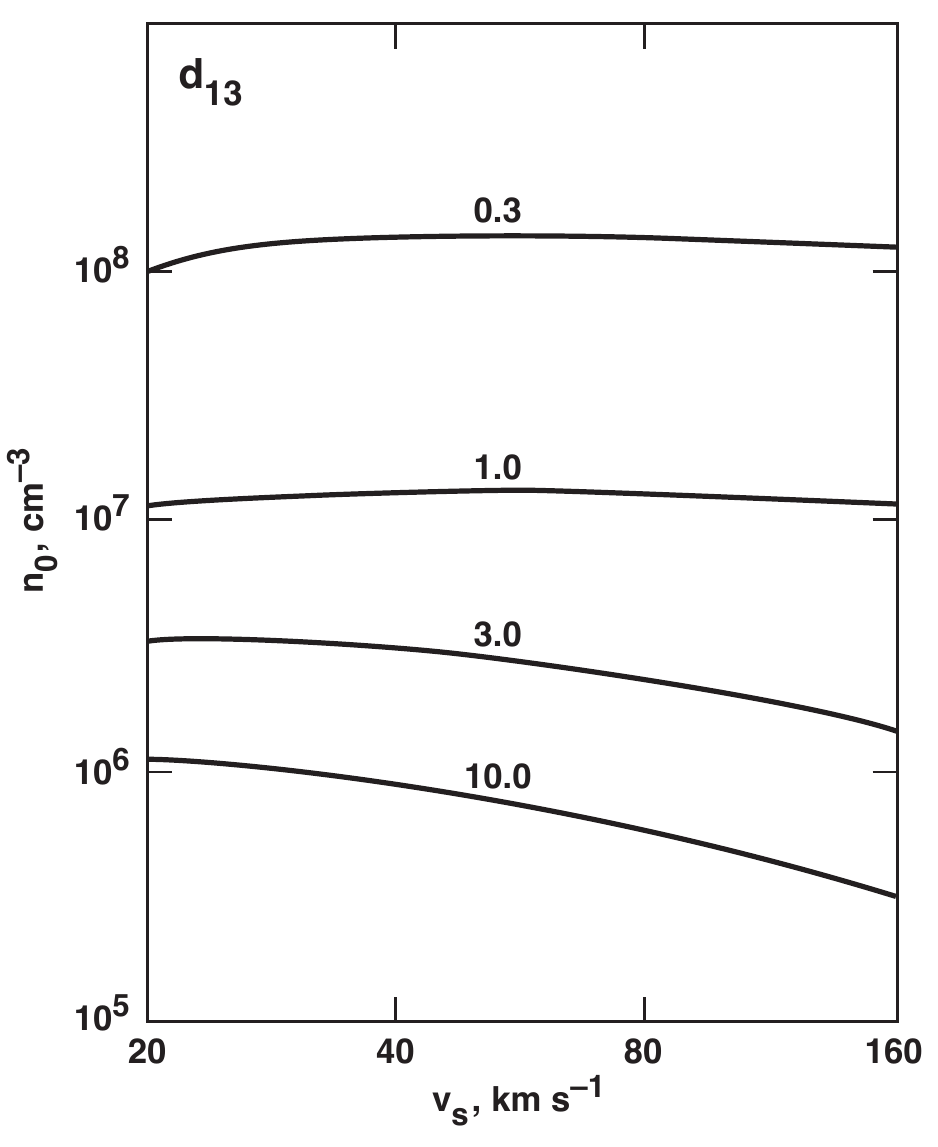}
\caption{The thickness of the masing region, $d_{13} \equiv d/(10^{13}$
cm), is plotted
as a function of $n_0$ and $v_s$.  The parameter $d$ is measured from the shock
front
downstream to the point where $x({\rm H_2})=0.375$. We discuss in text how $d$
is a quite
accurate measure of the maser spot size in the dimension parallel to the shock
velocity.
}
\end{figure}

\clearpage
\begin{figure}[ht!]
\label{fig:xi}
\plotone{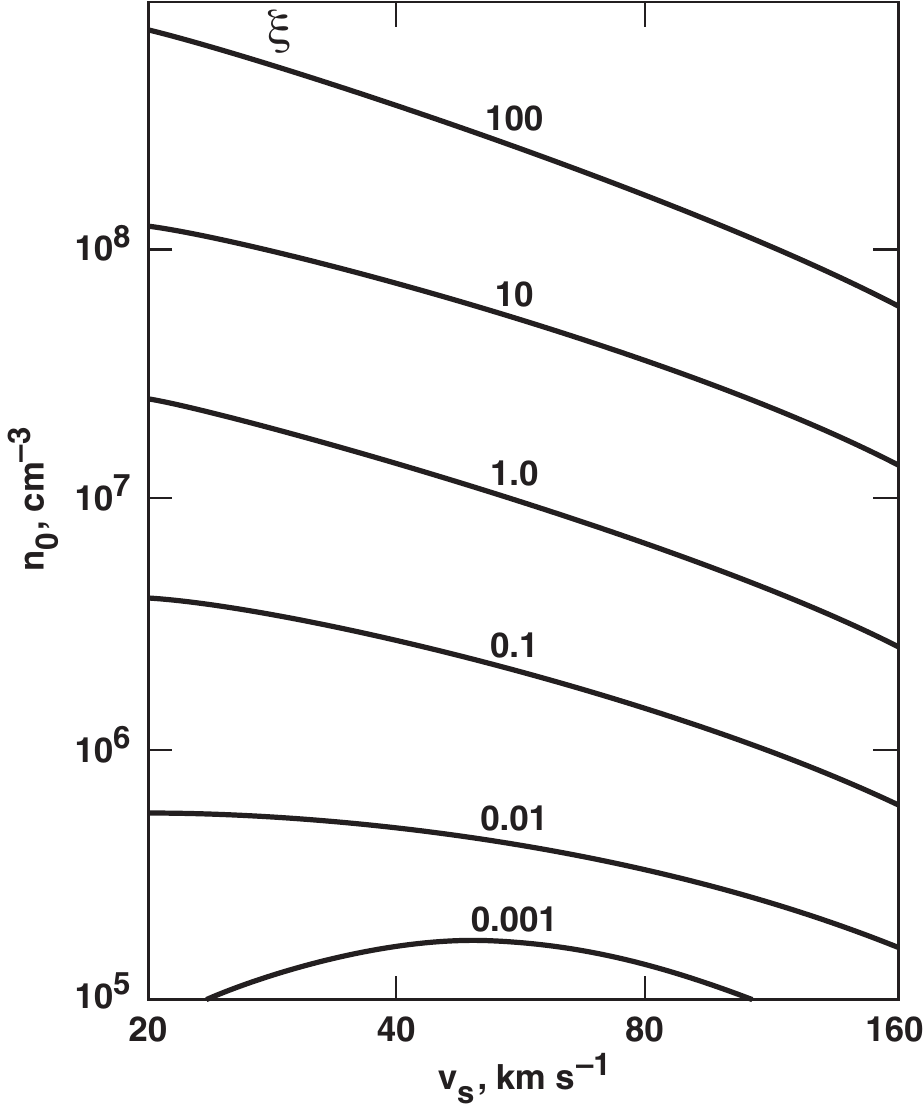}
\caption{The shock ``maser emission measure'' $\xi$ is plotted as a
function
of $n_0$ and $v_s$.  In saturated masers the H$_2$O maser luminosity and
brightness
temperature scale as functions of $\xi$ (see text, \S2.1), but at high $n_0$ the
high
values of $\xi$ do not lead to bright masers because the inversion can be
quenched as
the maser levels are driven to LTE. }
\end{figure}

\clearpage
\begin{figure}[ht!]
\label{fig:va}
\plotone{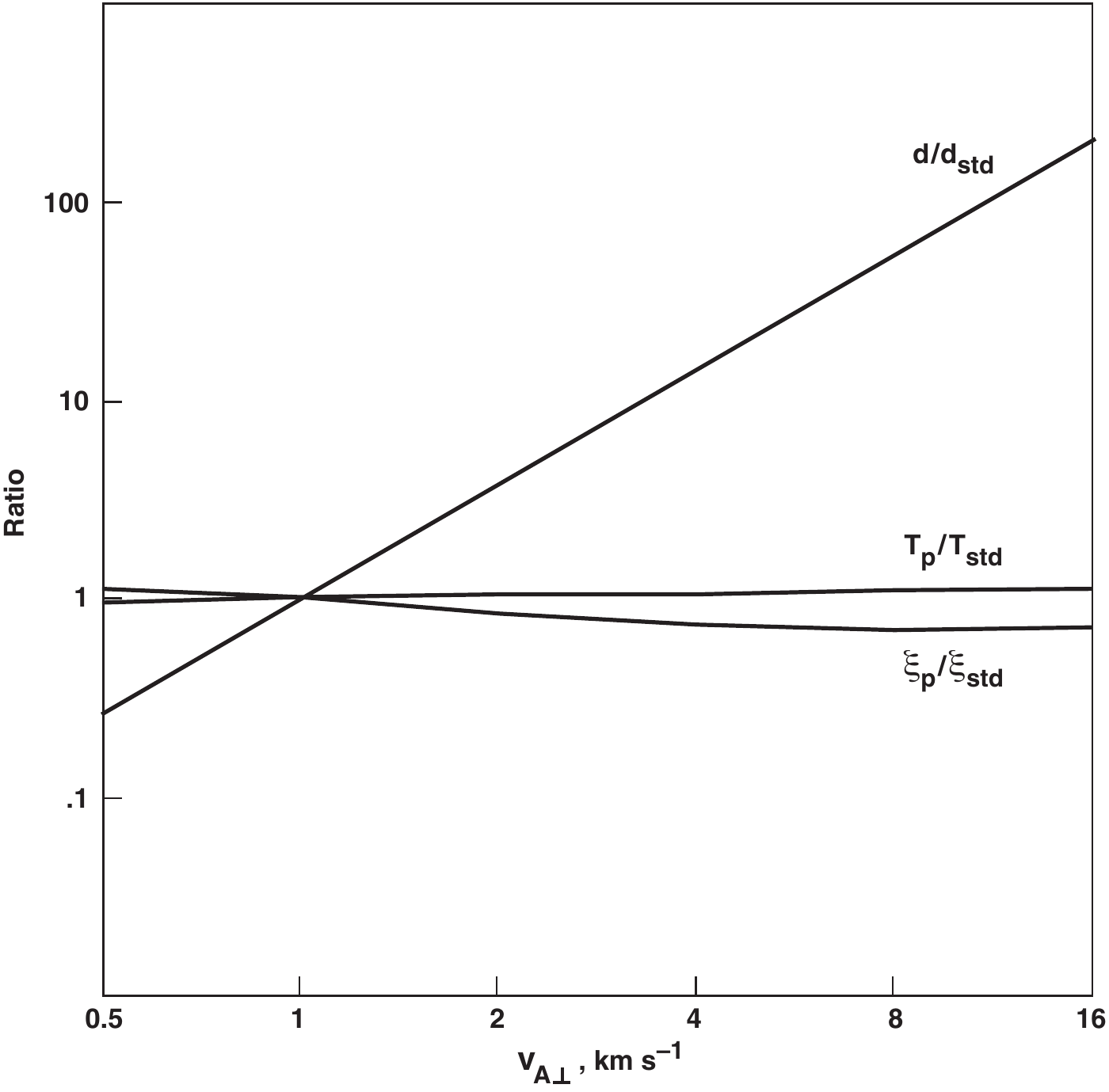}
\caption{The variation of $d$, $T_p$, and $\xi$ as functions of $v_{A\perp}$
is
plotted for the standard case (see Fig. 3). The values $d_{\rm std}$, $T_{p,\rm std}$,
and
$\xi _{\rm std}$ refer to the standard case $v_{A\perp,5}=1$. As predicted (see Section 3.1), $d$
varies
as $v_{A\perp}^2$, whereas $T_p$ and $\xi$ are insensitive to $v_{A\perp}$. }
\end{figure}

\clearpage
\begin{figure}[ht!]
\label{fig:gamH2}
\plotone{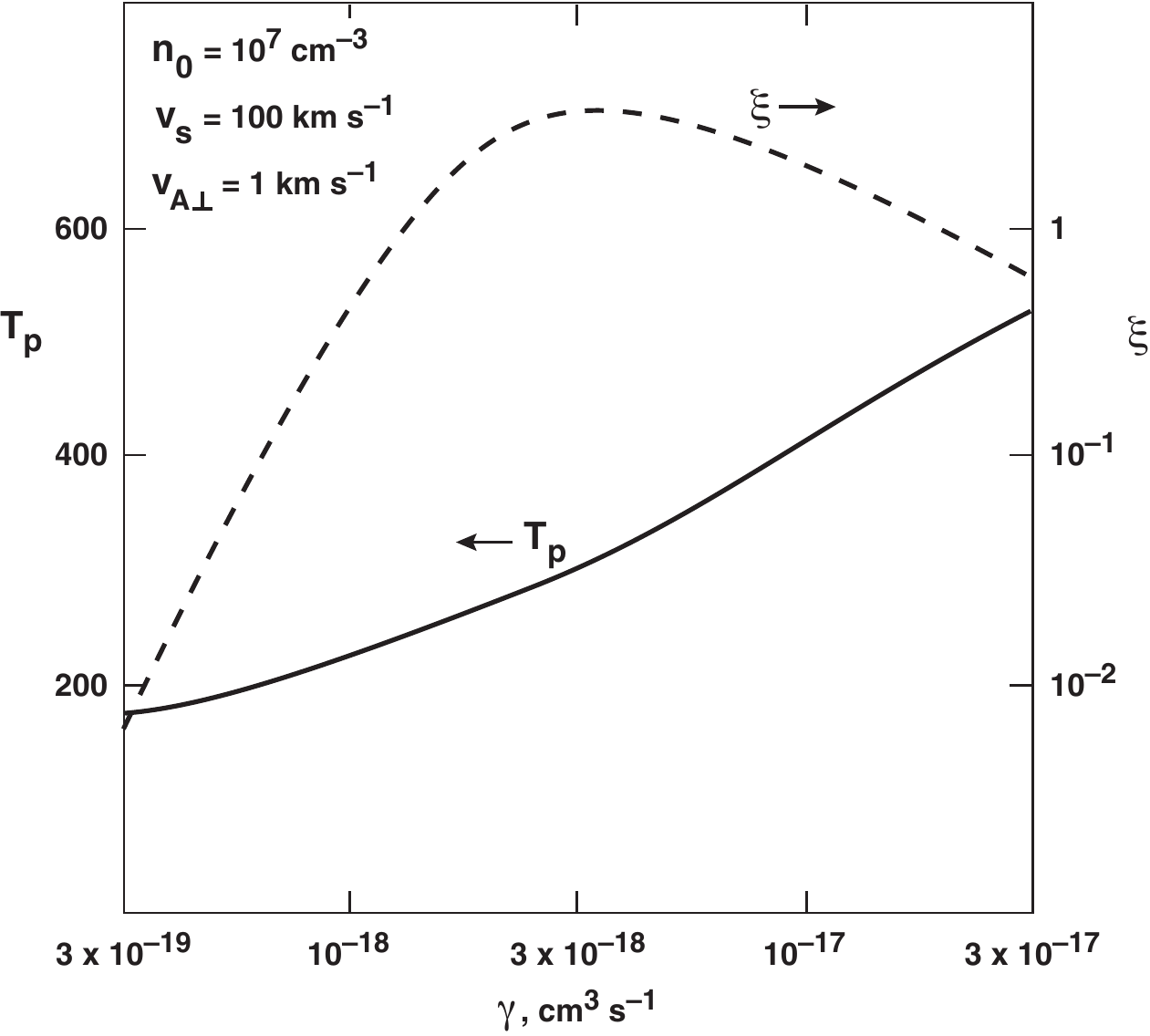}
\caption{The sensitivity of $T_p$ and $\xi$ to the rate coefficient
$\gamma$
for H$_2$ formation. In the parameter study shown in Figures 3 -- 8, $\gamma$ is a
function
of $T$ and $T_{\rm gr}$ (HM79).  In this figure, $\gamma$ is held constant in a
given run,
but varied from $3\times 10^{-19}$ to $3\times 10^{-17}$ cm$^3$ s$^{-1}$ over a
number
of runs. $T_p$ is plotted on the left $y$ axis, $\xi$ on the right $y$ axis.
All runs
assumed the standard parameters $n_{0,7}=v_{s7}=v_{A\perp,5}=1$.  These results also
test the
sensitivity of the model to the assumed heating due to H$_2$ formation. }
\end{figure}

\clearpage
\begin{figure}[ht!]
\label{fig:asat}
\plotone{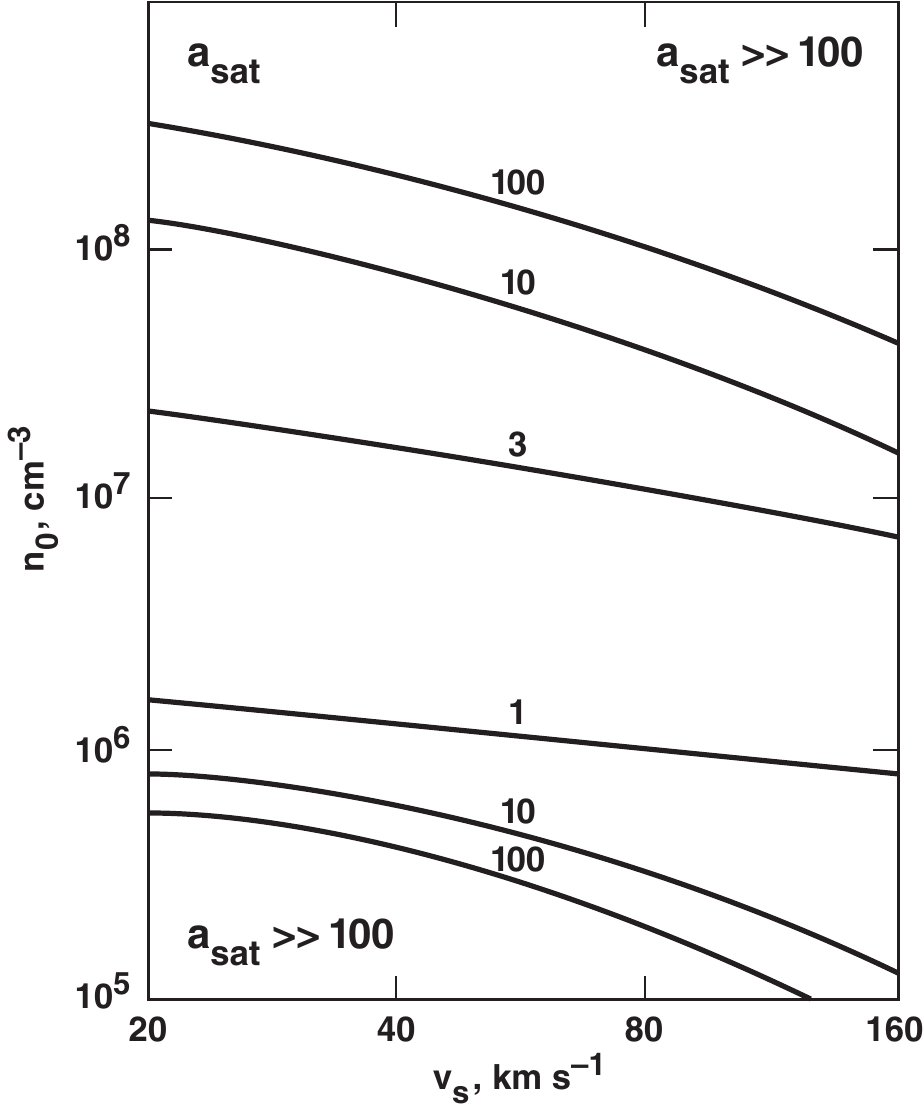}
\caption{The saturation aspect ratio $a_{\rm sat}$ is plotted as a
function
of $n_0$ and $v_s$.  High aspect ratios ($a\ga30-100$),
or high coherence lengths relative to
$d$,
are not often achieved in interstellar J shocks. }
\end{figure}

\clearpage
\begin{figure}[ht!]
\label{fig:s}
%\plotone{fig12.pdf}
\centering
\includegraphics[width=0.9\hsize,clip]{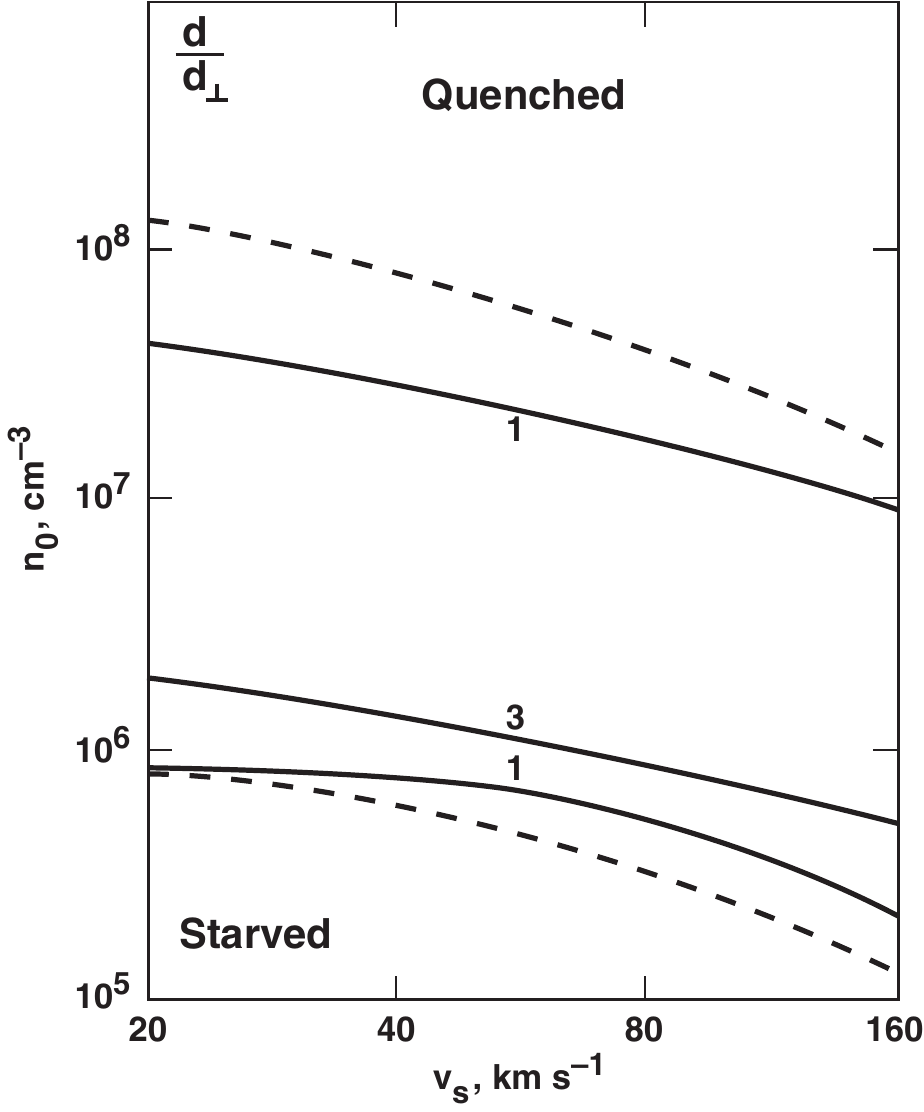}

\caption{The shape parameter for thin masers (see Appendix B), $ d/d_\perp$,
plotted as a function of $n_0$ and $v_s$ for $a=10$.
Here $d$ is  parallel to the shock velocity, and $d_\perp$ is perpendicular to
the shock velocity and in the plane of the shock. At high preshock densities,
$n_0\ga 10^8$ cm$^{-3}$, the maser is collisionally quenched as levels approach
LTE.   At low preshock densities, $n_0 \la 10^6$~cm\eee, there are too few
collisions to excited states to feed the maser (collisionally ``starved").  The
dashed lines represent $a_{\rm sat}=10$ from Figure 11.   Above the top line
and below the bottom line,  $a_{\rm sat} > 10$ and this $a=10$ maser is
unsaturated.  Here, our equations for $d/d_\perp$ fail, and the maser is very
weak (see text).}
\end{figure}

\clearpage
\begin{figure}[ht!]
\label{fig:T0}
\plotone{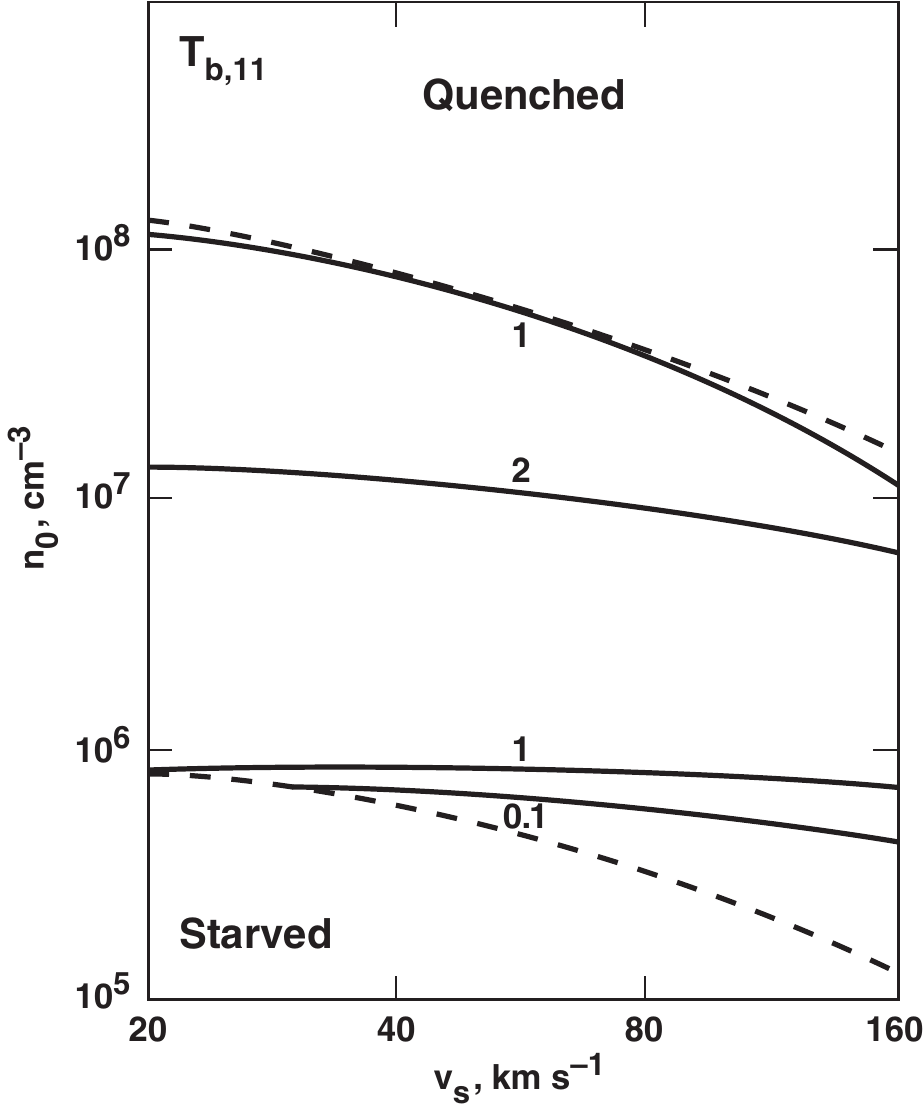}
\caption{The brightness temperature, $T_{b,11} \equiv T_b/
(10^{11}$ K),
is plotted versus $n_0$ and $v_s$
for $a=10$.  The dashed lines represent $a_{\rm sat}=10$ (Figure 11), so that the $a=10$ maser
is unsaturated and $T_b$ falls extremely rapidly outside them.     Note that these boundaries
would expand somewhat (see Figure 11) for $a> 10$ since that would allow larger 
values of $a_{\rm sat}$.
  We assume in this figure
that $v_{A\perp}= 1$ km s$^{-1}$ and  $\Delta v_D = 1$ km s$^{-1}$.
The brightness temperature $T_b$ scales as $v_{A\perp} \Delta v_D^{-1/2} a^3$
(see Table 3) for saturated masers.
 }
\end{figure}

\clearpage
\begin{figure}[ht!]
\label{fig:Lm}
\plotone{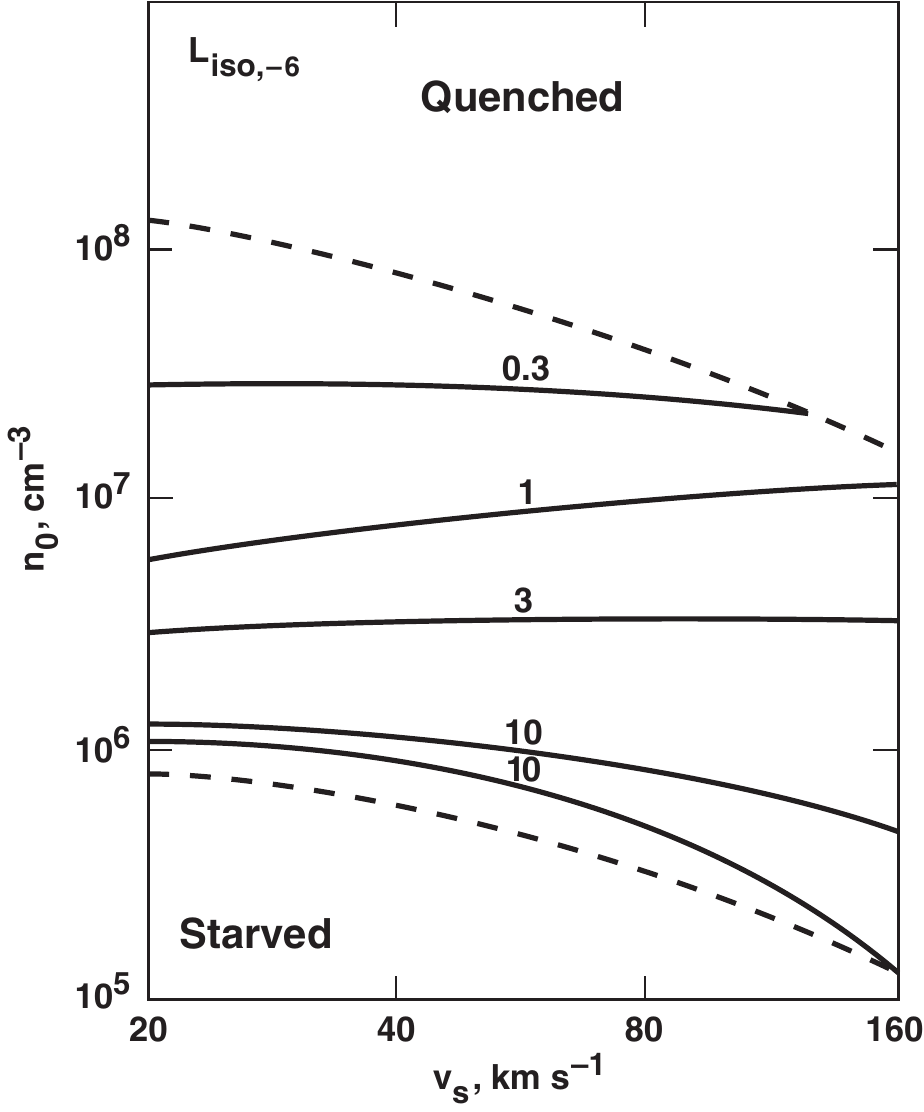}
\caption{The ``isotropic'' luminosity, $L_{\rm iso,-6}\equiv L_{\rm iso}/ (10^{-6}$ $L_\odot)$, of
the 22 GHz maser
 is plotted against $n_0$ and $v_s$ for the standard values $v_{A\perp}= 1$ km s$^{-1}$,
 $\Delta v_D = 1$ km s$^{-1}$, and $a= 10$.
  The dashed lines represent $a_{\rm sat}=10$, so that the $a=10$ maser plotted here is unsaturated
 outside them, and $L_{\rm iso}$ falls rapidly there.
 }
\end{figure}

\clearpage
\begin{figure}[ht!]
\label{fig:maser}
  \centering
  \includegraphics[width=0.8\hsize,clip]{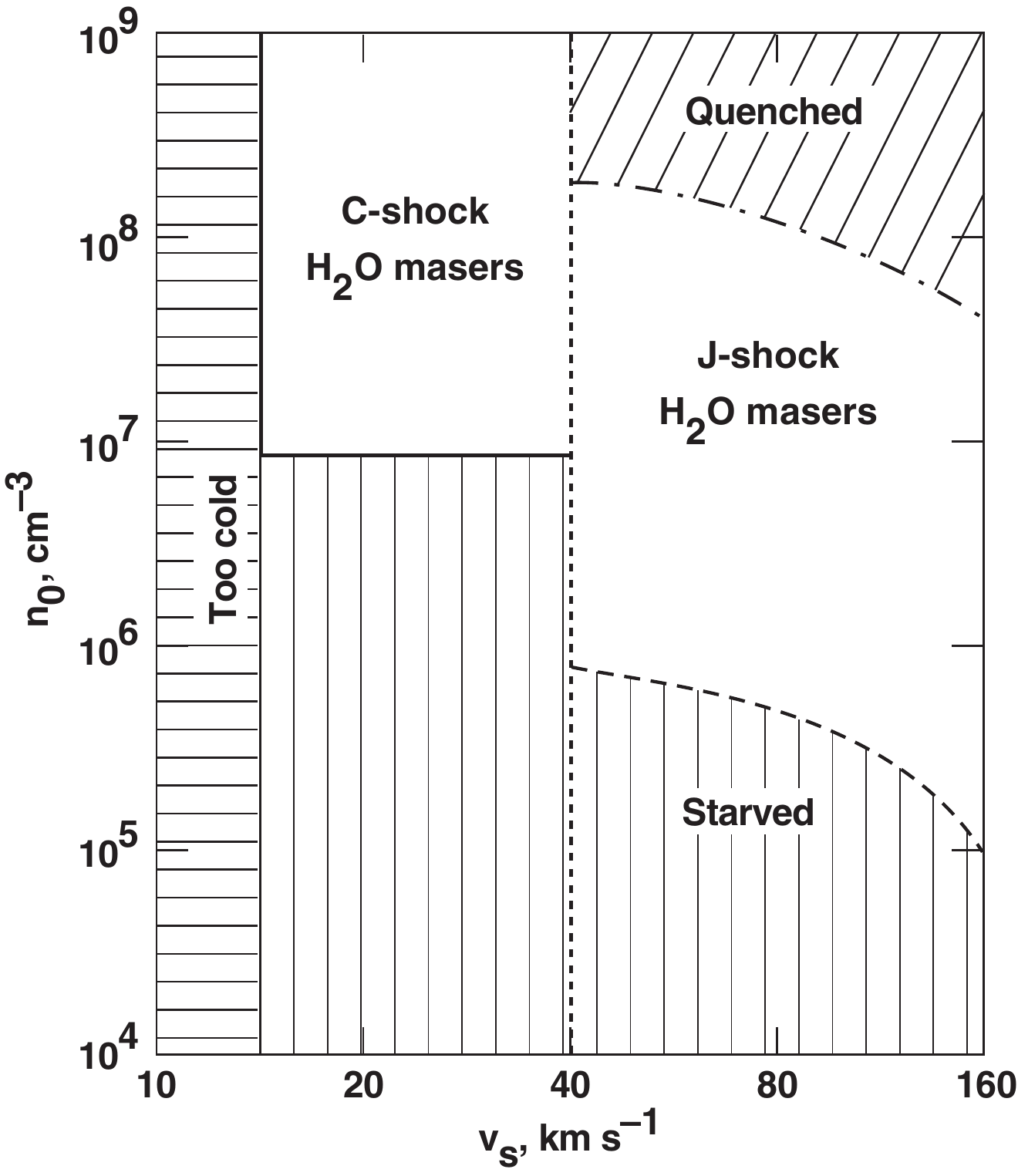}
\caption{The parameter space ($n_0$, $v_s$) that produces J-shock masers is
shown.  For velocities $v_s \la 40$ km s$^{-1}$, the shock is likely not a
J shock but a C shock (left of dotted vertical line).  This boundary could be at lower
$v_s$ if the ionization fraction in the preshock gas exceeds $\sim 10^{-7}$.
Likewise, the low-velocity boundary of C shocks marked ``too cold" could be
at lower values of $v_s$ if the ionization fraction is higher (see text).  J shocks
with $v_s \la 15$ km s$^{-1}$ are too cold to form masers, because they do
not destroy sufficient H$_2$ to form a significant warm plateau heated by H$_2$ re-formation.
For velocities $v_s \ga 160$ km s$^{-1}$, the shock will significantly destroy
grains, which prevents the re-formation of H$_2$ in the postshock gas and the
production of warm H$_2$O.   For
preshock densities $n_0 \la 10^6$ cm$^{-3}$
in J shocks and $\la 10^7$
cm$^{-3}$ in C shocks the masing region requires
extremely high aspect ratios, $a\ga100$, in order to saturate and become bright
enough to detect.  For densities $n_0 \ga 10^8$ cm$^{-3}$ in J shocks, the postshock
densities and optical depths are so high that the maser level thermalizes, and
the inversion is quenched.
}
\end{figure}

\clearpage


\begin{thebibliography}{}

\bibitem[Alves et al.(2012)]{2012A&A...542A..14A} Alves, F.~O., Vlemmings, W.~H.~T., Girart, J.~M., \& Torrelles, J.~M.\ 2012, \aap,542, A14 
 \bibitem[Bragg et al.(2000)]{2000ApJ...535...73B} Bragg, A.~E., Greenhill,
L.~J., Moran, J.~M., \& Henkel, C.\ 2000, \apj, 535, 73
 \Ref Brogan, C., Johnson, K., \& Darling, J.\ 2010, \apjl, 716, L51
 \bibitem[Cazaux et 
al.(2011)]{2011A&A...535A..27C} Cazaux, S., Morisset, S., Spaans, M., \& Allouche, A.\ 2011, \aap, 535, A27 
 \Ref Cernicharo, J., Thum, C., Hein, H., John, D., Garcia, P., \&
 Mattioco, F. 1990, A\&A, 231, L15
 \Ref Claussen, M.T., Marvel, K.B., Wootten, A., Wilking, B. 1998, ApJL, 507, L79
 \bibitem[Claussen et al.(2009)]{2009ApJ...691..219C} Claussen, M.~J., Sahai, R., \& Morris, M.~R.\ 2009, \apj, 691, 219 
 \Ref Claussen, M.T., Wilking, B.A., Benson, P.J., et al. 1996, ApJS, 106, 111
 \Ref Collison, A.J., \& Watson, W.D. 1995, ApJ, 452, L103
 \bibitem[Crutcher et al.(2010)]{2010ApJ...725..466C} Crutcher, R.~M.,
 Wandelt, B., Heiles, C., Falgarone, E., \& Troland, T.~H.\ 2010, \apj, 725, 466
 \bibitem[Cuppen et al.(2010)]{2010MNRAS.406L..11C} Cuppen, H.~M., 
Kristensen, L.~E., \& Gavardi, E.\ 2010, \mnras, 406, L11 
 \Ref Darling, J., Brogan, C., \& Johnson, K.\ 2008, \apjl, 685, L39
 \bibitem[Day et al.(2010)]{2010ApJ...713..986D} Day, F.~M., Pihlstr{\"o}m, Y.~M., Claussen, M.~J., \& Sahai, R.\ 2010, \apj, 713, 986
 \bibitem[Deguchi \& Watson(1989)]{1989ApJ...340L..17D} Deguchi, S., \& Watson,
     W.~D.\ 1989, \apjl, 340, L17
 \Ref de Jong, T. 1973, A\&A, 26, 297
 \Ref Draine, B.T. 1980, ApJ, 241, 1021
 \bibitem[Draine(1995)]{1995Ap&SS.233..111D} Draine, B.~T.\ 1995, \apss, 233, 111 
 \bibitem[Draine
 \& McKee(1993)]{1993ARA&A..31..373D} Draine, B.~T., \& McKee, C.~F.\ 1993, \araa, 31, 373
 \Ref Dubernet, M.-L. et al 2006, A\&A 460, 323
 \Ref Elitzur, M. 1990, ApJ, 363, 638
 \Ref Elitzur, M. 1992, {\it Astronomical Masers}, (Dordrecht: Kluwer) (E92)
 \Ref Elitzur, M. 1994, ApJ, 422, 751
 \bibitem[Elitzur(1995)]{1995RMxAC...1...85E} Elitzur, M.\ 1995, Revista
 Mexicana de Astronomia y Astrofisica Conference Series, 1, 85
 \Ref Elitzur, M., Hollenbach, D.J., \& McKee, C.F. 1989, ApJ, 346, 983 (EHM)
 \Ref Elitzur, M., Hollenbach, D.J., \& McKee, C.F. 1992, ApJ, 394, 221 (EHM92)
 \Ref Elitzur, M., McKee, C.F., \& Hollenbach, D.J. 1991, ApJ,367, 333
 \bibitem[Falgarone et
 al.(2008)]{2008A&A...487..247F} Falgarone, E., Troland, T.~H., Crutcher, R.~M., \& Paubert, G.\ 2008, \aap, 487, 247
 \Ref Felli, M., Palagi, F., \& Tofani, G. 1992, A\&A, 255, 293
 \Ref Fiebig, D. \& G\"usten, R. 1989, A\&A Letters, 214, 333
 \bibitem[Flower
 \& Pineau Des For{\^e}ts(2010)]{2010MNRAS.406.1745F} Flower, D.~R., \& Pineau Des For{\^e}ts, G.\ 2010, \mnras, 406, 1745
 \Ref Furuya, R.S., Kitamura, Y., Wootten, A., Claussen, M., Kawabe, R. 2001,
     ApJL, 559, L143
 \bibitem[Furuya et al.(2003)]{2003ApJS..144...71F} Furuya, R.~S., Kitamura, Y.,
     Wootten, A., Claussen, M.~J., \& Kawabe, R.\ 2003, \apjs, 144, 71
 \Ref Genzel, R., Reid, M., Moran, J.M., Downes, D.  1981, ApJ, 244, 844 \Ref
 Genzel, R. 1986, in {\it Masers, Molecules and Mass Outflows in Star
      Forming Regions}, ed. A.D. Haschick (Westford, MA:Haystack Observatory),
      p233
 \bibitem[Goddi et al.(2011)]{2011A&A...535L...8G} Goddi, C., Moscadelli, L., \&
     Sanna, A.\ 2011, \aap, 535, L8
 \bibitem[Guillet et 
al.(2009)]{2009A&A...497..145G} Guillet, V., Jones, A.~P., \& Pineau Des For{\^e}ts, G.\ 2009, \aap, 497, 145 
 \Ref Green, S., Maluendes, S., McLean, A.D. 1993, ApJS, 85, 181
 \Ref Gwinn, C.R., 1994a, ApJ, 429, 241
 \Ref Gwinn, C.R., 1994b, ApJ, 429, 253
 \Ref Gwinn, C.R., 1994c, ApJL, 431, L123
 \Ref Gwinn, C. R., Moran, J.M., Reid, M.J.1992, ApJ, 393, 149
 \Ref Heiles, C., Goodman, A.A., McKee, C.F., \& Zweibel, E.G. 1993, in {\it
      Protostars and Planets III}, ed. E. Levy, J. Lunine \& M. Matthews
      (Tucson: Univ. of Arizona Press), p279.
 \Ref Hollenbach, D.J., Chernoff, D., \& McKee, C.F. 1989, in {\it Infrared
      Spectroscopy in Astronomy}, ed. B. Kaldeich, ESA SP-290, p245.
 \Ref Hollenbach, D.J., Elitzur, M., \& McKee, C.F. 1993, in {\it Astrophysical
 Masers},
 ed. , p.159
 \Ref Hollenbach, D.J. \& McKee, C.F. 1979, ApJ Suppl, 41, 555 (HM79)
 \Ref Hollenbach, D.J. \& McKee, C.F. 1989, ApJ, 342, 306 (HM89)
 \Ref Hollenbach, D.J., McKee, C.F., \& Chernoff, D. 1987, in {\it
      Star Forming Regions}, ed. M. Peimbert \& J. Jugaku (Dordrecht: Reidel),
      p334
 \bibitem[Humphreys et al.(2008)]{2008ApJ...672..800H} Humphreys, E.~M.~L.,
Reid, M.~J., Greenhill, L.~J., Moran, J.~M., \& Argon, A.~L.\ 2008, \apj, 672,
800
\bibitem[Imai et al.(2013)]{2013MNRAS.432L..16I} Imai, H., Katayama, Y., Ellingsen, S.~P., \& Hagiwara, Y.\ 2013, \mnras, 432, L16 
 \Ref Imai, H., Obara, K., Diamond, P.~J., Omodaka, T., \& Sasao, T.\ 2002, Nature,
 417, 829
 \Ref Jones, A.P., Tielens, A.G.G.M., \& Hollenbach, D.J. 1996, ApJ, 469, 740
 \bibitem[Kartje et al.(1999)]{1999ApJ...513..180K} Kartje, J.~F.,
 K{\"o}nigl, A., \& Elitzur, M.\ 1999, \apj, 513, 180
 \Ref Kaufman, M.J. \& Neufeld, D.A. 1996, ApJ, 456, 250
 \Ref Kylafis, N. \& Norman, C. 1986, ApJL, 300, L73
 \bibitem[Lekht et al.(2007)]{2007ARep...51..967L} Lekht, E.~E., Slysh, V.~I., \& Krasnov, V.~V.\ 2007, Astronomy Reports, 51, 967 
 \Ref Liljestrom, T. \& Gwinn, C.R. 2000, ApJ, 534, 781
 \bibitem[Mac Low et al.(1994)]{1994ApJ...427..914M} Mac Low, M.-M., Elitzur,
    M., Stone, J.~M., \& Konigl, A.\ 1994, \apj, 427, 914
 \Ref Maoz, E., \& McKee, C.F. 1998, ApJ, 494, 218
 \Ref Marvel, K.~B., B.~A.~Wilking, M.~J.~Claussen, \& A.~Wootten, 2008, \apj,
      685, 285
 \Ref Melnick, G.J., Menten, K.M., Phillips, T.G., \& Hunter, T. 1993, ApJ,
      416, L37
 \Ref Menten, K.M., Melnick G.J., \& Phillips, T.G. 1990, ApJ, 350, L41
 \Ref Menten, K.M., Melnick, G.J., Phillips, T.G., \& Neufeld, D.A. 1990,
 ApJ, 363, L27
 \Ref Miranda, L.~F., G{\'o}mez, Y., Anglada, G., \& Torrelles, J.~M.\ 2001,
 \nat, 414, 284
 \bibitem[Moscadelli et al.(2006)]{2006A&A...446..985M} Moscadelli, L., Testi,
     L., Furuya, R.~S., et al.\ 2006, \aap, 446, 985
  \bibitem[Moscadelli et 
al.(2013)]{2013A&A...549A.122M} Moscadelli, L., Li, J.~J., Cesaroni, R., et al.\ 2013, \aap, 549, A122 
 \bibitem[Neufeld \& Dalgarno(1989)]{1989ApJ...340..869N} Neufeld, D.~A., \&
     Dalgarno, A.\ 1989, \apj, 340, 869
 \Ref Neufeld, D.A. \& Hollenbach, D.J. 1994, ApJ, 428, 170
 \bibitem[Neufeld et al.(1994)]{1994ApJ...436L.127N} Neufeld, D.~A.,
 Maloney, P.~R., \& Conger, S.\ 1994, \apjl, 436, L127
 \Ref Neufeld, D. \& Melnick, G. 1990, ApJ, 352, L9
 \bibitem[Patel et al.(2007)]{2007ApJ...658L..55P} Patel, N.~A., Curiel, S.,
     Zhang, Q., et al.\ 2007, \apjl, 658, L55
 \Ref Peck, A.~B., Henkel, C., Ulvestad, J.~S., et al. \ 2003, \apj, 590, 149
 \bibitem[Pollack et al.(1994)]{1994ApJ...421..615P} Pollack, J.~B.,
 Hollenbach, D., Beckwith, S., et al.\ 1994, \apj, 421, 615
 \Ref Richards, A.~M.~S., Elitzur, M., \& Yates, J.~A.\ 2011, \aap, 525, A56
 \bibitem[Sarma et al.(2008)]{2008ApJ...674..295S} Sarma, A.~P., Troland, T.~H., Romney, J.~D., \& Huynh, T.~H.\ 2008, \apj, 674, 295 
 \Ref Schmeld, I.K., Strelnitski, V.S., \& Muzylev, V.V. 1976, AZh, 53, 728
 \bibitem[Smith
 \& Brand(1990)]{1990MNRAS.242..495S} Smith, M.~D., \& Brand, P.~W.~J.~L.\ 1990, \mnras, 242, 495
 \bibitem[Smith 
\& Rosen(2003)]{2003MNRAS.339..133S} Smith, M.~D., \& Rosen, A.\ 2003, \mnras, 339, 133
 \Ref Strelnitski, V.S. 1973, AZh, 50, 1133
 \Ref Strelnitski, V.S. 1980, PAZh, 6, 354
 \Ref Strelnitski, V.S. 1984, MNRAS, 207, 339
 \Ref Tarchi, A., Castangia, P., Columbano, A., Panessa, F., \& Braatz, J. A.
    2011, \aap, 532, A125
 \Ref Tarter, J. \& Welch, W.J. 1986, ApJ, 305, 469
 \Ref Tielens, A.G.G.M. \& Allamandola, L. 1987, in {\it Interstellar Processes},
 ed. D. Hollenbach \& H. Thronson, (Dordrect: Reidel), p.397
 \Ref Torrelles, J.M., Patel, N.A., Gomez, J.F., et al.  2001a, ApJ,
 560, 853
 \bibitem[Torrelles et al.(2001)]{2001Natur.411..277T} Torrelles, J.~M.,
 Patel, N.~A., G{\'o}mez, J.~F., et al.\ 2001b, \nat, 411, 277
 \bibitem[Uscanga et al.(2008)]{2008MNRAS.390.1127U} Uscanga, L., G{\'o}mez, Y., Raga, A.~C., et al.\ 2008, \mnras, 390, 1127 
 \Ref Walker, R.C., Matsakis, D.N., \& Garcia-Barreto, J.A. 1982, ApJ,
      255, 128
  \bibitem[Walsh et al.(2011)]{2011MNRAS.416.1764W} Walsh, A.~J., Breen, S.~L., Britton, T., et al.\ 2011, \mnras, 416, 1764 
 \bibitem[Welch et al.(1987)]{1987Sci...238.1550W} Welch, W.~J., Dreher, J.~W.,
    Jackson, J.~M., Terebey, S., \& Vogel, S.~N.\ 1987, Science, 238, 1550
 \bibitem[Yates et al.(1997)]{1997MNRAS.285..303Y} Yates, J.~A., Field, D., \&
     Gray, M.~D.\ 1997, \mnras, 285, 303


\end{thebibliography}
\end{document}